\documentclass[aps,prb,reprint,superscriptaddress]{revtex4-2}

\usepackage{graphicx}
\usepackage{times}
\usepackage{physics}
\usepackage{braket}
\usepackage{comment}
\usepackage{placeins}
\usepackage[title]{appendix}
\usepackage{dsfont}
\usepackage[dvipsnames]{xcolor}
\usepackage{enumitem}
\usepackage{soul}
\usepackage{relsize}
\usepackage{amsfonts}
\usepackage{amssymb}
\usepackage{amsmath}
\graphicspath{{"../figures/"}}
\definecolor{goodgreen}{rgb}{0.1,0.5,0}
\usepackage[colorlinks,urlcolor=goodgreen,citecolor=blue,linkcolor=goodred]{hyperref}
\definecolor{goodred}{rgb}{0.7,0,0}

\usepackage{hyperref}
\newcommand{\orcid}[1]{\href{https://orcid.org/#1}{\includegraphics[width=8pt]{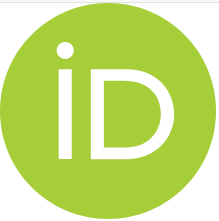}}}
\begin{document}

\title{Sub-Thouless energy scale in interface-dominated unconventional superconductor junctions}

\author{Léon Goedegebuur\orcid{0009-0001-0413-5418}}
\email{l.goedegebuur@utwente.nl}
\affiliation{MESA+ Institute for Nanotechnology, University of Twente, The Netherlands}
\author{Tim Kokkeler\orcid{0000-0001-8681-3376}}
\email{tim.h.kokkeler@jyu.fi}
\affiliation{Department of Physics and Nanoscience Center, University of Jyväskylä, Finland}
\author{Yukio Tanaka\orcid{0000-0003-1537-4788}}
\email{tanaka.yukio.j2@f.mail.nagoya-u.ac.jp}

\affiliation{Department of Applied Physics, Nagoya University, 464-8603 Nagoya, Japan}
\affiliation{Research Center for Crystalline Materials Engineering, Nagoya University, 464-8603 Nagoya Japan}

\begin{abstract}

\noindent We derive the conductance in unconventional superconductor - normal metal - normal metal junctions, taking into account the resistance between the normal metals. 
We show that applying Tanaka--Nazarov boundary conditions at both superconductor and normal metal interfaces reveals a new characteristic low-energy scale which is sub-Thouless in the experimentally relevant strong-interface regime. This scale governs, among other quantities, the curvature of the local density of states (LDOS) and the width of conductance dips near the zero-bias conductance peak in triplet superconductor junctions, thereby replacing the Thouless energy as the dominant low-energy parameter. We derive universal $s$-wave proximitized diffusive metal expressions for the zero-bias LDOS and differential conductance which solely depend on the interface and bulk resistances, independent of the superconducting gap and Thouless energy scale. This framework provides a possible origin of the anomalous low-bias conductance features observed in unconventional superconductors, offering a refined mechanism for interpreting interface and correlated electron physics in experimental hybrid junctions.

\end{abstract}
\maketitle
\section{Introduction}
\label{sec:intro}
\noindent A standard experimental probe for unconventional superconductivity is the conductance of a 
normal-metal/superconductor junctions~\cite{blonder1982transition, TanakaKashiwaya1995}. The differential conductance of such junctions can reveal features of the pairing symmetry through characteristic features such as a zero-bias conductance peak (ZBCP)~\cite{TanakaKashiwaya1995, kashiwaya2000tunnelling} due to surface Andreev bound states (SABS) \cite{hu1994midgap}, whose existence and dispersion carry information about the structure of the pair potential. For example, for one-dimensional $p$-wave superconductors the zero-energy surface Andreev bound states (ZESABS) are dispersionless and the ZBCP is sharp, while for
helical $p$-wave superconductors the SABS disperse linearly, and the ZBCP has a dome-like profile~\cite{yamashiro1998theory, Iniotakis2007, TanakaYokoyamaBalatskyNagaosa2009}. These
distinctions have been studied in detail
theoretically~\cite{TanakaNazarov, TanakaKashiwaya2004, Kokkeler2023, tanaka2005theory}.
Additional bumps and modulations in the conductance on the scale of the
Thouless energy $E_{\text{Th}} = \hbar D/L^2$ are well understood within
the quasiclassical proximity-effect framework~\cite{golubov2005density,
Kokkeler2023, TanakaKashiwaya2004}. Yet experiment continues to pose new puzzles. In
epitaxial CoSi$_2$/TiSi$_2$ junctions, conductance minima flanking the ZBCP on an energy scale significantly smaller than the Thouless energy have been observed~\cite{chiu2021observation}. Similar sub-Thouless low-bias features have been reported in
point-contact junctions on UTe$_2$~\cite{MonctonUTe22024}, a
candidate bulk $p$-wave
superconductor~\cite{AokiIshidaFlouquet2019, RanUTe22019} and in
layered AuSn$_4$ where two-fold symmetric ZBCPs have been attributed to
Rashba-driven $s+p$-wave mixing~\cite{AuSn42023}. These additional
conductance minima and sub-Thouless structures are not consistent with existing theoretical predictions and suggest that a degree of freedom is missing from the standard treatment.\\

\noindent This prompts a closer examination of the transport equations describing
these junctions. These equations are typically formulated within the
quasiclassical theory of superconductivity, which reduces the
microscopic Gor'kov equations~\cite{Gorkov} to a tractable set of
equations for an energy-resolved Green's function, which in the clean limit are called the
Eilenberger equation~\cite{Eilenberger,larkin1975nonlinear} and, in the dirty limit, the
Usadel equation~\cite{usadel1970generalized} (see
Refs.~\cite{Belzig1999, chandrasekhar2008proximity} for reviews). The
extension to Keldysh space~\cite{Keldysh, SchmidSchon1975,
HeikkilaSilaevVirtanenBergeret2019} allows for the computation of
nonequilibrium quantities such as the differential conductance. For
hybrid structures the quasiclassical equations need to be supplemented
by boundary conditions to describe the interfaces. For junctions with
unconventional superconductors, the Tanaka--Nazarov boundary
conditions~\cite{TanakaKashiwaya2004} are typically used. They
generalize the Nazarov circuit theory~\cite{Nazarov1999} to incorporate
angle-resolved and triplet pair potentials as well as their resulting
surface Andreev bound states~\cite{kashiwaya1995origin,
TanakaKashiwaya2004}.\\

\noindent Within this framework, recent work has explored the proximity effect in junctions with a variety of unconventional superconductors~\cite{KanasugiYanase2022,
TanakaKokkelerGolubov2022, kokkeler2023spin, Kokkeler2023,
kokkeler2024countingstatisticsunconventionalsuperconductor}. A common
simplification in these treatments is the use of a continuity condition
at the normal-metal lead~\cite{kuprianov1988influence, Kokkeler2023},
which treats the normal contact as a perfect reservoir with vanishing
interface resistance. In many standard SN junction geometries this is
well justified: the diffusive region and the normal reservoir are often
the same material, at times simply a single metal with a constriction or
a lateral protrusion serving as a weak link. In such cases there is no
physical interface on the normal side and the continuity condition serves
as a good approximation. \\

\noindent The situation changes when the junction is fabricated from structurally different materials or when the diffusive region is grown as a bridge on top of two separate electrodes. In these devices the normal-side contact is formed across a grain boundary or a lattice-mismatched interface and the resulting boundary resistance can be comparable to or larger than the bulk resistance of the diffusive region. The CoSi$_2$/TiSi$_2$ junctions of Ref. \cite{chiu2021observation} are a concrete example. As shown by the SEM images in the Supplementary Materials of \cite{chiu2021observation}, near the CoSi$_2$ interface the TiSi$_2$ layer forms a flat-surface region that changes into a ripple-like surface morphology further away from the contact, i.e. the CoSi$_2$ interface \cite{ChiuLinPrivate}. Such a structural transition is indicative of strain grain boundaries and especially boundary grooving (which is a precursor state of the agglomeration effect during high-temperature formation of the TiSi$_2$ electrode, see Figs. S1E and S1F in Ref. \cite{chiu2021observation}), which give rise to a significant boundary resistance on the normal side \cite{ChiuLinPrivate}. In such systems, the continuity approximation cannot be used. \\

\begin{figure}[h]
  \centering
  \includegraphics[width=1\linewidth]{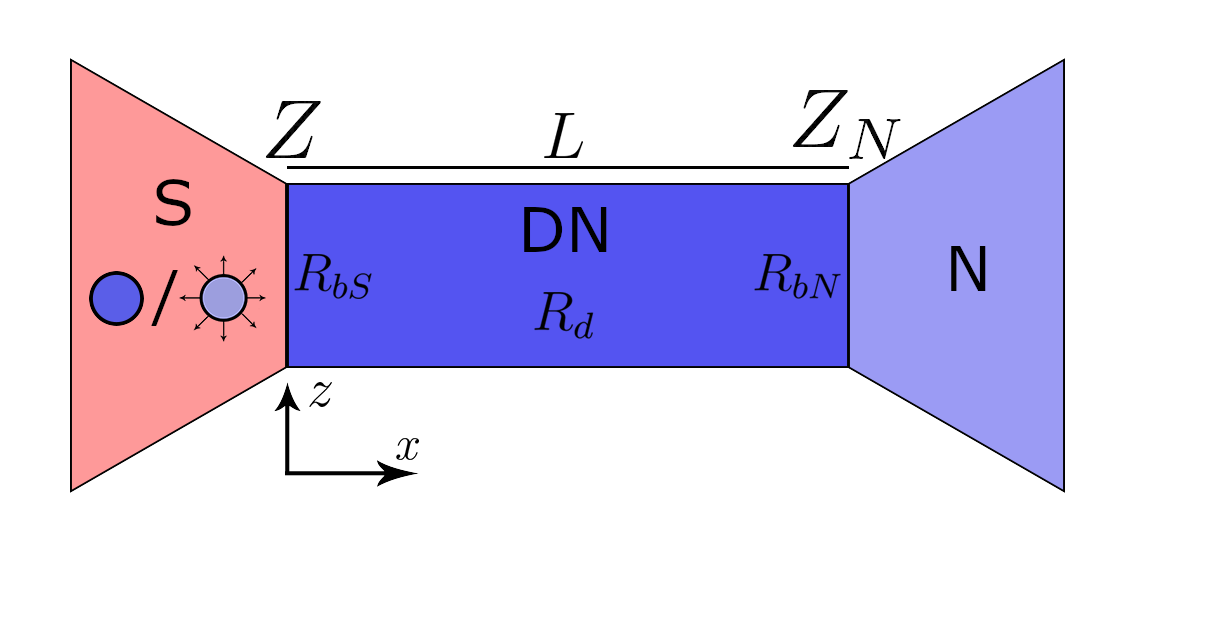}
  \caption{Schematic of the S/DN/N junction considered in
  this work. The superconducting electrode (left) carries a general pair
  potential $\hat{\Delta}$, here illustrated as a singlet $s$-wave or helical $p$-wave state. The central diffusive bar (middle) is a
  disordered normal metal of length $L$. The right contact is a ballistic
  normal-metal electrode held at bias voltage $V$. The S/DN and DN/N interfaces are characterized by interface resistances $R_{\text{bS}}$ and $R_{\text{bN}}$ and Blonder-Tinkham-Klapwijk parameters $Z$ and $Z_N$ respectively.}
  \label{fig:SFN junction}
\end{figure}

\noindent In this work we incorporate a finite boundary resistance with
the normal-metal electrode by applying the Tanaka--Nazarov boundary
conditions at both interfaces of the superconductor / dirty normal metal
/ normal metal electrode (S/DN/N) junction, illustrated in
Fig.~\ref{fig:SFN junction}. The diffusive region is characterized by its
resistance $R_d$ and by the Thouless energy $E_{\text{Th}} = \hbar D/L^2$,
set by the diffusion constant $D$ and the length $L$ of the bar. Each interface carries a boundary resistance: $R_{\text{bS}}$ at the S/DN interface and $R_{\text{bN}}$ at the DN/N interface. We show that incorporating a finite normal metal interface resistance reveals a characteristic energy
\begin{equation}
\frac{E^\ast}{\Delta_0}
= \sqrt{\frac{2}{3}}\;
\frac{R_{\text{bN}}^{2}+R_{\text{bS}}^{2}}
     {R_\text{bN}^2}\;
\frac{E_{\text{Th}}/\Delta_0}
     {R_{\text{bS}}/R_d+E_{\text{Th}}/\Delta_0}\;.
\end{equation}
This scale lies well below the Thouless energy when both boundary resistances exceed the bulk resistance $R_\text{d}$ of the diffusive region and the normal-side boundary resistance is not much smaller than the superconducting-side one. The various energy scales of the hybrid junction, and the roles they play,
are collected in Table~\ref{tab:energy_scales}. \\

\begin{table}[t]
\centering
\caption{Relevant energy scales in S/DN/N junctions.}
\label{tab:energy_scales}
\begin{tabular}{l p{7.0cm}}
\hline\hline
Scale & Role \\
\hline
$\Delta_0$  &
    Superconducting gap and upper energy scale for subgap features.
    The coherence peak occurs at $|eV|=\Delta_0$ for both $s$- and (helical) $p$-wave pairing. \\[6pt]
$E_{\text{Th}}$ &
    Thouless energy of the diffusive region. Below $E_{\text{Th}}$,
    the energy-dependent coherence length $\xi_E=\sqrt{\hbar D/|E|}$ exceeds
    the junction length, so superconducting pair amplitudes can extend
    across the entire diffusive region.  \\[6pt]
$E^\ast$  &
    Width of the sub-Thouless
    spectral feature: proximity dip
    ($s$-wave) or ZESABS-induced ZBCP (odd-parity). \\[6pt]
$k_B T$  &
    Thermal effects smear features below $k_B T$.
    Low-energy structure $E_{\text{eff}}$ survives
    only when $E_{\text{eff}} \gtrsim k_B T$. \\
$T_1 \Delta_0$  &
    Energy scale associated with the smearing of a conductance peak
    arising from a single SABS, set by the transparency $T_1$ of the dominant transport channel. \\[6pt]
\hline\hline
\end{tabular}
\end{table}

\noindent In the remainder of this work we explore the effect of the finite interface resistance of the normal electrode on the conductance in a normal metal contacted by a superconductor and a normal metal electrode. We begin with an analytical treatment of the
$s$-wave case in the short-junction, strong-interface regime, deriving
closed-form expressions for the retarded Green's function, the
low-energy LDOS and the characteristic scale $E^\ast$. We then turn to
numerical results for helical $p$-wave and three-dimensional
Balian--Werthamer pairings~\cite{balian1963superconductivity} and close with a
direct comparison to recent experimental conductance data on a triplet
superconductor candidate~\cite{chiu2021observation}.

\section{Model and formalism}
\label{sec:theory}
\subsection{Model and formalism}
\label{sec:model}

\noindent We consider a one-dimensional junction consisting of an unconventional superconductor ($S$), a diffusive normal metal bar ($DN$) of length $L$ and a normal metal electrode ($N$), stacked along the transport axis $x$ with the diffusive region occupying $0 \le x \le L$. The voltage difference between the superconductor and normal metal electrode is denoted by $V$ and the diffusive bar is assumed to be in the dirty limit, $\ell_e \ll \xi$, where $\ell_e$ is the elastic mean-free path and $\xi \sim \sqrt{\hbar D/\Delta_0}$ the superconducting coherence length. In this regime the quasiclassical Green's function formalism, originating from the coarse-graining of the microscopic Gor'kov equations~\cite{Gorkov} through the Eilenberger~\cite{Eilenberger} and Usadel~\cite{usadel1970generalized} equations, provides the natural language for the problem~\cite{Belzig1999, chandrasekhar2008proximity}. The junction geometry corresponds to Fig.~\ref{fig:SFN junction} and its parameters are summarized in Table~\ref{tab:params}.

\begin{table}[h]
  \centering
  \renewcommand{\arraystretch}{1.25}
  \begin{tabular}{ll}
    \hline\hline
    Symbol & Meaning \\
    \hline
    $E_{\text{Th}} = \hbar D/L^2$ & Thouless energy of the diffusive bar \\
    $R_d$ & Normal-state bulk resistance of the diffusive region \\
    $R_{bS}$, $R_{bN}$ & Interface resistances at S/DN, DN/N \\
    $\gamma_{\text{BS}} = R_{bS}/R_d$ & Dimensionless S/DN resistance ratio \\
    $\gamma_{\text{BN}} = R_{bN}/R_d$ & Dimensionless DN/N resistance ratio \\
    $z, z_N$ & BTK barrier strength at S/DN and DN/N \\
    $r = \Delta_t/\Delta_s$ & Triplet-to-singlet mixing parameter \\
    $\chi_t$ & Relative phase between $\Delta_s$ and $\Delta_t$ \\
    \hline\hline
  \end{tabular}
  \caption{Physical and dimensionless parameters used throughout this work.}
  \label{tab:params}
\end{table}

\noindent The  quasiclassical Green's function $\bar{G}(x, E)$ takes values in the tensor product of Keldysh, Nambu and spin spaces. Because of causality it has an upper %
triangular form in Keldysh space~\cite{Keldysh, SchmidSchon1975, HeikkilaSilaevVirtanenBergeret2019},
\begin{equation}
\bar{G} = \begin{pmatrix} \check{G}^\text{R} & \check{G}^\text{K} \\ 0 & \check{G}^\text{A} \end{pmatrix},
\label{eq:Gbar}
\end{equation}
where the retarded ($\check{G}^\text{R}$) and advanced ($\check{G}^\text{A}$) components encode the spectral properties and the Keldysh component ($\check{G}^\text{K}$) encodes the occupation of these states, given by the temperature-dependent equilibrium distribution in equilibrium and by nonequilibrium distribution functions when a bias is applied. The retarded and advanced components obey the nonlinear normalization $(\check{G}^{R,A})^2 = \check{1}_4$, with the advanced component related to the retarded one by the chronology symmetry
\begin{equation}
\check{G}^\text{A} = -\tau_3\,(\check{G}^\text{R})^\dagger\,\tau_3,
\label{eq:GA_GR}
\end{equation}
where $\tau_i$ denote Pauli matrices in Nambu space. The Keldysh component is  determined by $\check{G}^{R,A}$ and a distribution-function matrix $\check{f}$,
\begin{equation}
\check{G}^\text{K} = \check{G}^\text{R}\,\check{f} - \check{f}\,\check{G}^\text{A},
\label{eq:GK_param}
\end{equation}
with, in the spin-resolved case~\cite{HeikkilaSilaevVirtanenBergeret2019},
\begin{equation}
\check{f} = f_L \otimes \hat{1}_4 + f_T \otimes \tau_3 + \sum_{i=1}^3 (f_{Ti} + f_{Li}\tau_3) \otimes \sigma_i,
\label{eq:fcheck}
\end{equation}
where $\sigma_i$ are the Pauli matrices in spin space. Here $f_L$ (longitudinal) represents the energy-symmetric part associated with heating or effective temperature, while $f_T$ (transverse) describes particle–hole imbalance. For the spin components the labelling follows the same convention:
$f_{Ti}$ correspond to spin imbalance (spin-polarized occupation), whereas $f_{Li}$ describe spin-energy modes
linked to spin-dependent heating~\cite{HeikkilaSilaevVirtanenBergeret2019}. In thermal equilibrium all transverse components as well as $f_{Li}$ vanish and
$f_L = \tanh{\frac{E}{2k_BT}}$ is related to the Fermi function. Out of equilibrium, these distribution components evolve according
to kinetic equations derived from the Keldysh part of the quasiclassical equations~\cite{Belzig1999, Yanase2008}. In the dirty limit, $\bar{G}$ satisfies the Usadel equation,
\begin{equation}
D\,\partial_x(\bar{G}\,\partial_x\bar{G}) + i\,[E\tau_3,\,\bar{G}] = 0.
\label{eq:usadel}
\end{equation}
The electric current through the junction is obtained from the Keldysh component,
\begin{equation}
I = \frac{\sigma_D A}{16e}\int_{-\infty}^{\infty} dE\;\operatorname{Tr}\!\left[\tau_3\,(\check{G}\,\partial_x\check{G})^\text{K}\right],
\label{eq:current}
\end{equation}
where $\sigma_D=2e^2N_0D$ is the Drude conductivity, with $N_0$ being the density of states per spin, and $A$ the cross section of the diffusive bar. Throughout this work $\sigma_\text{N}=1/R_\text{N}$ denotes the normal-state conductance of the  junction, with $R_\text{N}=R_\text{bS}+R_\text{d}+R_\text{bN}=R_\text{d}(1+\gamma_{\text{BS}}+\gamma_{\text{BN}})$ and $R_\text{d}=L/(\sigma_DA)$,

\subsection{Interface boundary conditions}
\label{sec:bc}

\noindent The defining element of our model is the application of the Tanaka--Nazarov boundary conditions at both the S/DN and the DN/N interfaces. This lifts the asymmetric treatment employed in the bulk of the literature~\cite{Kokkeler2023, golubov2005density, TanakaKashiwaya2004}, which typically reduces the $N$-side interface to the continuity limit~\cite{kuprianov1988influence}.

\noindent At the $S/DN$ interface ($x=0$) the boundary condition reads
\begin{equation}
\check{G}\,\partial_x\check{G}\bigg|_{x=0} = \frac{1}{\gamma_{\text{BS}}\,L\langle \tilde{T}\rangle}\,\bigl\langle\check{S}_S(\phi)\bigr\rangle\;,
\label{eq:bc_S}
\end{equation}
where $\gamma_{\text{BS}} = R_{\text{bS}} / R_\text{d}$ is the ratio of the S/DN interface resistance and the resistance of the DN, the angular bracket denotes averaging over the injection angle $\phi$, i.e. $\langle \cdot \rangle = \frac{1}{2}\int_{-\pi/2}^{\pi/2} (\cdot)\, \cos\phi\, d\phi$ and the matrix $\check{S}_S(\phi)$ encodes the angle-resolved scattering into the superconducting electrode,
\begin{equation}
\check{S}_S(\phi) = T_1\bigl(1 + T_1^2 + T_1[\check{C}_S,\check{G}]_+\bigr)^{-1}\,[\check{C}_S,\check{G}]_-\;,
\label{eq:SS}
\end{equation}
with $[\cdot\,,\cdot]_\pm$ denoting anticommutator ($+$) and commutator ($-$). The effective transparency $\tilde{T}(\phi)$ follows from the Blonder--Tinkham--Klapwijk (BTK) model~\cite{blonder1982transition},
\begin{equation}
\tilde{T}(\phi) = \frac{\cos^2\phi}{\cos^2\phi + z^2}\;,
\label{eq:T_BTK}
\end{equation}
with $z$ the dimensionless barrier-strength parameter and $T_1 = \tilde{T}/(2-\tilde{T}+2\sqrt{1-\tilde{T}})$. The matrix $\check{C}_S$ is built from the bulk Green's function $\check{G}_S$ of the superconductor through
\begin{equation}
\check{C}_S = \check{H}_{S,+}^{-1}\,(\check{1} - \check{H}_{S,-}),\qquad
\check{H}_{S,\pm} = \tfrac{1}{2}\bigl[\check{G}_S^s(\phi) \pm \check{G}_S^s(\pi-\phi)\bigr],
\label{eq:CS}
\end{equation}
and is thus determined by the pairing symmetry~\cite{TanakaKokkelerGolubov2022, TanakaKashiwaya2004}. Here, we focus on \textit{s}-wave superconductors, for which $\check{G}_S(\phi)$ takes the angle-independent BCS form, as well as \textit{p}-wave superconductors with pair potential $\check{\Delta}(\phi)\vec{d}(\phi)\cdot\vec{\sigma}$, in which case  
\begin{equation}
\check{G}_\text{S}(\varphi)=\frac{1}{\sqrt{\Delta_0^2-E^2}}\Bigl(-iE\,\tau_3+\Delta_0\,\mathbf{d}(\varphi)\cdot\boldsymbol{\sigma}\,\tau_1\Bigr)\;.
\end{equation}

\noindent At the DN/N interface ($x=L$) we apply the analogous Tanaka--Nazarov boundary condition, which in this case reduces to the Nazarov boundary condition,
\begin{equation}
\check{G}\,\partial_x\check{G}\bigg|_{x=L} = -\,\frac{1}{\gamma_{\text{BN}}\,L}\,\bigl\langle\check{S}_N(\phi)\bigr\rangle,
\label{eq:bc_N}
\end{equation}
where $\gamma_{\text{BN}} = R_{\text{bN}}/R_\text{d}$ is the ratio of the DN/N interface resistance and the resistance of the DN. The scattering matrix has the same structure as~\eqref{eq:SS},
\begin{equation}
\check{S}_N(\phi) = 
T_{1,N}\,\bigl(1 + T_{1,N}^2 + T_{1,N}[\check{G}_N,\check{G}]_+\bigr)^{-1}\,[\check{G}_N,\check{G}]_-\;,
\label{eq:SN}
\end{equation}
with $\tilde{T}_N(\phi) = \cos^2\phi/(\cos^2\phi + z_N^2)$ the BTK transparency at barrier strength $z_N$ and $T_{1,N} = \tilde{T}_N/(2 - \tilde{T}_N + 2\sqrt{1 - \tilde{T}_N})$. Because the normal electrode has no pairing and is isotropic, its bulk Green's function $\check{G}_N = \tau_3 \otimes \hat{1}_{\text{spin}}$ is momentum independent, so the connector reduces to $\check{C}_N = \check{G}_N$ directly and no angular decomposition of the form in Eq.~\eqref{eq:CS} is needed. The minus sign difference between Eq.~\eqref{eq:bc_S} and Eq.~\eqref{eq:bc_N} reflects the outward-normal convention at the right interface: the normal direction at $x=0$ points along $-\hat{\mathbf{x}}$ (outward from the bar), while at $x=L$ it points along $+\hat{\mathbf{x}}$.
In Eq.~\eqref{eq:bc_N} the scattering matrix $\check{S}_N(\phi)$ has the same structural form as Eq.~\eqref{eq:SS}, with $\check{C}_S$ replaced by the bulk Green's function of the normal reservoir, $\check{G}_N = \tau_3\otimes\hat{1}_{\text{spin}}$. \\

\subsection{Unconventional pair potentials}
\label{sec:pairpotential}
 
\noindent To describe unconventional superconducting electrodes we take a spin-triplet pair potential
\begin{equation}
\hat{\Delta}(\varphi) = \Delta_0\,\mathbf{d}(\varphi)\cdot\boldsymbol{\sigma}\;,
\label{eq:pairpotential}
\end{equation}
where $\mathbf{d}$ is the $d$-vector encoding the spin structure of the triplet pair potential as a function of the direction of the quasiparticle momentum, and $\boldsymbol{\sigma}$ are the Pauli matrices in spin space~\cite{TanakaKashiwaya2004,Kokkeler2023}. For the three-dimensional state, $\theta$ denotes the polar angle between the momentum and the interface normal and $\varphi$ the azimuthal angle around it. For the two-dimensional helical state momentum is confined to a plane and $\varphi$ denotes the angle between the momentum and the interface normal. Two odd-parity states appear in this work,
\begin{align}
    \text{Helical:}\quad
        & \mathbf{d}(\varphi)
          = \cos\varphi\,\hat{\boldsymbol{a}}
          + \sin\varphi\,\hat{\boldsymbol{b}}\;, \label{eq:dvectorhelical}\\
    \text{3D BW:}\quad
        & \mathbf{d}(\varphi)
          = \sin\theta \cos \varphi \,\hat{\boldsymbol{a}} + \sin \theta \sin \varphi \,\hat{\boldsymbol{b}} + \cos \theta \,\hat{\boldsymbol{c}}\;,\label{eq:dvectorBW}
\end{align}
with $\hat{\boldsymbol{a}}$, $\hat{\boldsymbol{b}}$ and $\hat{\boldsymbol{c}}$ mutually orthogonal unit vectors in spin space. In both states the $d$-vector is parallel to the momentum, with the component along the interface normal given by $\cos\varphi\,\hat{\boldsymbol{a}}$ for the helical state and by $\cos\theta\,\hat{\boldsymbol{c}}$ for the BW state. The orientation of the spin axes $\hat{\boldsymbol{a}}$, $\hat{\boldsymbol{b}}$ and $\hat{\boldsymbol{c}}$ relative to the interface is irrelevant for the present results. Neither the diffusive bar nor the normal electrode contains spin-orbit coupling or magnetism, so that a global rotation in spin space leaves all observables unchanged and only the relative orientation of $\mathbf{d}$ at different momenta matters.
Both the helical and BW states preserve time-reversal and are topological~\cite{Topology1,tanaka2024theory}. Their gap magnitude $|\mathbf{d}(\varphi)| = \Delta_0$ is isotropic, yet specular reflection at the interface reverses the normal component of the momentum, $\varphi\to\pi-\varphi$, so that the reflected quasiparticle experiences $\mathbf{d}(\pi-\varphi)$ instead of $\mathbf{d}(\varphi)$. At normal incidence this reverses the sign of the pair potential and produces a ZESABS, while at oblique incidence the bound state lies at the finite energy $\pm\Delta_0\sin\varphi$.
 
\noindent The three-dimensional Balian--Werthamer state is the equilibrium phase of superfluid $^3$He~\cite{balian1963superconductivity} and is a fully gapped, odd-parity order parameter with $\mathbf{d}$ pointing along the  momentum direction on the entire Fermi sphere~\cite{sigrist1991phenomenological}, $\mathbf{d}(\vec{k}) = \hat{\boldsymbol{k}}$. The helical $p$-wave state can be seen as a two-dimensional variant of the BW state and the two share many properties \cite{BuchholtzZwicknagl1981}. Both host a ZESABS at normal incidence that contributes to the zero-bias conductance. The SABS at oblique incidence lie at nonzero energy and contribute at finite bias~\cite{YamakagePhysRevB, asano2003theoretical}.  \\
 
\noindent The difference between the two lies in the fraction of transport channels that host the zero-energy state. In the two-dimensional helical case the angular measure $d\varphi$ is the same for all angles, so the relative importance of each channel is set by the transparency alone, which favors normal incidence; helical $p$-wave superconductors therefore feature a dome-shaped zero-bias conductance peak (ZBCP)~\cite{yamashiro1998theory, TanakaKashiwaya2004}. In the three-dimensional BW case the solid-angle element $\sin\theta\,d\theta\,d\varphi$ vanishes at normal incidence, so the normal-incidence channels that carry the ZESABS occupy a vanishing fraction of the Fermi surface. The zero-energy contribution is therefore suppressed relative to the finite-energy states, which can turn the ZBCP into a zero-bias conductance dip~\cite{kokkeler2023proximity}.

\section{Analytical results: $s$-wave S(DN)N junction}
\label{sec:analytics}
\noindent The diffusive normal region between two resistive interfaces is a
double-barrier structure and several features of its low-energy transport
are known from earlier work. The finite-transparency boundary condition
that governs each interface was formulated in Nazarov's circuit theory of
Andreev reflection~\cite{Nazarov1994, Nazarov1999} and later generalized to include unconventional superconductors \cite{tanaka2004theory}, while the nonequilibrium
proximity effect in double-barrier superconducting junctions was analysed
by Volkov et al.~\cite{VolkovZaitsevKlapwijk1993}. The
zero-bias conductance of two barriers in series and its dependence on the
ratio of the two barrier resistances alone is a result of the
random-matrix theory of quantum transport~\cite{Beenakker1997}. Our purpose in this section is to
use the closed-form solution of the double-barrier Usadel equation to
extract the characteristic energy scale $E^\ast$ that controls the
low-energy spectral and transport features of the S/DN/N junction and to
establish the width--amplitude factorisation that carries over to the
unconventional-pairing results of Secs.~\ref{sec:helical_p}
and~\ref{sec:3DBW}, which are the central results of this work.

\subsection{Closed-form retarded Green's function and LDOS}
\label{sec:gR_closed}

\noindent To gain insight into
the spectral properties of the junction, we seek analytical expressions for the retarded Green's function in an S(DN)N junction with \textit{s}-wave superconductors. Since the superconductor has purely singlet pairing and no exchange field is present, all Green's functions are diagonal in spin space and the problem reduces from the Keldysh--Nambu--spin structure ($\check{G}$) to a Keldysh--Nambu structure ($\bar{G}$). The retarded sector then reduces further to a $2 \times 2$ Nambu matrix ($\hat{g}^\text{R}$), which is the object we work with below. We work in the dominant-interface limit, in which both boundary resistances are large compared with the resistance of the bar, $\gamma_{\text{BS}},\gamma_{\text{BN}}\gg1$, and we restrict to low quasiparticle energies $|E|\ll\Delta_0$. In this limit the Green's function changes mainly at the two interfaces, while its variation inside the bar is smaller by a factor of order $1/\gamma_{\text{B}}$. If in addition $|E|\ll E_{\text{Th}}$, the energy-dependent phase accumulated across the bar is small as well, so that the Green's function can be taken as position independent within the wire, $\bar{G}(E,x)\approx\bar{G}(E)$. Both conditions are required: for transparent interfaces the Green's function varies across the bar even at $E=0$. \\

\noindent Integrating the Usadel equation, Eq.~\eqref{eq:usadel}, over the length $L$ of the diffusive region and inserting the strong-interface ($z_{\text{S}}, z_{\text{N}} \gg 1$) Kupriyanov--Lukichev form of the Tanaka--Nazarov boundary conditions, Eqs.~\eqref{eq:bc_S} and~\eqref{eq:bc_N}, the boundary terms and the energy term combine into a single commutator condition on the position-independent retarded Green's function,
\begin{equation}
[\mathcal{M},\hat{g}^\text{R}]_-=0\;,\qquad
\mathcal{M}(E)\equiv\frac{1}{\gamma_{\text{BS}}}\hat{G}_S^\text{R}+\frac{1}{\gamma_{\text{BN}}}\hat{G}_N^\text{R}-i\frac{E}{E_{\text{Th}}}\tau_3\;,
\label{eq:M_def}
\end{equation}
where $\hat{g}^\text{R}$ is the $2 \times 2$ Nambu retarded Green's function and $E_{\text{Th}} = \hbar D/L^2$. The details of this reduction are given in Appendix~\ref{app:LDOS_cartesian}. The normalization $(\hat{g}^\text{R})^2 = \hat{1}$ fixes the solution to be proportional to $\mathcal{M}$,
\begin{equation}
\hat{g}^\text{R}(E)=\frac{\mathcal{M}(E)}{\sqrt{\tfrac{1}{2}\operatorname{Tr}\mathcal{M}^2(E)}}\;,
\label{eq:gR_formal}
\end{equation}
with the branch selected by $\operatorname{Re}\operatorname{Tr}\tau_3\hat{g}^\text{R} \geq 0$. In the low-energy window $|E| \ll \Delta_0$ the superconducting reservoir reduces to $\hat{G}_S^\text{R} \simeq -i(E/\Delta_0)\tau_3 + \tau_1$ and the normal reservoir is $\hat{G}_N^\text{R} = \tau_3$, so that $\mathcal{M} = \mathcal{G}(E)\,\tau_3 + \gamma_{\text{BS}}^{-1}\tau_1$ with the complex scalar
\begin{equation}
\mathcal{G}(E)\equiv\frac{1}{\gamma_{\text{BN}}}-i\!\left(\frac{1}{\gamma_{\text{BS}}}\frac{E}{\Delta_0}+\frac{E}{E_{\text{Th}}}\right)\;.
\label{eq:zeta_def}
\end{equation}
Its imaginary part collects the two contributions, the finite dwell time in the bar ($\propto 1/E_{\text{Th}}$) and the energy dependence of the S reservoir ($\propto 1/(\gamma_{\text{BS}}\Delta_0)$). Since $\mathcal{M}^2 = (\mathcal{G}^2 + \gamma_{\text{BS}}^{-2})\hat{1}$, the LDOS follows directly from the $\tau_3$ component of Eq.~\eqref{eq:gR_formal},
\begin{equation}
\frac{N(E)}{N_0}=\tfrac{1}{2}\operatorname{Re}\operatorname{Tr}\!\left[\tau_3\,\hat{g}^\text{R}(E)\right]
= \operatorname{Re}\!\left[\frac{\mathcal{G}(E)}{\sqrt{\mathcal{G}^2(E)+\gamma_{\text{BS}}^{-2}}}\right]\;.
\label{eq:LDOS_compact}
\end{equation}
\noindent Setting $E=0$ gives $\mathcal{G}(0)=\gamma_{\text{BN}}^{-1}$ and hence
\begin{equation}
\frac{N(0)}{N_0}
=\frac{\gamma_{\text{BN}}^{-1}}{\sqrt{\gamma_{\text{BN}}^{-2}+\gamma_{\text{BS}}^{-2}}}
=\frac{\gamma_{\text{BS}}}{\sqrt{\gamma_{\text{BS}}^{2}+\gamma_{\text{BN}}^{2}}}\;.
\label{eq:LDOS_zero}
\end{equation}
Equation~\eqref{eq:LDOS_zero} interpolates between a gapped spectrum, $N(0)/N_0= 0$, for a transparent S-interface ($\gamma_{\text{BS}} = 0$) and the bare normal-metal result, $N(0)/N_0 = 1$, for a transparent (DN)N-interface ($\gamma_{\text{BN}} = 0$), the latter being the continuity limit employed in earlier works~\cite{Kokkeler2023, golubov2005density, tanaka2024theory}. Between these limits $N(0)$ is a continuous function of the interface-resistance ratio alone: neither $\Delta_0$ nor $E_{\text{Th}}$ enters Eq.~\eqref{eq:LDOS_zero}. The independence of $\Delta_0$ follows from the fact that the $E = 0$ LDOS of the bulk superconductor vanishes for any finite gap, so the pair-breaking scale drops out at zero energy. The independence of $E_{\text{Th}}$ (and thereby of $L$) has a related origin: $E_{\text{Th}}$ enters only through the energy term of the Usadel equation, which vanishes at $E=0$. Within the position-independent approximation only the two boundary conductances then remain, so that in the absence of inelastic scattering the zero-energy LDOS of the S/DN/N junction depends only on their ratio.

\subsection{Low-energy expansion and characteristic energy scale}
\label{sec:expansion}

\noindent Expanding Eq.~\eqref{eq:LDOS_compact} to order $E^2$ around zero energy illuminates whether the density of states has a zero energy peak or dip as well as its width. The expansion, carried out in full in Appendix~\ref{app:LDOS_cartesian}, gives
\begin{equation}
\frac{N(E)}{N_0}
=\frac{N(0)}{N_0}\bigl[1+\Big(\frac{E}{E^*}\Big)^2\bigr]+\mathcal{O}(E^4)\;,
\label{eq:LDOS_O2}
\end{equation}
with, to leading order in $E_{\text{Th}}/(\gamma_{\text{BS}}\Delta_0)$
\begin{equation}
\frac{E^\ast}{\Delta_0}
= \sqrt{\frac{2}{3}}\;
\frac{\gamma_{\text{BN}}^{2}+\gamma_{\text{BS}}^{2}}
     {\gamma_{\text{BN}}^{2}}\;
\frac{E_{\text{Th}}/\Delta_0}
     {\gamma_{\text{BS}}+E_{\text{Th}}/\Delta_0}.
\label{eq:Estar_general}
\end{equation}
Thus, the DOS has a minimum at zero energy, with a width of order $E^*$.
In the large symmetric barrier limit $\gamma_{\text{BS}}=\gamma_{\text{BN}}\equiv\gamma_B\gg E_\text{Th}/\Delta_0$ (i.e.\ $R_{\text{bN}} = R_{\text{bS}} \equiv R_b \gg R_d$) this collapses to the compact form
\begin{equation}
E^\ast\simeq\sqrt{\frac{8}{3}}\,\frac{E_{\text{Th}}}{\gamma_B}\sim\frac{R_d}{R_{b}}\,E_{\text{Th}}.
\label{eq:Estar_PIB}
\end{equation}
\noindent The physical meaning of $E^\ast$ follows from the dwell time of a quasiparticle in the bar. The Thouless energy $E_{\text{Th}}=\hbar D/L^2$ is the inverse time needed to diffuse across the bar once. With resistive interfaces a quasiparticle is reflected many times before it escapes into one of the electrodes, and for $\gamma_{\text{BS}}=\gamma_{\text{BN}}\equiv\gamma_{\text{B}}$ its dwell time is enhanced by the ratio of the interface resistance to the bar resistance, $\tau_\text{dw}\sim\gamma_{\text{B}}\hbar/E_{\text{Th}}$. Up to the numerical prefactor, the symmetric result Eq.~\eqref{eq:Estar_PIB} is the corresponding inverse dwell time, $E^\ast\simeq\sqrt{8/3}\,E_{\text{Th}}/\gamma_{\text{B}}\sim\hbar/\tau_\text{dw}$: quasiparticles with $|E|\lesssim E^\ast$ remain phase coherent with their Andreev-reflected partners during their entire stay in the bar, so that the proximity effect, and with it the low-energy structure of the LDOS and the conductance, is confined to this energy window. This is the diffusive analogue of the tunneling model of the proximity effect, in which the induced gap in a thin normal layer is set by the escape rate through the barrier~\cite{McMillan1968}, and of the dwell-time scale of a cavity coupled to reservoirs through tunnel contacts~\cite{Beenakker1997}. For $\gamma_{\text{BN}}\gg\gamma_{\text{BS}}$ the quasiparticles escape mainly through the S interface, $\tau_\text{dw}\sim\gamma_{\text{BS}}\hbar/E_{\text{Th}}$, in agreement with Eq.~\eqref{eq:Estar_BNdom}. For $\gamma_{\text{BS}}\gg\gamma_{\text{BN}}$ this interpretation does not apply, since the proximity effect is then weak and $E^\ast$ only sets the curvature of a shallow dip, as discussed below.\\ 

\noindent Two additional limits of Eq.~\eqref{eq:Estar_general} are instructive to show how $E^{\ast}$ explicitly depends on either $\gamma_{\text{BN}}$ or $\gamma_{\text{BS}}$. When the DN/N interface is much more resistive than the S/DN interface ($\gamma_{\text{BN}}\gg\gamma_{\text{BS}}$), the $(\gamma_{\text{BN}}^2+\gamma_{\text{BS}}^2)/\gamma_{\text{BN}}^2$ prefactor reduces to unity and the crossover scale simplifies to
\begin{equation}
\frac{E^\ast}{\Delta_0}\simeq\sqrt{\frac{2}{3}}\;\frac{E_{\text{Th}}/\Delta_0}{\gamma_{\text{BS}}+E_{\text{Th}}/\Delta_0},
\qquad \gamma_{\text{BN}}\gg\gamma_{\text{BS}}.
\label{eq:Estar_BNdom}
\end{equation}
In this regime $E^\ast$ depends only on the S/DN interface transparency through $\gamma_{\text{BS}}$ and on the Thouless energy. In this case, the DN/N resistance is large enough such that the effect of the normal state electrode on the LDOS is negligible. For an opaque S/DN contact ($\gamma_{\text{BS}}\gg E_{\text{Th}}/\Delta_0$) the scale reduces to $E^\ast\simeq \sqrt{2/3}\,E_{\text{Th}}/\gamma_{\text{BS}}$. \\

\noindent In the opposite limit, where the S/DN interface dominates the resistance ($\gamma_{\text{BS}}\gg\gamma_{\text{BN}}$), the prefactor grows as $\gamma_{\text{BS}}^2/\gamma_{\text{BN}}^2$ and the crossover scale becomes
\begin{equation}
\frac{E^\ast}{\Delta_0}
\simeq\sqrt{\frac{2}{3}}\;\frac{E_{\text{Th}}/\Delta_0}{\gamma_{\text{BN}}^{2}}\gamma_{\text{BS}}\;,
\qquad \gamma_{\text{BS}}\gg\gamma_{\text{BN}}\;.
\label{eq:Estar_BSdom}
\end{equation}
 A more opaque normal contact (larger $\gamma_{\text{BN}}$) suppresses $E^\ast$ and confines the parabolic plateau to lower energies. Physically, the DN/N interface acts as a bottleneck that confines quasiparticles in the DN region, strengthening the proximity-induced gap suppression. Equation~\eqref{eq:Estar_BSdom} shows that this confinement effect can push $E^\ast$ well below the Thouless energy even when the S/DN contact itself is highly resistive.

\subsection{Differential conductance}
\label{sec:conductance}

\noindent The spectral functions of Sec.~\ref{sec:gR_closed} fix the local density of states, but charge transport requires the Keldysh component, which encodes not only the spectral structure but also the nonequilibrium quasiparticle distribution in the diffusive bar. We denote by $\hat{g}_3^\text{R}$ and $\hat{g}_1^\text{R}$ the $\tau_3$ and $\tau_1$ components of the retarded Green's function, Eq.~\eqref{eq:gR_formal}. Since $\mathcal{G}$ multiplies $\tau_3$ and $\gamma_{\text{BS}}^{-1}$ multiplies $\tau_1$ in the matrix $M$, they follow directly from the normalised form, Eqs.~(\ref{eq:gR_formal},\ref{eq:zeta_def}), as
\begin{equation}
\hat{g}_3^\text{R}(E)=\frac{\mathcal{G}(E)}{\sqrt{\mathcal{G}^2(E)+\gamma_{\text{BS}}^{-2}}}\;,\qquad
\hat{g}_1^\text{R}(E)=\frac{\gamma_{\text{BS}}^{-1}}{\sqrt{\mathcal{G}^2(E)+\gamma_{\text{BS}}^{-2}}}\;,
\label{eq:g13}
\end{equation}
with the branch of the square root fixed by the retarded condition $\mathrm{Re}\,\hat{g}_3^\text{R}>0$, so that $\bigl(\hat{g}_3^\text{R}\bigr)^2+\bigl(\hat{g}_1^\text{R}\bigr)^2=1$. The real part $\mathrm{Re}\,\hat{g}_3^\text{R}(E)=N(E)/N_0$ is the LDOS of Eq.~\eqref{eq:LDOS_compact}, while $\mathrm{Re}\,\hat{g}_1^\text{R}(E)$ is the corresponding spectral density of the induced pair amplitude. \\

\noindent The Keldysh Green's function is parametrised in terms of distribution functions as in Eq.~\eqref{eq:GK_param}. This separates $\bar{G}^\text{K}$ into the available states, described by the retarded and advanced Green's functions, and their occupation, described by a longitudinal ($f_L$) and a transverse ($f_T$) distribution function that represent the energy and charge nonequilibrium modes respectively. Solving the Keldysh part of the boundary-integrated Usadel equation, detailed in Appendix~\ref{app:keldysh}, yields two compact results in the subgap window $|E|<\Delta_0$. First, the superconducting reservoir injects no Keldysh weight, $\bar{G}_S^\text{K}=0$, because its retarded and advanced components coincide below the gap. The only nonequilibrium source is therefore the normal electrode. Second, the longitudinal distribution in the bar equals that of the normal lead, $f_L=f_L^N$, while the transverse distribution that carries the charge current reads
\begin{equation}
f_T(E,V)=f_T^N(E,V)\,
\frac{\gamma_{\text{BS}}\,\mathrm{Re}\,\hat{g}_3^\text{R}(E)}
{\gamma_{\text{BN}}\,\mathrm{Re}\,\hat{g}_1^\text{R}(E)+\gamma_{\text{BS}}\,\mathrm{Re}\,\hat{g}_3^\text{R}(E)}\;.
\label{eq:fT_bulk}
\end{equation}

\noindent Inserting the Keldysh component into the matrix-current expression~\eqref{eq:current} and evaluating the $\tau_3$ trace at the DN/N boundary gives the charge current
\begin{equation}
I=\frac{1}{e}\int_{-\infty}^{+\infty}\!dE\;
\frac{\mathrm{Re}\,\hat{g}_1^\text{R}(E)\,\mathrm{Re}\,\hat{g}_3^\text{R}(E)}
{R_\text{bN}\,\mathrm{Re}\,\hat{g}_1^\text{R}(E)+R_\text{bS}\,\mathrm{Re}\,\hat{g}_3^\text{R}(E)}\,
f_T^N(E,V)\;,
\label{eq:current_main}
\end{equation}
where all voltage dependence resides in the transverse reservoir distribution $f_T^N$. The numerator is the product of the normal ($\tau_3$) and anomalous ($\tau_1$) spectral densities. Subgap charge transfer requires both to be present in the bar at the same energy: an electron tunnelling in from the N electrode needs an available state, and its conversion into a Cooper pair at the S/DN interface needs a finite pair amplitude. The denominator weights the two contributions by the corresponding interface resistances and expresses the series connection of the two boundaries. Since
\begin{equation}
f_T^N(E,V)=\frac{1}{2}\left[\tanh\frac{E+eV}{2T}-\tanh\frac{E-eV}{2T}\right]\,,
\end{equation}
and the integrand is even in $E$, differentiation at $T=0$ gives the differential conductance
\begin{equation}
\frac{\partial I}{\partial V}\bigg|_{T=0}
=\frac{\mathrm{Re}\,\hat{g}_1^\text{R}(eV)\,\mathrm{Re}\,\hat{g}_3^\text{R}(eV)}
{R_\text{bN}\,\mathrm{Re}\,\hat{g}_1^\text{R}(eV)+R_\text{bS}\,\mathrm{Re}\,\hat{g}_3^\text{R}(eV)}\;,
\label{eq:didv}
\end{equation}
so that at zero temperature the conductance at bias $V$ is set by the spectral functions at $E=eV$ alone.\\

\noindent At zero energy $\mathcal{G}(0)=\gamma_{\text{BN}}^{-1}$ is real and Eq.~\eqref{eq:g13} gives
\begin{equation}
\mathrm{Re}\,\hat{g}_3^\text{R}(0)=\frac{\gamma_{\text{BS}}}{\sqrt{\gamma_{\text{BS}}^2+\gamma_{\text{BN}}^2}}\;,\qquad
\mathrm{Re}\,\hat{g}_1^\text{R}(0)=\frac{\gamma_{\text{BN}}}{\sqrt{\gamma_{\text{BS}}^2+\gamma_{\text{BN}}^2}}\;,
\label{eq:g13_zero}
\end{equation}
which satisfy the normalisation $\bigl(\hat{g}_3^\text{R}\bigr)^2+\bigl(\hat{g}_1^\text{R}\bigr)^2=1$ and reproduce the zero-energy LDOS, Eq.~\eqref{eq:LDOS_zero}. Substitution into Eq.~\eqref{eq:didv} yields the dimensionless zero-bias conductance
\begin{equation}
\frac{\sigma(eV=0)}{\sigma_\text{N}}
=\bigl(1+\gamma_{\text{BS}}+\gamma_{\text{BN}}\bigr)
\frac{\gamma_{\text{BS}}\,\gamma_{\text{BN}}}{\bigl(\gamma_{\text{BS}}^2+\gamma_{\text{BN}}^2\bigr)^{3/2}}\;.
\label{eq:zbc}
\end{equation}
The role of a finite boundary resistance at the normal side was explored in Ref. ~\cite{tanaka2005anomalous} for DN/triplet-superconductor junctions, where the zero-energy peak height was shown to depend on the ratio $R_d/R_{b}$. While related double-barrier proximity structures have been studied previously, we are not aware of a closed-form expression for the normalized zero-bias conductance that makes the dependence on both $\gamma_{\text{BS}}$ and $\gamma_{\text{BN}}$ explicit in the present Tanaka--Nazarov framework.\\

\noindent At fixed $\gamma_{\text{BN}}$ the zero-bias conductance is not monotonic in $\gamma_{\text{BS}}$. For $\gamma_{\text{BS}},\gamma_{\text{BN}}\gg 1$, where $1+\gamma_{\text{BS}}+\gamma_{\text{BN}}\simeq\gamma_{\text{BS}}+\gamma_{\text{BN}}$, Eq.~\eqref{eq:zbc} depends only on the ratio $\gamma_{\text{BS}}/\gamma_{\text{BN}}$ and is maximal for equal boundary resistances, $\gamma_{\text{BS}}=\gamma_{\text{BN}}$, where $\sigma(0)/\sigma_\text{N}=1/\sqrt{2}$. This maximum results from a competition between the two factors in the numerator of Eq.~\eqref{eq:didv}. For $\gamma_{\text{BS}}\ll\gamma_{\text{BN}}$ the S interface is nearly transparent and the bar is strongly proximitised. Pair correlations are abundant, $\mathrm{Re}\,\hat{g}_1^\text{R}(0)\to 1$, but the zero-energy density of states is suppressed, $\mathrm{Re}\,\hat{g}_3^\text{R}(0)\approx\gamma_{\text{BS}}/\gamma_{\text{BN}}$, so that few states are available for electrons tunnelling in from the N electrode. For $\gamma_{\text{BS}}\gg\gamma_{\text{BN}}$ the situation is reversed. The density of states recovers its normal value, $\mathrm{Re}\,\hat{g}_3^\text{R}(0)\to 1$, but the induced pair amplitude is small, $\mathrm{Re}\,\hat{g}_1^\text{R}(0)\approx\gamma_{\text{BN}}/\gamma_{\text{BS}}$, and electrons entering the bar can hardly be converted into Cooper pairs at the S interface. In both limits the conductance vanishes, in the first because quasiparticle states are missing and in the second because pair correlations are missing. The maximum marks the balance. \\

\noindent The scale $E^\ast$ also sets the width of the low-bias conductance feature. For $E^\ast\ll|E|\ll\Delta_0$ the LDOS approaches its normal value, $\mathrm{Re}\,\hat{g}_3^\text{R}(E)\to 1$, while the pair amplitude decays as $\mathrm{Re}\,\hat{g}_1^\text{R}(E)\propto 1/E^2$. The proximity-induced contribution to Eq.~\eqref{eq:didv} is therefore appreciable only for $|eV|\lesssim E^\ast$ and is small at larger subgap bias. Within this window the lineshape evolves from a minimum at zero bias, flanked by maxima at $|eV|\sim E^\ast$, for $\gamma_{\text{BS}}\lesssim\gamma_{\text{BN}}$, to a single maximum at zero bias for $\gamma_{\text{BS}}\gtrsim\gamma_{\text{BN}}$ (Sec.~\ref{sec:numerics}). The width of the feature is thus fixed by $E^\ast$ rather than by $E_{\text{Th}}$, consistent with the role of $E^\ast$ in the curvature of the LDOS.\\

\noindent Equations~\eqref{eq:current_main} and~\eqref{eq:didv} complete the analytical low-energy description of the $s$-wave junction. The LDOS and the conductance depend differently on the two boundary resistances. The zero-energy LDOS, Eq.~\eqref{eq:LDOS_zero}, depends on both but is not symmetric under $\gamma_{\text{BS}}\leftrightarrow\gamma_{\text{BN}}$, whereas the zero-bias conductance, Eq.~\eqref{eq:zbc}, is symmetric in the two. A measurement of both quantities therefore determines $\gamma_{\text{BS}}$ and $\gamma_{\text{BN}}$ separately, which the conductance alone cannot. The LDOS of the bar can be measured by scanning tunneling microscopy, provided that the tip-sample conductance is much smaller than both interface conductances, so that the tip probes the bar without altering its distribution function or acting as an additional reservoir.

\noindent The analytical treatment rests on three approximations whose range of validity should be kept in mind when interpreting the transport. First, the closed-form results employ the position-independent bulk limit, $\check{G}(E,x)\approx\check{g}(E)$. This approximation is controlled by the energy-dependent coherence length $\xi_E=\sqrt{D/|E|}$ and is valid when $|E|\ll E_{\text{Th}}$, irrespective of the ratio $L/\xi_\Delta$ where $\xi_\Delta$ is the superconductive coherence length. At higher energies, spatial variations of the Green's function become important and the single-matrix ansatz breaks down, necessitating the numerical solution. Second, the analytical formula for the conductance is derived from a low-energy expansion of the S-reservoir Green's function, so the closed-form expressions are restricted to $|eV|/\Delta_0\ll1$. Third, the analytical treatment assumes that inelastic scattering is absent, such that coherence is maintained over the junction length $L$. Extending the analytical treatment towards the gap edge would require retaining the BCS spectral function in $\mathcal{M}$. Here $\mathcal{M}$ is the matrix of Eq.~\eqref{eq:M_def}, the sum of the reservoir Green's functions weighted by the inverse interface resistances and the energy term of the bar. In the position-independent limit the Usadel equation with its boundary conditions reduces to the condition that the Green's function of the bar commutes with $\mathcal{M}$, which fixes it to be the normalized $\mathcal{M}$. An analytical description of the ZESABS-induced anomaly for odd-parity pairing, incorporating the odd-frequency spin-triplet correlations, is likely tractable within the same framework. In the next section we lift these constraints and solve the Usadel equation numerically.

\section{Numerical results}
\label{sec:numerics}

\subsection{\texorpdfstring{$s$}{s}-wave}
\label{sec:swave_numerics}

\noindent We begin by validating the analytical results of Secs.~\ref{sec:gR_closed}--
\ref{sec:conductance} against the numerics, which build on the
quasiclassical treatment of diffusive proximity junctions developed in
Ref.~\cite{Tanaka2003}. Then, we use the numerical data to explore regimes
where the closed-form expressions break down and new structure emerges.
Throughout this subsection the superconductor has isotropic pairing, i.e.\
$s$-wave order ($\Delta_t/\Delta_s = 0$). Throughout the entire section unless stated otherwise, a Dynes-type broadening $\eta$ is included in the quasiparticle energy as $E \to E + i\eta$, with $\eta/\Delta_0 = 10^{-8}$. \\

\begin{figure}[t]
    \centering
    \includegraphics[width=\columnwidth]{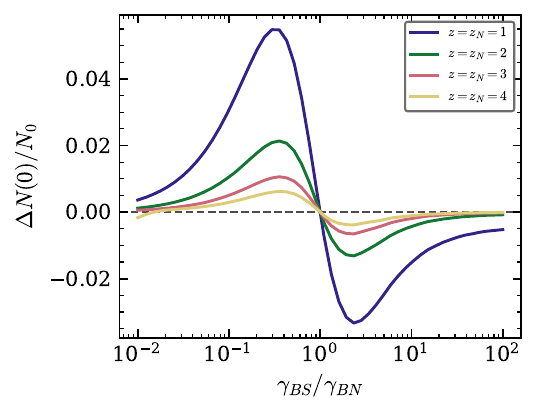}
    \caption{Difference between the numerical zero-energy LDOS
    $N(0)/N_0$ at $x/L=0.5$ and the analytical prediction of
    Eq.~\eqref{eq:LDOS_zero}, plotted as a function of the
    interface-resistance ratio $\gamma_{\text{BS}}/\gamma_{\text{BN}}$
    for $z = z_{\text{N}} \in \{1,\, 2,\, 3,\, 4\}$. Fixed parameters: $E_{\text{Th}}/\Delta_0 = 0.4$, $\gamma_{\text{BS}}=10$.} 
    \label{fig:zbc_ldos_collapse}
\end{figure}

\noindent  In the previous section, we calculated the zero energy density of states analytically in the low-transparency limit, and found that it depends only on $\gamma_{\text{BS}}/\gamma_{\text{BN}}$. To quantify the corrections of higher order in the transmission eigenvalues, we solved the Usadel equation with Tanaka-Nazarov boundary conditions numerically. 
Figure~\ref{fig:zbc_ldos_collapse} shows the difference between the zero energy density of states following from the numerical and analytical calculations.   At $z = z_{\text{N}} = 4$ the residual is of order $10^{-3}$. As $z$ decreases the deviation
grows systematically, reaching order $10^{-2}$ around $z = 1$. At each $z$
value, data at different $E_{\text{Th}}$ collapse onto the same residual
curve, confirming that the deviation is controlled by $z$ alone and not
by the Thouless energy. \\

\noindent The residual, i.e.\ the difference between the numerical and analytical LDOS, vanishes identically at the symmetric point
$\gamma_{\text{BS}} = \gamma_{\text{BN}}$ (with $z_{\text{S}} = z_{\text{N}}$)
for every barrier strength. This can be shown exactly, without relying on the
position-independent approximation or on an expansion in the transparency.
At $E = 0$ the energy term in the Usadel equation~\eqref{eq:usadel} vanishes,
so that $\partial_x(\hat{g}\,\partial_x\hat{g}) = 0$ and the matrix current is
conserved along the bar. Moreover, the electrode Green's functions reduce to
$\hat{G}_S(0) = \tau_1$ and $\hat{G}_N = \tau_3$ and the problem is 
specified by the two boundary conditions. Consider now the transformation that
combines the spatial reflection $x \to L - x$ with the exchange
$\tau_1 \leftrightarrow \tau_3$, implemented by the unitary matrix
$U = (\tau_1 + \tau_3)/\sqrt{2}$, which satisfies $U\tau_1U = \tau_3$ and
$U\tau_3U = \tau_1$. This transformation exchanges the two electrodes,
$\hat{G}_S \leftrightarrow \hat{G}_N$ and the two interfaces. For
$\gamma_{\text{BS}} = \gamma_{\text{BN}}$ and $z_{\text{S}} = z_{\text{N}}$ the
boundary value problem at $E = 0$ is therefore invariant under it and since
its solution is unique, the solution itself is invariant,
$\hat{g}(L - x) = U\hat{g}(x)U$. At the midpoint $x = L/2$ this requires
$\hat{g}(L/2)$ to have equal $\tau_1$ and $\tau_3$ components, which together
with the normalization $\hat{g}^2 = \hat{1}$ fixes
$\hat{g}(L/2) = (\tau_1 + \tau_3)/\sqrt{2}$ and hence $N(0)/N_0 = 1/\sqrt{2}$,
independent of $z$, $E_{\text{Th}}$ and the value of $\gamma_{\text{BS}} = \gamma_{\text{BN}}$.
This coincides with the analytical calculation, Eq.~\eqref{eq:LDOS_zero} at the
symmetric point, so that the residual at $x/L = 0.5$ vanishes exactly.\\

\noindent The position-independent description shows in addition why the residual grows away from the symmetric point. The matrix $\hat{g}(0) = (\tau_1 + \tau_3)/\sqrt{2}$ satisfies
\begin{equation}
[\hat{G}_S, \hat{g}]_- = [\hat{g}, \hat{G}_N]_-\;,
\label{eq:comm_symmetric}
\end{equation}
which is the Kupriyanov--Lukichev condition at the symmetric point. For
traceless $2\times 2$ matrices the anticommutator is proportional to the
identity, $[\hat{A}, \hat{B}]_+ = \operatorname{Tr}(\hat{A}\hat{B})\,\hat{1}$, so that by
direct evaluation
\begin{equation}
[\hat{G}_S, \hat{g}]_+ = [\hat{g}, \hat{G}_N]_+ = \sqrt{2}\,\hat{1}\;.
\label{eq:anticomm_symmetric}
\end{equation}
The denominator of the Tanaka--Nazarov matrix current,
$[\hat{G}_S, \hat{g}]_-/(1 + T_1^2 + T_1[\hat{G}_S, \hat{g}]_+)$, is therefore a
scalar and finite transparency only rescales the boundary term at each
interface. Since at $E = 0$ the solution depends only on the ratio of the two
boundary terms and Eqs.~\eqref{eq:comm_symmetric}
and~\eqref{eq:anticomm_symmetric} make both the numerators and the denominators
equal,
\begin{equation}
\frac{[\hat{G}_S, \hat{g}]_-}{1 + T_1^2 + T_1[\hat{G}_S, \hat{g}]_+}
=
\frac{[\hat{g}, \hat{G}_N]_-}{1 + T_1^2 + T_1[\hat{G}_N, \hat{g}]_+}\;,
\label{eq:TN_equal}
\end{equation}
the ratio is unaffected by the transparency and the solution for the problem with Kupriyanov--Lukichev boundary conditions
also solves the problem with Tanaka--Nazarov boundary conditions. Away from the
symmetric point this no longer holds. In general
$[\hat{G}_S, \hat{g}]_+ = 2\hat{g}_1\hat{1}$ and $[\hat{G}_N, \hat{g}]_+ = 2\hat{g}_3\hat{1}$,
with $\hat{g}_1 \neq \hat{g}_3$ for $\gamma_{\text{BS}} \neq \gamma_{\text{BN}}$, so the two
boundary terms are rescaled by different factors. The effective ratio of the
interface resistances then deviates from $\gamma_{\text{BS}}/\gamma_{\text{BN}}$,
and a residual appears that grows with the interface asymmetry and with
the channel transparency, i.e.\ with decreasing $z$.\\

\noindent At finite energy $E \neq 0$ the exact cancellation is lost for two
reasons. First, $\hat{G}_S$ acquires a $\tau_3$ component and the energy term
$-iE\tau_3$ in the Usadel equation is not invariant under
$\tau_1 \leftrightarrow \tau_3$, so the symmetry argument no longer applies.
Second, the energy term introduces the absolute scale $E_{\text{Th}}$. The
solution is then no longer fixed by the electrodes and the ratio of the
boundary terms alone, but depends on the magnitude of the boundary terms
relative to $E/E_{\text{Th}}$ and on the spatial profile of $\hat{g}(x)$ obtained
from the Usadel equation. The rescaling of the boundary terms by finite
transparency then changes the result even at the symmetric point.\\

\begin{figure}[t]
    \centering
    \includegraphics[width=\columnwidth]{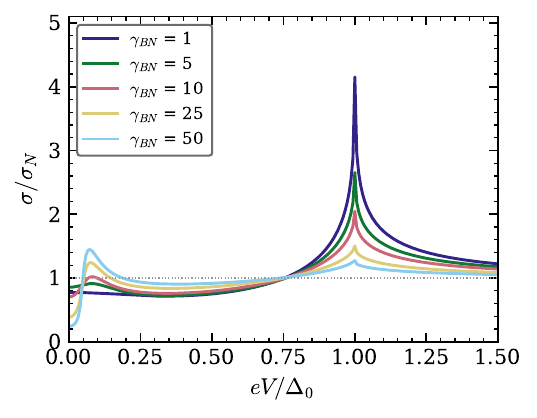}
    \caption{Normalized differential conductance $\sigma(eV)/\sigma_N$ at
    fixed $\gamma_{\text{BS}}=5$ for several values of $\gamma_{\text{BN}}$, with
    $z = z_N = 1.0$. The sub-Thouless peak near $eV = 0$ grows as
    $\gamma_{\text{BN}}$ increases while its position remains approximately fixed.
    Fixed parameters: $E_{\text{Th}}/\Delta_0 = 0.4$, $\gamma_{\text{BS}} = 10$.}
    \label{fig:didv_sweep_gBN}
\end{figure}

\noindent Figure~\ref{fig:didv_sweep_gBN} displays the differential conductance as a
function of bias for fixed $\gamma_{\text{BS}}$ and increasing $\gamma_{\text{BN}}$, at
$z = z_N = 1$. With increasing $\gamma_{\text{BN}}$, that is, with increasing DN/N interface resistance, the properties of the DN become more similar to those of the superconductor, that is, the proximity gap deepens (lower $N(0)/N_0$). The spectral
weight displaced from the gap reappears as a coherence peak near the
characteristic energy $E^\ast$. This spectral redistribution translates
directly into the conductance: the sub-gap peak sharpens and its amplitude
increases, while the peak position remains approximately constant. The
invariance of the peak position confirms the analytical prediction that
$E^\ast$ depends on $\gamma_{\text{BS}}$ and $E_{\text{Th}}$ but it is insensitive to
$\gamma_{\text{BN}}$  (Eq.~\eqref{eq:Estar_general} in the regime
$\gamma_{\text{BN}} \gg \gamma_{\text{BS}}$). Physically, the N-side resistance sets the residual density of states within the
proximity-induced pseudogap
and therefore the height
of the conductance peak, while the width of that peak is mostly determined
by the S-side resistance through the dephasing rate
$\alpha = 1/(\gamma_{\text{BS}}\Delta_0) + 1/E_{\text{Th}}$.
At $z = 1$ a finite sub-gap background is visible that is absent from the
closed form; this is the direct Andreev channel at finite interface
transparency, which the KL continuity condition does not capture.

\begin{figure}[t]
    \centering
    \includegraphics[width=\columnwidth]{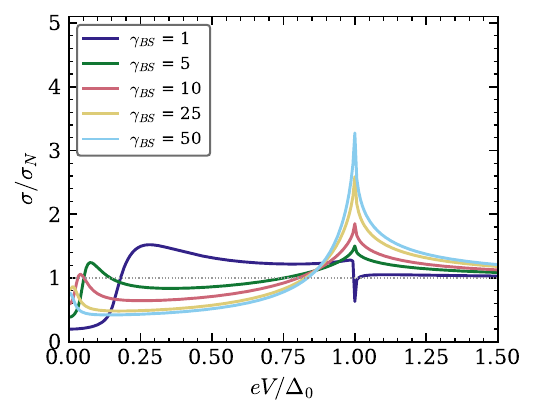}
    \caption{Normalized differential conductance $\sigma(eV)/\sigma_N$ at
    fixed $\gamma_{\text{BN}}=5$ for several values of $\gamma_{\text{BS}}$, with
    $z = z_N = 1.0$. The sub-Thouless feature broadens as $\gamma_{\text{BS}}$
    decreases, in accordance with $E^\ast \propto 1/\gamma_{\text{BS}}$. Fixed
    parameters: $E_{\text{Th}}/\Delta_0 = 0.4$, $\gamma_{\text{BN}} = 25$.}
    \label{fig:didv_sweep_gBS}
\end{figure}

\noindent Figure~\ref{fig:didv_sweep_gBS} shows the complementary sweep: fixed
$\gamma_{\text{BN}}$ with varying $\gamma_{\text{BS}}$, again at
$z = z_{\text{N}} = 1$. The dominant effect is a change in the width of
the sub-gap feature. At large $\gamma_{\text{BS}}$ the proximity coupling
is weak, $E^\ast \propto E_{\text{Th}}/\gamma_{\text{BS}}$ is small and
the conductance enhancement is confined to a narrow window around zero
bias. As $\gamma_{\text{BS}}$ decreases the S/DN interface becomes more
transparent, the proximity-induced correlations extend to higher energies
and the sub-Thouless feature broadens. At the gap edge the coherence peak
near $eV = \Delta_0$ sharpens and grows with increasing $\gamma_{\text{BS}}$:
as the S interface becomes more opaque, less spectral weight is
transferred into the subgap region and it accumulates at the gap edge
instead, producing a more pronounced coherence peak.
The peak height changes comparatively little across the sweep because
$N(0)/N_0 = \gamma_{\text{BS}}/\sqrt{\gamma_{\text{BS}}^2 +
\gamma_{\text{BN}}^2}$ is a weak function of $\gamma_{\text{BS}}$ when
$\gamma_{\text{BN}} \gg \gamma_{\text{BS}}$.\\

\begin{figure}[t]
    \centering
    \includegraphics[width=\columnwidth]{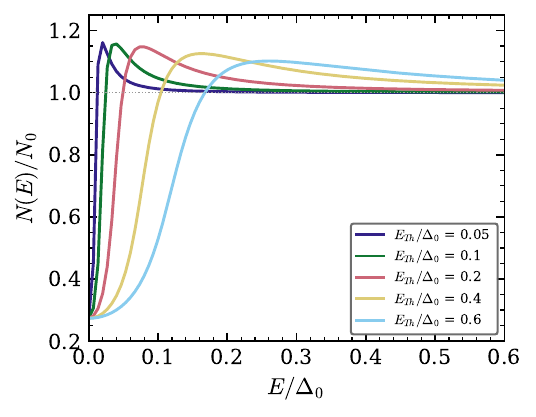}
    \caption{LDOS $N(E)/N_0$ for the $s$-wave case for several values of
    $E_{\text{Th}}/\Delta_0$ at fixed interface parameters. The lineshape broadens
    with increasing $E_{\text{Th}}$ while $N(0)/N_0$ remains constant. Fixed
    parameters: $z = z_N = 1$, $\gamma_{\text{BS}} = 3$, $\gamma_{\text{BN}} = 15$.}
    \label{fig:swave_ldos_Eth}
\end{figure}

\begin{figure}[t]
    \centering
    \includegraphics[width=\columnwidth]{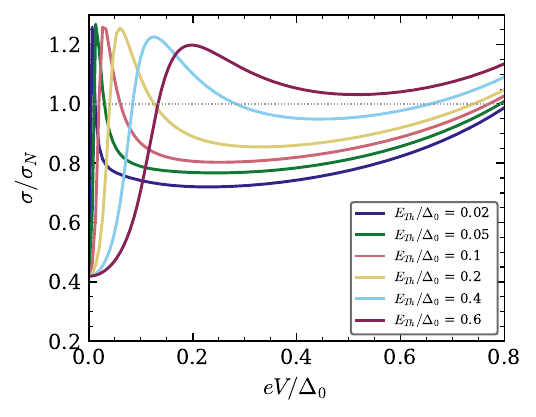}
    \caption{Normalized differential conductance $\sigma(eV)/\sigma_N$ for
    the $s$-wave case for several values of $E_{\text{Th}}/\Delta_0$ at fixed
    interface parameters. The subgap feature broadens with increasing
    $E_{\text{Th}}$ while $\sigma(0)/\sigma_N$ is independent of $E_{\text{Th}}$. Fixed
    parameters: $z = z_N = 1$, $\gamma_{\text{BS}} = 3$, $\gamma_{\text{BN}} = 15$.}
    \label{fig:swave_didv_Eth}
\end{figure}

\noindent As shown in Figs.~\ref{fig:swave_ldos_Eth} and~\ref{fig:swave_didv_Eth}, the width of both the LDOS coherence feature and the conductance peak
scales linearly with $E_{\text{Th}}$, as predicted by
$E^\ast \propto E_{\text{Th}}/\gamma_{\text{BS}}$. At small $E_{\text{Th}}$ the spectral
features are narrow and confined to a window $|eV| \lesssim E^\ast \ll \Delta_0$;
as $E_{\text{Th}}$ increases the features broaden proportionally.
In the position-independent (short-junction) approximation the zero-energy and
zero-bias values are independent of $E_{\text{Th}}$, a property we confirm
numerically. 
This is a consequence of the fact that at $E = 0$ all components of the matrix current $-Dg\nabla g$ are conserved in the absence of inelastic relaxation. Therefore,
the zero-energy LDOS depends only on the ratio
$\gamma_{\text{BS}}/\gamma_{\text{BN}}$ (Eq.~\eqref{eq:LDOS_zero}). \\

\noindent Figure~\ref{fig:swave_didv_Eth} shows that the width of the subgap feature in the conductance grows approximately linearly with $E_{\text{Th}}$. This agrees with Eq.~\eqref{eq:Estar_BNdom}, which for $\gamma_{\text{BN}}\gg\gamma_{\text{BS}}$ and $E_{\text{Th}}\ll\gamma_{\text{BS}}\Delta_0$ gives $E^\ast\simeq\sqrt{2/3}\,E_{\text{Th}}/\gamma_{\text{BS}}$. At small $E_{\text{Th}}$ the feature is narrow and confined to a window $|eV|\lesssim E^\ast\ll\Delta_0$. With increasing $E_{\text{Th}}$ it broadens accordingly, while its zero-bias value $\sigma(0)/\sigma_\text{N}$ does not change, since Eq.~\eqref{eq:zbc} does not depend on $E_{\text{Th}}$. This
scaling confirms that the Thouless energy enters the problem exclusively
through the dephasing rate $1/E_{\text{Th}}$ in the imaginary part of
$\mathcal{G}$ (Eq.~\eqref{eq:zeta_def}) and does not affect the
spectral balance at zero energy.\\

\noindent Together, Figs.~\ref{fig:didv_sweep_gBN} through~\ref{fig:swave_didv_Eth}
demonstrate that the sub-Thouless energy scale $E^\ast$ identified
analytically in Sec.~\ref{sec:expansion} is directly observable in the
differential conductance and that the two interface resistances can, in
principle, be extracted independently from a joint measurement of the peak
width and peak height. This factorization into amplitude and energy-scale
control is not generic to proximity systems; it arises specifically from
the presence of two separate interfaces with distinct physical roles (pair
injection versus normal-state drain) and is absent in the single-interface
SN geometry studied in earlier work~\cite{Kokkeler2023,
golubov2005density}. The single-interface limits of this behavior are known from the theory of
conventional proximity junctions~\cite{VolkovZaitsevKlapwijk1993,golubov2004current,Belzig1999}.\\

\subsection{Helical \texorpdfstring{$p$}{p}-wave}
\label{sec:helical_p}

\noindent We now turn to a superconductor with helical $p$-wave symmetry, with pair potential defined in Eqs.~(\ref{eq:pairpotential},\ref{eq:dvectorhelical}).
This produces a gapped bulk spectrum with $|\Delta(\phi)| = \Delta_0$
for all $\phi$, the same magnitude as in the $s$-wave case. The crucial
difference lies in the parity: the pair potential changes
sign under reversal of the transverse momentum component and this sign
change generates SABS at the S/DN interface
that have no counterpart in the isotropic case \cite{TanakaKashiwaya2004, tanaka2005anomalous, tanaka2005theory}. The ZESABS among them produce a zero-energy peak in the local spectral weight at the interface, whose
influence on the bulk of the diffusive bar depends on how strongly the
interface couples to the wire. \\

\begin{figure}[t]
    \centering
    \includegraphics[width=\columnwidth]{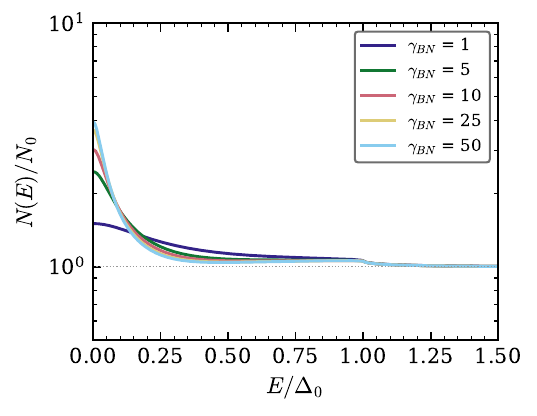}
    \caption{LDOS $N(E)/N_0$ in the helical $p$-wave case at fixed
    $\gamma_{\text{BS}}$ for several values of $\gamma_{\text{BN}}$. A zero-energy peak
    replaces the $s$-wave proximity dip. Fixed parameters:
    $z = z_N = 1$, $E_{\text{Th}}/\Delta_0 = 0.4$, $\gamma_{\text{BS}} = 10$.}
    \label{fig:helical_ldos_gBN}
\end{figure}

\noindent Figure~\ref{fig:helical_ldos_gBN} shows the LDOS for fixed
$\gamma_{\text{BS}}$ and increasing $\gamma_{\text{BN}}$. The most
striking contrast with the $s$-wave results of the preceding subsection
is the behavior at zero energy: instead of a proximity-induced
suppression below unity, the LDOS exhibits a peak at $E = 0$ that
exceeds the normal-state value. This zero-energy anomaly originates from the ZESABS at the superconductor interface, which induce odd-frequency
spin-triplet correlations in the diffusive
bar~\cite{tanaka2005anomalous,tanaka2005theory,tanaka2021theoryodd, TanakaGolubov2007, TanakaSatoNagaosa2012}.
As $\gamma_{\text{BN}}$ increases the peak sharpens and grows.\\

\noindent The role of the normal-side interface resistance in this
anomaly was first addressed by Tanaka et al.~\cite{tanaka2005anomalous},
who considered the same triplet superconductor S/DN/N geometry with resistive interfaces on both
sides of the bar and Tanaka--Nazarov boundary conditions at each of them.
Solving the position-dependent Usadel equation numerically for a chiral
$p$-wave superconductor, they found that the height of the zero-energy
peak increases as $R_\text{d}/R_{\text{b}}'$ decreases, where
$R_{\text{b}}'$ corresponds to our $R_{\text{bN}}$ and they attributed
this to a stronger confinement of the odd-frequency pair amplitude in the
bar. Analytically, they considered the limit
$R_\text{d}/R_{\text{b}}' \to 0$, in which the bar is decoupled from the
normal reservoir and the LDOS diverges as $|E|^{-1/2}$ at low
energy, regularized only by a Dynes broadening. The present
work extends this picture in three respects. First, we keep $\gamma_{\text{BN}}$
finite throughout, so that the coupling to the normal reservoir itself cuts
off the divergence and sets a finite peak height and width. Second, the
closed-form description of Secs.~\ref{sec:gR_closed}--\ref{sec:expansion}
makes the dependence on $\gamma_{\text{BN}}$ explicit for the $s$-wave case,
which serves as a reference for the odd-parity results. Third, we compute
the differential conductance and show
how the zero-energy peak in the LDOS translates into a ZBCP.\\

\noindent The growth of the zero-energy value specifically, rather than a
uniform enhancement across all subgap energies, can be understood from the
energy dependence of the coherence length $\xi_E = \sqrt{\hbar D/|E|}$. At
zero energy $\xi_E$ diverges, so the pair correlations induced by the ZESABS extend over the entire bar and reach the DN/N interface without
attenuation. Increasing $\gamma_{\text{BN}}$ reduces the leakage of these
correlations into the normal reservoir, so that they accumulate in the bar
and the zero-energy LDOS grows. At finite energy $\xi_E$ is finite, the pair
correlations decay along the bar and a smaller fraction reaches the DN/N
interface. The effect of increasing $\gamma_{\text{BN}}$ is therefore weaker at
finite $E$ and the enhancement is concentrated around zero energy. The width
of the peak is set by a sub-Thouless scale that, as in the $s$-wave case,
decreases with increasing boundary resistance and increases with
$E_{\text{Th}}$. Since the closed form of Sec.~\ref{sec:expansion} is derived
for $s$-wave pairing, its expression for $E^\ast$ applies to the helical case
only qualitatively.\\

\noindent This confinement picture requires phase coherence across the bar; inelastic
scattering would cut off the divergence of the coherence length at $E = 0$. The
same phase-coherent penetration of zero-energy weight underlies the quantization
of the zero-voltage conductance in diffusive normal-metal/triplet-superconductor
junctions~\cite{TanakaKashiwaya2004}, shown to be robust against disorder in
Ref.~\cite{Ikegaya2016}. Our result is the finite-$\gamma_{\text{BN}}$
counterpart of that effect, where the N interface controls the amplitude of the
zero-energy peak. \\

\begin{figure}[t]
    \centering
    \includegraphics[width=\columnwidth]{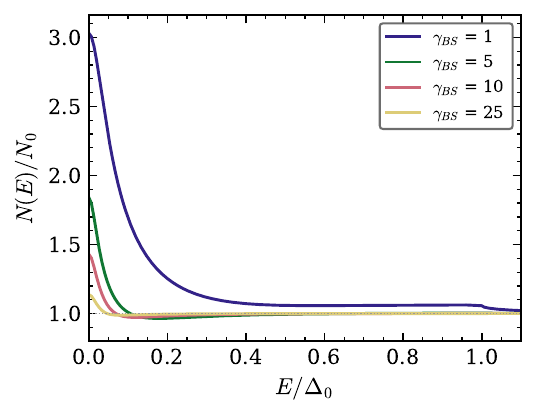}
    \caption{LDOS $N(E)/N_0$ in the helical $p$-wave case at fixed
    $\gamma_{\text{BN}}$ for several values of $\gamma_{\text{BS}}$. Reducing
    $\gamma_{\text{BS}}$ broadens the zero-energy peak, mirroring the $s$-wave
    behavior of $E^\ast$. Fixed parameters: $z = z_N = 1$,
    $E_{\text{Th}}/\Delta_0 = 0.4$, $\gamma_{\text{BN}} = 30$.}
    \label{fig:helical_ldos_gBS}
\end{figure}

\noindent Figure~\ref{fig:helical_ldos_gBS} displays the complementary sweep over
$\gamma_{\text{BS}}$ at fixed $\gamma_{\text{BN}}$. Decreasing $\gamma_{\text{BS}}$ (making
the S interface more transparent) broadens the zero-energy peak, in direct
analogy with the broadening of the sub-Thouless conductance feature in the
$s$-wave case (Fig.~\ref{fig:didv_sweep_gBS}). The peak height varies more
weakly with $\gamma_{\text{BS}}$ than with $\gamma_{\text{BN}}$
(compare Fig.~\ref{fig:helical_ldos_gBN}), consistent with the picture
that the width is primarily set by $\gamma_{\text{BS}}$ and the amplitude
primarily by $\gamma_{\text{BN}}$. This approximate separation of roles
persists across the change in pairing symmetry, indicating that it
originates from the two-interface structure of the S/DN/N junction rather
than from the specific form of the pair potential.\\

\begin{figure}[t]
    \centering
    \includegraphics[width=\columnwidth]{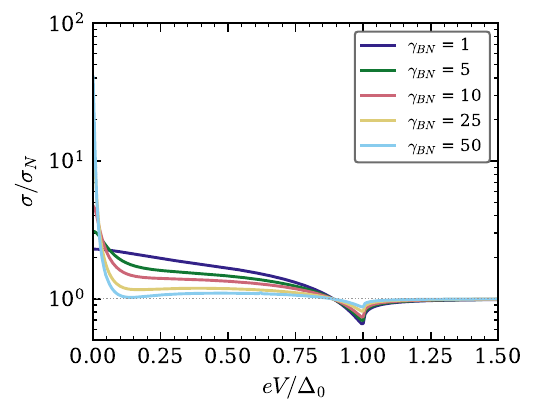}
    \caption{Normalized differential conductance $\sigma(eV)/\sigma_N$ in the
    helical $p$-wave case at fixed $\gamma_{\text{BS}}$ for several values of
    $\gamma_{\text{BN}}$. A zero-bias conductance peak (ZBCP) develops whose height
    grows with $\gamma_{\text{BN}}$. Fixed parameters: $z = z_N = 1$,
    $E_{\text{Th}}/\Delta_0 = 0.4$, $\gamma_{\text{BS}} = 10$.}
    \label{fig:helical_didv_gBN}
\end{figure}

\noindent Figure~\ref{fig:helical_didv_gBN} shows the differential conductance for
the helical state. The ZESABS-induced spectral peak at zero energy translates
into a zero-bias conductance peak (ZBCP) that exceeds the normal-state
conductance, $\sigma(0)/\sigma_N > 1$. As $\gamma_{\text{BN}}$ increases, the ZBCP sharpens
and grows, following the same pattern as the LDOS: the N interface controls
the amplitude of the anomaly. The sub-gap structure flanking the ZBCP,
including dips below unity near $|eV| \sim E^\ast$, arises from the crossover between the ZESABS-enhanced low-energy region and the depleted spectral weight
at intermediate energies. The width of both the peak and the flanking dips
scales with $E^\ast$, reinforcing the role of the sub-Thouless scale as the
characteristic low-energy scale of the S/DN/N junction.\\

\begin{figure}[t]
    \centering
    \includegraphics[width=\columnwidth]{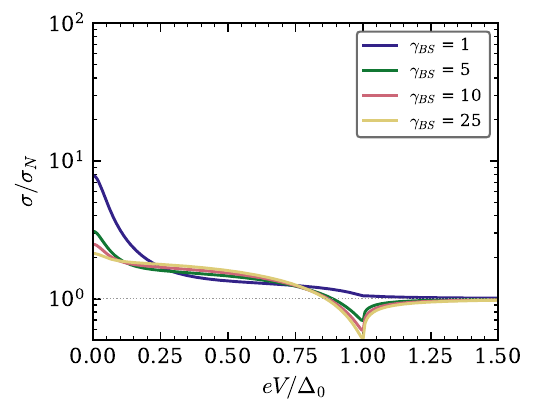}
    \caption{Normalized differential conductance $\sigma(eV)/\sigma_N$ in the
    helical $p$-wave case at fixed $\gamma_{\text{BN}}$ for several values of
    $\gamma_{\text{BS}}$. The ZBCP broadens as $\gamma_{\text{BS}}$ decreases. Fixed
    parameters: $z = z_N = 1$, $E_{\text{Th}}/\Delta_0 = 0.4$,
    $\gamma_{\text{BN}} = 30$.}
    \label{fig:helical_didv_gBS}
\end{figure}

\noindent Figure~\ref{fig:helical_didv_gBS} shows the effect of varying
$\gamma_{\text{BS}}$ at fixed $\gamma_{\text{BN}}$ on the conductance. Both the width
and the height of the ZBCP depend strongly on $\gamma_{\text{BS}}$. At large
$\gamma_{\text{BS}}$ the peak is narrow and confined to $|eV| \lesssim E^\ast \ll E_{\text{Th}}$.
Decreasing $\gamma_{\text{BS}}$ broadens the peak and moves the dips on either side
of it to larger bias, in line with the increase of $E^\ast$ with decreasing
$\gamma_{\text{BS}}$ found for the $s$-wave case. In the regime
$\gamma_{\text{BS}}\ll\gamma_{\text{BN}}$ and $\gamma_{\text{BS}}\Delta_0\gg E_{\text{Th}}$
the $s$-wave expression reduces to $E^\ast\simeq\sqrt{2/3}\,E_{\text{Th}}/\gamma_{\text{BS}}$.
At the same time the peak height decreases by a factor of about eight
when $\gamma_{\text{BS}}$ is increased from 1 to 25, 
so that the change in amplitude is at least as pronounced as the change in
width. This trend is opposite to the $s$-wave case, where for
$\gamma_{\text{BS}}<\gamma_{\text{BN}}/\sqrt{2}$ the zero-bias conductance, Eq.~\eqref{eq:zbc}
increases with $\gamma_{\text{BS}}$. The analytical description of Sec.~\ref{sec:expansion} does
not carry over directly to the helical case. In the $s$-wave case the
anomalous part of $\mathcal{M}$ is the $\tau_1$ component, which represents
even-frequency spin-singlet pairing. For the helical superconductor the
pair amplitude induced in the bar is an odd-frequency spin-triplet $s$-wave
amplitude~\cite{tanaka2005anomalous, TanakaGolubov2007}, which is shifted in phase
by $\pi/2$ with respect to the even-frequency amplitude. A closed form for the helical case
thus requires $\mathcal{M}$ to be constructed from $\tau_3$ and $\tau_2$, with the angle average of the SABS contribution entering the anomalous part and
the identification of $E^\ast$ above should be regarded as qualitative.\\

\begin{figure}[t]
    \centering
    \includegraphics[width=\columnwidth]{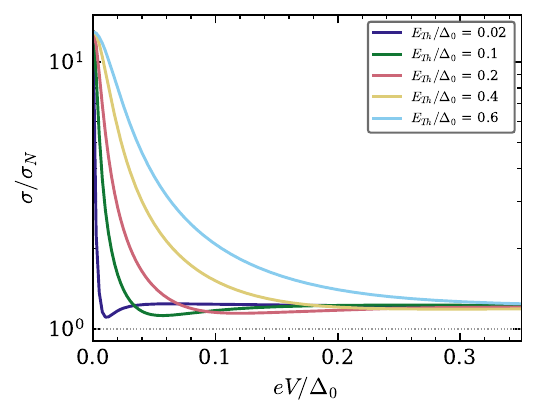}
    \caption{Normalized differential conductance $\sigma(eV)/\sigma_N$ in the
    helical $p$-wave case for several values of $E_{\text{Th}}/\Delta_0$ at fixed
    interface parameters. The ZBCP broadens with increasing $E_{\text{Th}}$ while its
    height is unchanged. Fixed parameters:
    $z = z_N = 1$, $\gamma_{\text{BS}} = 3$, $\gamma_{\text{BN}} = 15$.}
    \label{fig:helical_didv_Eth}
\end{figure}

\noindent Figure~\ref{fig:helical_didv_Eth} varies the Thouless energy at fixed
interface parameters. Increasing $E_{\text{Th}}$ broadens the ZBCP, while the
zero-bias conductance is unchanged. Both observations agree with the
$s$-wave description. The term containing the Thouless energy in Eq.~\eqref{eq:usadel} vanishes at
$E=0$, so the zero-bias conductance is independent of $E_{\text{Th}}$, while the
width scales as $E^\ast\propto(1/E_{\text{Th}}+1/\gamma_{\text{BS}}\Delta_0)^{-1}$. The width therefore grows linearly with $E_{\text{Th}}$ as long as $E_{\text{Th}}\ll\gamma_{\text{BS}}\Delta_0$, i.e.\ as long as the electron-hole phase accumulated in the bar dominates over the energy dependence of the superconducting reservoir. For $\gamma_{\text{BS}}=3$ this condition is only approximately fulfilled over the range of Fig.~\ref{fig:helical_didv_Eth}, so that the broadening is slightly sublinear at the largest $E_{\text{Th}}$.\\

\noindent The helical $p$-wave results show a systematic correspondence with
the $s$-wave case. The width of the low-energy feature is set by a
sub-Thouless scale that decreases with increasing boundary resistance. The barrier strength $z$, however, plays a different role. In the $s$-wave case a decreasing $z$ only takes the interfaces out of the tunneling regime in which the Tanaka--Nazarov boundary conditions reduce to the Kupriyanov--Lukichev form. For the helical state $z$ in addition sets the relative weight of the surface Andreev bound states at different incidence angles. The transparency $T(\varphi,z)=\cos^2\varphi/(\cos^2\varphi+z^2)$ favors normal incidence the more strongly the larger $z$, so that the ZESABS dominate over the SABS at finite energy $\pm\Delta_0\sin\varphi$ for opaque interfaces, while all bound states contribute for transparent ones.
Neither the width nor the height of the feature is controlled by a single
boundary resistance, as both depend on $\gamma_{\text{BS}}$ and $\gamma_{\text{BN}}$. What
changes is the frequency symmetry of the pairs induced in the bar. For an $s$-wave electrode these are even-frequency spin-singlet $s$-wave pairs, which suppress the zero-energy LDOS and produce a subgap conductance dip. For a helical electrode the ZESABS at the S interface induce odd-frequency spin-triplet $s$-wave pairs, which enhance the zero-energy LDOS~\cite{tanaka2005anomalous, TanakaGolubov2007}. As a result the
subgap dip is replaced by a ZESABS-induced ZBCP. \\

In the next section, we discuss the BW state, the three-dimensional counterpart of the helical state, with the $d$-vector parallel to the momentum. \\

\subsection{Three-dimensional Balian--Werthamer state}
\label{sec:3DBW}
\noindent The Balian--Werthamer (BW) state is the equilibrium phase of superfluid
$^3$He-B~\cite{balian1963superconductivity} and the leading candidate for the bulk order parameter in several
heavy-fermion superconductors~\cite{3DBWHeavyFermion, 3DBWHeavyFermionExperiment}.
Surface Andreev bound states in the BW state were first identified by
Buchholtz and Zwicknagl~\cite{BuchholtzZwicknagl1981}, who showed that a
zero-energy state exists at normal incidence on the surface of a $p$-wave
superconductor. Subsequent work on superfluid $^3$He-B established
the connection between these surface states and Majorana
quasiparticles~\cite{Aoki2005He3, Choi2006He3, Nagato1998, Nagai2008,
ChungZhang2009, Nagato2009, Volovik2009a, Volovik2009b}.
For the BW state we choose $\hat{a}=\hat{y}$, $\hat{b}=\hat{z}$ and $\hat{c}=\hat{x}$ in Eq.~\eqref{eq:dvectorBW}, with $\hat{x}$ the interface normal, so that $\mathbf{d}(\mathbf{k})=\hat{\mathbf{k}}$ points along the quasiparticle momentum at every point on the Fermi surface. As discussed in Sec.~\ref{sec:pairpotential}, any other choice of mutually orthogonal $\hat{a}$, $\hat{b}$ and $\hat{c}$ is related to this one by a global spin rotation and gives the same results. The resulting gap magnitude $|\Delta(\boldsymbol{k})| = \Delta_0$
is isotropic, identical to the $s$-wave case, yet the order parameter
has odd parity and is topologically nontrivial~\cite{Topology1, Topology2}. The anomalous proximity effect between a diffusive normal metal and the
BW state of superfluid $^3$He-B has been analyzed
by Higashitani et al.~\cite{Higashitani2009}, whose results anticipate the zero-energy
enhancement we find in the S/DN/N geometry.\\

\noindent In the present geometry the junction interfaces are planar and the
Usadel equation is effectively one-dimensional ($x$ perpendicular to the
interfaces), so the three-dimensional Fermi surface enters only through
the angular integration in the Tanaka--Nazarov boundary condition. A
quasiparticle incident at polar angle $\theta$ from the interface normal
and azimuthal angle $\varphi$ in the interface plane experiences the pair
potential $\Delta_0\hat{\boldsymbol{k}}\cdot\boldsymbol{\sigma}$, with a
component of magnitude $\Delta_0|\cos\theta|$ along the interface normal and a component of magnitude
$\Delta_0|\sin\theta|$ in the interface plane. Specular reflection at the
interface reverses the normal component of the momentum while conserving
the parallel one. The reflected quasiparticle therefore experiences a pair
potential in which the normal component of $\boldsymbol{d}$ has changed sign
while the in-plane component is unchanged. At normal incidence the 
pair potential changes sign, which produces a ZESABS. At oblique incidence the sign change is only partial and the SABS lies at the finite energy
$E=\pm\Delta_0\sin\theta$~\cite{BuchholtzZwicknagl1981, Nagato2009}. This structure is the same as
in the helical case, of which the BW state is the three-dimensional
analogue. In both cases the ZESABS at the S interface induce odd-frequency
spin-triplet $s$-wave pairs in the diffusive bar~\cite{tanaka2005anomalous, TanakaGolubov2007, Higashitani2009},
which give rise to the anomalous proximity effect and an enhanced
zero-energy LDOS.\\

\noindent The two states differ in the weight with which the low-energy
bound states contribute to the angle average in the boundary condition.
Bound states below an energy $|E|\ll\Delta_0$ originate from channels close to normal incidence, with $\sin\varphi\lesssim|E|/\Delta_0$ in the helical case and $\sin\theta\lesssim|E|/\Delta_0$ in the BW case. In two dimensions the angular measure is $d\varphi$, so these channels carry a weight proportional to $|E|/\Delta_0$. In three dimensions the measure is
$\sin\theta\,d\theta\,d\varphi$, so the same channels carry a weight proportional to
$(|E|/\Delta_0)^2$. Compared with the helical state, the BW state
thus assigns less weight to the zero-energy bound state and relatively
more weight to the bound states at finite energy. Whether the conductance
shows a ZBCP or a zero-bias dip depends on the balance between these
contributions and therefore on the junction parameters. For
$\gamma_{\text{BN}}=0$ both a ZBCP and a dip have been found, depending on the
boundary resistance and the Thouless energy~\cite{kokkeler2023proximity}. In the present
S/DN/N geometry the zero-energy enhancement of the LDOS persists, as shown
below, but its height and the lineshape near $E^\ast$ differ from the helical
case.\\

\begin{figure}[t]
    \centering
    \includegraphics[width=\columnwidth]{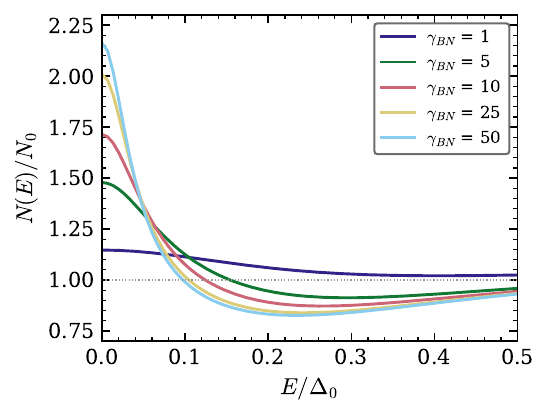}
    \caption{LDOS $N(E)/N_0$ in the BW case at fixed $\gamma_{\text{BS}}$ for
    several values of $\gamma_{\text{BN}}$. A zero-energy peak appears,
    qualitatively similar to the helical $p$-wave case
    (Fig.~\ref{fig:helical_ldos_gBN}). Fixed parameters: $z = z_{\text{N}} = 1$,
    $E_{\text{Th}}/\Delta_0 = 0.4$, $\gamma_{\text{BS}} = 1$.}
    \label{fig:bw_ldos_gBN}
\end{figure}

\noindent Figure~\ref{fig:bw_ldos_gBN} shows the LDOS for the BW state at fixed
$\gamma_{\text{BS}}$ and increasing $\gamma_{\text{BN}}$. As in the helical case (Fig.~\ref{fig:helical_ldos_gBN}), a zero-energy peak appears, with $N(0)/N_0$ growing
with $\gamma_{\text{BN}}$ as the N interface increasingly decouples the bar
from the normal reservoir. The qualitative agreement with
the helical result follows from the common origin of the peak: in both
states the ZESABS generate odd-frequency pairs in the
bar. The existence of the peak thus depends only on the sign change of the
pair potential upon reflection, whereas its height and the lineshape near
$E^\ast$ depend on the angular weights discussed above. Since the BW state
assigns less weight to the channels close to normal incidence, the
zero-energy enhancement is expected to be weaker than in the helical case
for the same junction parameters. \\

\begin{figure}[t]
    \centering
    \includegraphics[width=\columnwidth]{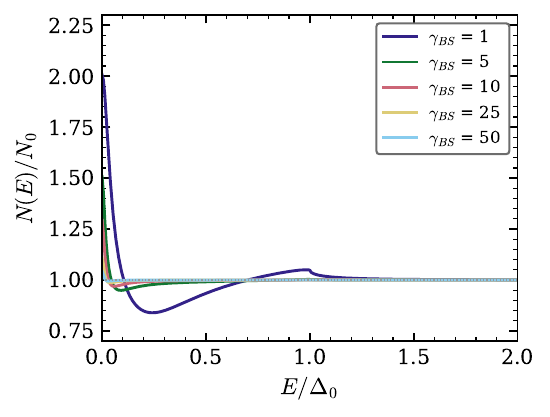}
    \caption{LDOS $N(E)/N_0$ in the BW case at fixed $\gamma_{\text{BN}}$ for
    several values of $\gamma_{\text{BS}}$. Reducing $\gamma_{\text{BS}}$ broadens the
    zero-energy peak, consistent with the increase of $E^\ast$ with decreasing
    $\gamma_{\text{BS}}$. Fixed parameters: $z = z_N = 1$, $E_{\text{Th}}/\Delta_0 = 0.4$,
    $\gamma_{\text{BN}} = 25$.}
    \label{fig:bw_ldos_gBS}
\end{figure}

\noindent Figure~\ref{fig:bw_ldos_gBS} sweeps $\gamma_{\text{BS}}$ at fixed $\gamma_{\text{BN}}$.
As in both the $s$-wave and helical cases, decreasing $\gamma_{\text{BS}}$ broadens
the low-energy feature. The energy scale of the broadening is
again consistent with $E^\ast \propto E_{\text{Th}}/\gamma_{\text{BS}}$, confirming that
the sub-Thouless scale is set by the same combination of Thouless energy
and S-interface resistance regardless of pairing symmetry. This robustness
is a consequence of the fact that $E^\ast$ originates from the
energy-dependent dephasing of proximity correlations in the diffusive bar
(the imaginary part of $\mathcal{G}$, Eq.~\eqref{eq:zeta_def}), a mechanism that is insensitive to whether those correlations are even-frequency spin-singlet or odd-frequency spin-triplet pairs. \\

\begin{figure}[t]
    \centering
    \includegraphics[width=\columnwidth]{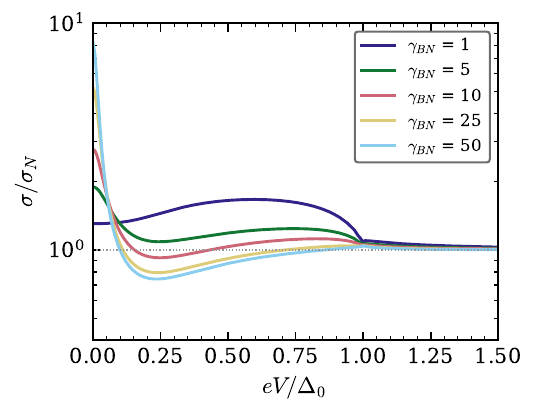}
    \caption{Normalized differential conductance $\sigma(eV)/\sigma_N$ in the
    BW case at fixed $\gamma_{\text{BS}}$ for several values of $\gamma_{\text{BN}}$. The
    ZBCP grows with $\gamma_{\text{BN}}$, mirroring the helical result
    (Fig.~\ref{fig:helical_didv_gBN}). Fixed parameters: $z = z_N = 1$,
    $E_{\text{Th}}/\Delta_0 = 0.4$, $\gamma_{\text{BS}} = 1$.}
    \label{fig:bw_didv_gBN}
\end{figure}

\noindent Figure~\ref{fig:bw_didv_gBN} translates the spectral features into the
differential conductance. A ZBCP exceeding the normal-state value develops
and sharpens with increasing $\gamma_{\text{BN}}$, reproducing the pattern
established for the helical state in Fig.~\ref{fig:helical_didv_gBN}. The
flanking dips near $|eV| \sim E^\ast$ are also present and become more prominent upon increasing the resistance of the interface with the normal metal electrode, confirming that the confinement-related
sub-Thouless scale $E^*$ governs the subgap conductance structure in the BW
case as well. As discussed above, the ZBCP height at a given $\gamma_{\text{BN}}$ may differ
quantitatively from the helical value because the effective ZESABS weight integrated over the Fermi surface differs between the two states. 
The difference depends on the barrier strength because $z$ governs the angle dependence of the transparency, 
$T(\theta,z)=\cos^2\theta/(\cos^2\theta+z^2)$. For small $z$ the transparency is angle-independent.
The angle average then samples the angular structure of the pair potential, and the different angular measures of the two states, $d\theta$ for the helical and $\sin\theta\,d\theta\,d\varphi$ for the BW state, lead to the largest differences. \\

\begin{figure}[t]
    \centering
    \includegraphics[width=\columnwidth]{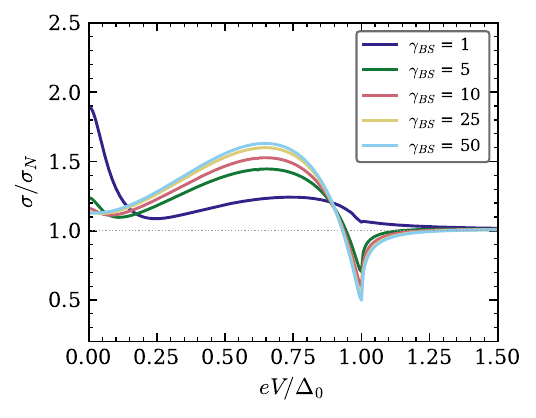}
    \caption{Normalized differential conductance $\sigma(eV)/\sigma_N$ in the
    BW case at fixed $\gamma_{\text{BN}}$ for several values of $\gamma_{\text{BS}}$. The
    ZBCP broadens as $\gamma_{\text{BS}}$ decreases. Fixed parameters:
    $z = z_N = 1$, $E_{\text{Th}}/\Delta_0 = 0.4$, $\gamma_{\text{BN}} = 30$.}
    \label{fig:bw_didv_gBS}
\end{figure}

\noindent Figure~\ref{fig:bw_didv_gBS} sweeps
$\gamma_{\text{BS}}$ at fixed $\gamma_{\text{BN}}$. The dominant effect is a
broadening of the ZBCP with decreasing $\gamma_{\text{BS}}$, tracking
$E^\ast$. The peak height also decreases with increasing
$\gamma_{\text{BS}}$, as in the helical case
(Fig.~\ref{fig:helical_didv_gBS}), but the width variation is the stronger
effect. Lastly, increasing the SN/N interface resistance results in a diminishing differential conductance at the pair amplitude $|eV|=\Delta_0$ by surpressing the proximity effect in DN.\\

\noindent The BW results show that the main features reported for the helical $p$-wave state in Sec.~\ref{sec:helical_p} are not specific to a particular choice of $d$-vector but follow from odd-parity pairing in the S/DN/N geometry. In both states the ZESABS at the S interface induce odd-frequency pairs in the bar, which enhance the zero-energy LDOS, and the width of the low-energy feature is set by a sub-Thouless scale that decreases with increasing boundary resistance.
 
\subsection{Comparison with experiment: CoSi\texorpdfstring{$_2$}{2}/TiSi\texorpdfstring{$_2$}{2} junctions}
\label{sec:chiu_comparison}
 
\noindent Conductance minima flanking the ZBCP have been measured experimentally in high-transparency
CoSi$_2$/TiSi$_2$ junctions~\cite{chiu2021observation}. In this Section we compare our calculations for helical \textit{p}-wave superconductors with their observations. In Ref.~\cite{chiu2021observation} three junctions (J1--J3) were characterized, each displaying a zero-bias
conductance peak with varying degrees of contrast. Junction J1 shows the
most pronounced low-bias structure and is the focus of the comparison
below. Two conductance features are of interest: (a) a narrow ZBCP and
(b) conductance dips flanking the ZBCP near zero bias.\\

\noindent Several junction parameters are constrained by the device geometry and
independent transport characterization~\cite{chiu2021observation}.  The diffusive TiSi$_2$ region
has length $L_{\text{DN}} \approx 350$~nm, resistivity
$\rho \approx 4.6\;\mu\Omega\,$cm and diffusion constant
$D \approx 51\;\text{cm}^2/\text{s}$. The Thouless energy is thus \begin{equation}
    E_{\text{Th}} = \frac{\hbar D}{L_{\text{DN}}^2}
    \approx 0.027\;\text{meV},
    \qquad
    \frac{E_{\text{Th}}}{\Delta_0} \approx 0.12\;.
\end{equation}
The barrier parameter at the S interface, $z_{\text{S}}\approx0.67$, is taken from Ref.~\cite{chiu2021observation}, where it was obtained from a fit with a model that contains a barrier at the S interface only and assumes continuity at the DN/N interface. Since that fit attributes all boundary scattering to the S side, this value is only an estimate within the present model. As discussed in Sec.~\ref{sec:intro}, the TiSi$_2$ near the contact changes from a flat surface into a ripple-like morphology over the first few hundred nanometers (see Fig. S1 in \cite{chiu2021observation}), which can produce a large effective boundary resistance on the normal side. On the other hand, the measured significant superconducting proximity effect in \cite{chiu2021observation} suggests a clean S/DN interface having a very low contact resistance $R_\text{bS}$ and therefore a small $\gamma_{\text{BS}}$. We therefore expect $R_{\text{bN}}\gg R_\text{d}$ in this experiment, which implies that confinement effects are important.\\

\begin{figure}[t]
    \centering
    \includegraphics[width=\columnwidth]{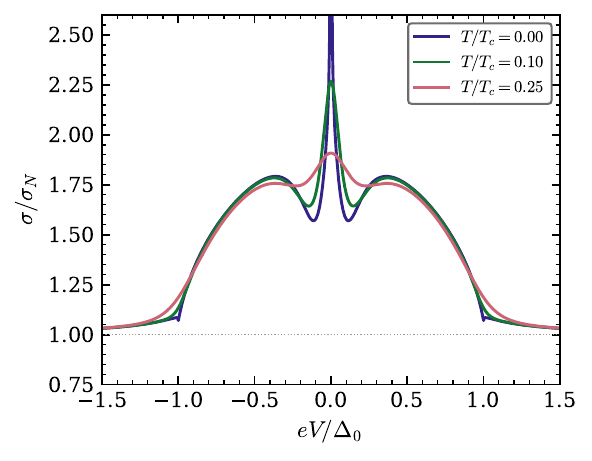}
    \caption{Comparison of the calculated differential conductance with junction J1 from Ref.~\cite{chiu2021observation}. The $T = 0$
    calculation uses a helical $p$-wave S/DN/N
    junction with $E_{\text{Th}}/\Delta_0 = 0.12$, $z_{\text{S}} = 0.67$,
    $z_{\text{N}} = 0.3$, $\gamma_{\text{BS}} = 1.4$ and
    $\gamma_{\text{BN}} = 15$. The
    experimental measurement temperature corresponds to
    $T/T_c \approx 0.25$.}
    \label{fig:chiu_reproduction}
\end{figure}

\noindent Figure~\ref{fig:chiu_reproduction} shows the differential conductance for
a helical $p$-wave S/DN/N junction with $\gamma_{\text{BS}} = 1.4$ and
$\gamma_{\text{BN}} = 15$. At $T=0$ the calculation produces both characteristic features of the experimental data, a narrow ZBCP and conductance dips flanking it near $|eV|\approx0.1\,\Delta_0$. Since $\gamma_{\text{BN}}\gg\gamma_{\text{BS}}$, Eq.~\eqref{eq:Estar_BNdom} estimates $E^\ast\approx0.065\,\Delta_0$, which is comparable to the width of the experimental ZBCP. This estimate relies on the strong-interface limit, which is not reached for $z_\text{S}=0.67$, and gives $E^\ast\approx0.5\,E_{\text{Th}}$. For junction J1 the two scales are therefore not well separated and the comparison does not allow us to attribute the width of the measured ZBCP to $E^\ast$ rather than to $E_{\text{Th}}$.\\

\noindent The overall comparison is encouraging. With $E_{\text{Th}}$ fixed by the device parameters, $z_{\text{S}}$ taken from Ref.~\cite{chiu2021observation} and only $\gamma_{\text{BS}}$,
$\gamma_{\text{BN}}$ and $z_{\text{N}}$ adjusted, the calculation is consistent with the qualitative structure of the experimental spectrum, namely the ZBCP and the flanking dips at low bias. \\ 

\noindent Even though our model is a noticeable improvement compared to previous approaches, it differs from the experiment in two respects. Firstly, the calculated ZBCP at the measurement temperature $T/T_c=0.25$ is lower than the measured one, whereas the calculation at $T/T_c=0.1$ comes closer to the data. At $T/T_c=0.25$ the thermal energy $k_\text{B}T\approx0.14\,\Delta_0$ exceeds the width of the calculated ZBCP, which is of the order of $E^\ast\approx0.065\,\Delta_0$, so that thermal smearing strongly reduces its height. The measured ZBCP is thus more robust against temperature than the calculated one, which indicates that its intrinsic width is larger than in our calculation. The junction parameters used here were chosen to reproduce the lineshape at $T=0$ and were not obtained from a systematic fit at the measurement temperature. Within the model a broader ZBCP requires a larger $E^\ast$, i.e.\ a smaller $\gamma_{\text{BS}}$ or a larger $E_{\text{Th}}$. The value of $E_{\text{Th}}$ follows from an estimate of $D$ and $L_\text{DN}$ and is uncertain, since $E_{\text{Th}}\propto L_\text{DN}^{-2}$ depends on the effective length of the diffusive region. We therefore cannot exclude that a different set of parameters reproduces the measured ZBCP at $T/T_c=0.25$, nor that an additional mechanism contributes to it. Next to this, while the theoretical model qualitatively captures the features close to zero bias, it does not reproduce the dips observed around $|eV| = \Delta_0$.\\

\noindent We identify a few elements absent from the present model that could account for this discrepancy. First, the N-side interface is described by a
single barrier with transparency $z_{\text{N}}$, whereas the TiSi$_2$ near the contact is structurally modified over an extended region~\cite{ChiuLinPrivate}. A spatially distributed
resistance or a position-dependent diffusion constant modifies the dwell time
in the bar and thereby the scale $E^\ast$, which a single effective
$\gamma_{\text{BN}}$ captures only approximately. Secondly, the
pairing state of CoSi$_2$ is not established. The helical
$p$-wave state serves here as a minimal odd-parity model and other triplet or
mixed-parity states produce lineshapes that differ in detail. Within these limitations, the comparison shows that the S/DN/N model with a
large normal-side boundary resistance captures the main low-energy features
of the experiment, while the remaining quantitative differences indicate that
the description of the device is not yet complete. Lastly, in our calculation we considered the proximity effect of the superconductor on the normal metal, but ignored the inverse proximity effect, treating the superconductor as an electrode.

\section{Conclusions and outlook}
\label{sec:conclusion}
\noindent We have examined the effect of a finite interface resistance between the diffusive region and the normal-metal reservoir in an S/DN/N hybrid junction. By applying Tanaka--Nazarov boundary conditions at both interfaces and solving the resulting Usadel equation analytically in the position-independent limit ($|E|/E_{\text{Th}}\ll1\;,z\gg 1\;,z_N \gg 1$) and numerically without these restrictions, we identify a sub-Thouless energy scale $E^\ast$
that governs the low-energy structure of both the local density of states
and the differential conductance.
This scale lies parametrically below the Thouless energy when the boundary resistances dominate. According to Eq.~\eqref{eq:Estar_general}, for $\gamma_{\text{BS}}\Delta_0\gg E_{\text{Th}}$, $E^\ast/E_{\text{Th}}=\sqrt{2/3}\,(\gamma_{\text{BS}}^2+\gamma_{\text{BN}}^2)/(\gamma_{\text{BS}}\gamma_{\text{BN}}^2)$, which is small provided that $\gamma_{\text{BS}}\gg 1$ and $\gamma_{\text{BN}}^2\gg\gamma_{\text{BS}}$. Whether the low-bias features appear well below $E_{\text{Th}}$ therefore depends not only on the magnitude of the two boundary resistances but also on their ratio. For comparable resistances, or a more resistive N side as in typical tunnel-contact geometries, the condition reduces to $R_\text{bS}\gg R_\text{d}$.\\

\noindent Our numerical results validate the closed form in the tunneling limit and quantify the corrections that appear at finite
interface transparency. Higher order tunneling processes such as Andreev reflection are present in
the subgap conductance at high transparency (low $z$), are absent from the Kupriyanov--Lukichev
closed form by construction and disappear systematically as the transparency decreases. For the $s$-wave case, the numerical zero-energy LDOS data at different $z$, $E_{\text{Th}}$ and interface asymmetry all fall onto the single analytical curve $N(0)/N_0=\gamma_{\text{BS}} / \sqrt{\gamma_{\text{BS}}^2 + \gamma_{\text{BN}}^2}$ predicted by Eq.~\eqref{eq:LDOS_zero} with residuals of order $10^{-2}$ (Fig.~\ref{fig:zbc_ldos_collapse}), confirming that the $s$-wave zero-energy spectral weight depends only on the ratio $\gamma_{\text{BS}}/\gamma_{\text{BN}} = R_{\text{bS}}/R_{\text{bN}}$ and is independent of both the superconducting gap and the Thouless energy. For odd-parity pairing, no such universal collapse exists: the zero-energy LDOS depends on $z$ through the angular structure of the SABS (Appendices~\ref{app:helical_zbc_ldos} and~\ref{app:bw_zbc_ldos}). The extension to helical $p$-wave and BW pairing shows
that the sub-Thouless scale is a geometric property of the S/DN/N junction rather than a symmetry-specific
feature: it carries over unchanged to odd-parity pairing, while the sign
of the zero-energy anomaly reverses, converting the $s$-wave subgap dip
into a ZESABS-induced peak. The BW results in particular demonstrate that
this qualitative reversal is generic to odd-parity states rather than
tied to any particular $\boldsymbol{d}$-vector orientation, and that
quantitative differences between candidate order parameters emerge only
in the detailed lineshape near $E^\ast$.\\

\noindent The predictions developed here can be tested in any S/DN/N junction where the normal-side boundary resistance is non-negligible. Natural candidates include epitaxial CoSi$_2$/TiSi$_2$ structures~\cite{chiu2021observation}, where the lattice-mismatched TiSi$_2$ region provides a structurally motivated $R_{\text{bN}}$, point-contact junctions on the candidate $p$-wave superconductor UTe$_2$~\cite{MonctonUTe22024, AokiIshidaFlouquet2019, RanUTe22019}, and layered AuSn$_4$ devices where Rashba-driven $s + p$ mixing has been reported~\cite{AuSn42023}. Our results for the helical state apply directly to chiral \textit{p}-wave superconductors as well. In the chiral state the $\mathbf{d}$-vector is fixed and the orbital phase of the pair potential winds once around the Fermi surface, which breaks time-reversal symmetry. The helical state consists of two such components with opposite chirality for opposite spin and preserves time-reversal symmetry. In both states the pair potential changes sign under reflection at normal incidence and both host ZESABS that induce odd-frequency spin-triplet $s$-wave pairs in the bar. Since the diffusive bar and the normal electrode are spin independent, the LDOS and the conductance do not change when the helical state is replaced by the chiral one~\cite{tanaka2005anomalous}. 

The sub-Thouless structure is not restricted to the short-junction regime: it appears in the
numerical solution as well, and in most of our calculations $E_{\text{Th}}$ is in fact smaller than $\Delta$, so the junction is not short. For comparison with the analytical expressions, the relevant condition is instead $|E|\ll E_{\text{Th}}$, or equivalently $\xi_E=\sqrt{D/|E|}\gg L$, which ensures that the Green's function is approximately position independent. A short junction, $L\ll\xi$, is therefore not required for the existence of the sub-Thouless feature, but is useful because it extends the energy range over which the position-independent analytical description can agree with the numerical solution. In addition, the interface resistances must be sufficiently large compared with the diffusive resistance for the sub-Thouless scale to be parametrically separated from $E_{\text{Th}}$. Outside this regime the numerical treatment remains applicable and captures the evolution of the feature beyond the range of the closed-form expressions. More generally, our results show that the boundary resistance on the normal side, which the continuity treatment discards, introduces a new low-energy scale in S/DN/N junctions and should be taken into account whenever tunneling spectra of unconventional superconductors are interpreted.

\begin{acknowledgments}
\noindent We thank A. Brinkman, S.P. Chiu and J.J. Lin for fruitful discussions. 

\noindent T.K. is supported by the Research Council of Finland through DYNCOR, Project Number 354735 and through the Finnish Quantum Flagship, Project Number 359240. His work is part of the Finnish Centre of Excellence in Quantum Materials (QMAT). Y. T. acknowledges support from JSPS with Grants-in-Aid for Scientific Research (KAKENHI Grant Nos. 23K17668, 24K00583, 24K00578, 24K00556, 25H00609 and 25H00613). L.G. thanks the Interfaces and Correlated Electrons group at the University of Twente for its hospitality during the first months of the research. \\

\noindent During the preparation of this work the authors used artificial intelligence to improve the language and readability of the manuscript, to assist with LaTeX formatting and the plotting scripts and to verify the bibliographic elements of the references. All output was reviewed, corrected and verified by the authors, who take full responsibility for the content of this publication. \\

\noindent The data that support the findings of this study are available from the corresponding author upon reasonable request.
\end{acknowledgments}

\begin{appendices}
\renewcommand{\thesubsection}{\thesection-\Roman{subsection}}
\makeatletter
\renewcommand{\p@subsection}{}
\makeatother

\section{Full derivation of LDOS in the low-energy approximation}
\label{app:LDOS_cartesian}

\noindent This appendix carries out the algebraic steps deferred from the main text. It contains (i) the derivation of the closed form of the retarded Green's function in the position-independent approximation, (ii) a  real expression for the LDOS, obtained by writing the complex square root in~\eqref{eq:LDOS_compact} in terms of its real and imaginary parts, and (iii) the low-energy expansion leading to~\eqref{eq:LDOS_O2}.\\

\noindent Integrating the Usadel equation~\eqref{eq:usadel} over the length $L$ of the diffusive region, under the position-independent approximation $\bar{G}(E,x)\approx\bar{G}(E)$ valid for $|E|\ll E_{\text{Th}}$, yields
\begin{equation}
\hbar D\left(\bar{G}\partial_x\bar{G}\big|_{x=L}-\bar{G}\partial_x\bar{G}\big|_{x=0}\right)+iL\,[E\tau_3,\bar{G}]_-=0\;.
\label{eq:app_usadel_integrated}
\end{equation}
In the strong-interface limit $z_{\text{S}}, z_{\text{N}}\gg 1$ the Tanaka--Nazarov boundary conditions~\eqref{eq:bc_S} and~\eqref{eq:bc_N} reduce to the Kupriyanov--Lukichev form~\cite{kuprianov1988influence, TanakaNazarov},
\begin{align}
\bar{G}\partial_x\bar{G}\big|_{x=0} &= \frac{1}{\gamma_{\text{BS}}L}[\bar{G}_S,\bar{G}]_-\;,\\
\bar{G}\partial_x\bar{G}\big|_{x=L} &= -\frac{1}{\gamma_{\text{BN}}L}[\bar{G}_N,\bar{G}]_-\;,
\label{eq:app_bc_KL}
\end{align}
where the relative sign reflects the outward-normal convention at the two interfaces. Substituting~\eqref{eq:app_bc_KL} into~\eqref{eq:app_usadel_integrated}, multiplying by $-L/(\hbar D)$ and using $E_{\text{Th}}=\hbar D/L^2$, the three commutators combine into one,
\begin{equation}
\left[\frac{1}{\gamma_{\text{BS}}}\bar{G}_S+\frac{1}{\gamma_{\text{BN}}}\bar{G}_N-i\frac{E}{E_{\text{Th}}}\tau_3,\;\bar{G}\right]_-=0\;.
\label{eq:app_commutator}
\end{equation}
Since the superconductor is a singlet with no exchange field, all objects are diagonal in spin space and the retarded sector reduces to a $2\times 2$ Nambu matrix $\hat{g}^\text{R}$. Equation~\eqref{eq:app_commutator} then reads $[\mathcal{M},\hat{g}^\text{R}]_-=0$ with
\begin{equation}
\mathcal{M}(E)\equiv\frac{1}{\gamma_{\text{BS}}}\hat{G}_S^\text{R}+\frac{1}{\gamma_{\text{BN}}}\hat{G}_N^\text{R}-i\frac{E}{E_{\text{Th}}}\tau_3\;.
\label{eq:app_M_def}
\end{equation}

\noindent Since $\mathcal{M}$ is a $2\times 2$ matrix that is not proportional to the identity, any matrix commuting with it is of the form $\hat{g}^\text{R}=a\hat{1}+\beta\mathcal{M}$. The matrix $\mathcal{M}$ is traceless, because $\hat{G}_S^\text{R}$, $\hat{G}_N^\text{R}$ and $\tau_3$ are, and the normalized Green's function is traceless as well. Taking the trace therefore gives $a=0$, so that $\hat{g}^\text{R}=\beta\mathcal{M}$. For a traceless $2\times 2$ matrix the Cayley--Hamilton theorem gives $\mathcal{M}^2=\tfrac{1}{2}\operatorname{Tr}\mathcal{M}^2\,\hat{1}$, and the normalization $(\hat{g}^\text{R})^2=\hat{1}$ fixes $\beta^2=1/(\tfrac{1}{2}\operatorname{Tr}\mathcal{M}^2)$, so that
\begin{equation}
\hat{g}^\text{R}(E)=\frac{\mathcal{M}(E)}{\sqrt{\tfrac{1}{2}\operatorname{Tr}\mathcal{M}^2(E)}}\;,
\label{eq:app_gR_formal}
\end{equation}
with the branch fixed by $\operatorname{Re}\operatorname{Tr}\tau_3\hat{g}^\text{R}\geq 0$. The reservoirs enter through $\hat{G}_N^\text{R}=\tau_3$ and the bulk BCS form
\begin{equation}
\hat{G}_S^\text{R}(E)=\frac{E\tau_3+i\Delta_0\tau_1}{\sqrt{E^2-\Delta_0^2+i0^+}}\;.
\label{eq:app_GS_BCS}
\end{equation}
For $|E|<\Delta_0$ we write $\sqrt{E^2-\Delta_0^2+i0^+}=i\Delta_0\sqrt{1-E^2/\Delta_0^2}$ and expand to second order in $E/\Delta_0$,
\begin{equation}
\hat{G}_S^\text{R}(E)\simeq -i\frac{E}{\Delta_0}\,\tau_3+\left(1+\frac{E^2}{2\Delta_0^2}\right)\tau_1+\mathcal{O}\!\left(\frac{E^3}{\Delta_0^3}\right).
\label{eq:app_GS_expanded}
\end{equation}
The $\tau_3$ component is odd in $E$ and receives its first correction at $\mathcal{O}(E^3)$, whereas the $\tau_1$ component is even in $E$ and is corrected already at $\mathcal{O}(E^2)$. In the main text this correction is dropped and only the linear form $\hat{G}_S^\text{R}\simeq-i(E/\Delta_0)\tau_3+\tau_1$ is retained. This requires motivation, since the scale $E^\ast$ is extracted from the $\mathcal{O}(E^2)$ term of the LDOS, to which the correction contributes at the same order in $E$. The justification is therefore not an order counting in $E$ but a parametric one. The $E^2$ terms that determine $E^\ast$ stem from $\mathcal{G}^2$ and are of order $\alpha^2E^2$, with $\alpha$ the rate defined in~\eqref{eq:app_alpha}, which contains the Thouless contribution $1/E_{\text{Th}}$. The correction to the $\tau_1$ component instead contributes a term of order $E^2/(\gamma_{\text{BS}}\Delta_0)^2$, which is smaller by a factor $(1+\gamma_{\text{BS}}\Delta_0/E_{\text{Th}})^{-2}$. In the strong-interface regime $\gamma_{\text{BS}}\Delta_0\gg E_{\text{Th}}$, and the correction is negligible. Below we keep it to make this statement quantitative. Inserting~\eqref{eq:app_GS_expanded} and $\hat{G}_N^\text{R}=\tau_3$ into~\eqref{eq:app_M_def},
\begin{equation}
\mathcal{M}(E)=\mathcal{G}(E)\,\tau_3+\mathcal{F}(E)\,\tau_1\;,
\label{eq:app_M_explicit}
\end{equation}
with the complex scalar
\begin{equation}
\mathcal{G}(E)\equiv\frac{1}{\gamma_{\text{BN}}}-i\!\left(\frac{1}{\gamma_{\text{BS}}}\frac{E}{\Delta_0}+\frac{E}{E_{\text{Th}}}\right)
\label{eq:app_zeta_def}
\end{equation}
and the real scalar
\begin{equation}
\mathcal{F}(E)\equiv\frac{1}{\gamma_{\text{BS}}}\left(1+\frac{E^2}{2\Delta_0^2}\right).
\label{eq:app_F_def}
\end{equation}
To linear order in $E$ one has $\mathcal{F}=\gamma_{\text{BS}}^{-1}$, which is the form used in the main text. The zero-energy values and the conductance are unaffected by the $E^2$ term in $\mathcal{F}$, which enters only the curvature $\kappa$. Using $\tau_3^2=\tau_1^2=\hat{1}$ and $\{\tau_1,\tau_3\}=\hat{0}$ gives $\mathcal{M}^2=(\mathcal{G}^2+\mathcal{F}^2)\hat{1}$, so that
\begin{equation}
\hat{g}^\text{R}(E)=\frac{\mathcal{G}(E)\,\tau_3+\mathcal{F}(E)\,\tau_1}{\sqrt{\mathcal{G}^2(E)+\mathcal{F}^2(E)}}\;,
\label{eq:app_gR_closed}
\end{equation}
and the LDOS follows from its $\tau_3$ component, reproducing the compact form~\eqref{eq:LDOS_compact} used below.\\

\noindent Starting from the compact form derived in Sec.~\ref{sec:gR_closed},
\begin{equation}
\frac{N(E)}{N_0}=\operatorname{Re}\!\left[\frac{\mathcal{G}(E)}{\sqrt{\mathcal{G}^2(E)+\mathcal{F}^2(E)}}\right],
\label{eq:app_LDOS_compact}
\end{equation}
we extract an explicitly real expression. We write the radicand in terms of its real and imaginary parts,
\begin{equation}
\mathcal{G}^2(E)+\mathcal{F}^2(E)=X(E)+iY(E),
\label{eq:app_radicand}
\end{equation}
and expand $\mathcal{G}^2=(\operatorname{Re}\mathcal{G})^2-(\operatorname{Im}\mathcal{G})^2+2i\,\operatorname{Re}\mathcal{G}\,\operatorname{Im}\mathcal{G}$, which gives
\begin{align}
X(E) &= (\operatorname{Re}\mathcal{G})^2 - (\operatorname{Im}\mathcal{G})^2 + \mathcal{F}^2, \label{eq:app_X}\\
Y(E) &= 2\,\operatorname{Re}\mathcal{G}\,\operatorname{Im}\mathcal{G}, \label{eq:app_Y}\\
R(E) &= \sqrt{X^2(E)+Y^2(E)}, \label{eq:app_R}
\end{align}
where $R=|X+iY|$ is the modulus of the radicand. The square root is written as $\sqrt{X+iY}=u+iv$ with real $u$ and $v$. Squaring and taking the modulus gives the three conditions $u^2-v^2=X$, $2uv=Y$ and $u^2+v^2=R$, which are solved by
\begin{equation}
u=\sqrt{\frac{R+X}{2}},\qquad v=\operatorname{sgn}(Y)\sqrt{\frac{R-X}{2}}.
\label{eq:app_uv}
\end{equation}
Since $R$ scales linearly with $X$ and $Y$, both $u$ and $v$ scale as the square root of the radicand, as they should. This is the principal branch, $u\geq 0$, which is the retarded branch since it ensures $N(E)/N_0\geq 0$. Rationalising with $u^2+v^2=R$, \begin{equation}
\frac{\mathcal{G}}{u+iv}=\frac{(\operatorname{Re}\mathcal{G})\,u+(\operatorname{Im}\mathcal{G})\,v}{R}+i\,\frac{(\operatorname{Im}\mathcal{G})\,u-(\operatorname{Re}\mathcal{G})\,v}{R}.
\label{eq:app_rationalize}
\end{equation} Taking the real part,
\begin{equation}
\boxed{\;\frac{N(E)}{N_0}=\frac{\operatorname{Re}\mathcal{G}\,u(E)+\operatorname{Im}\mathcal{G}\,v(E)}{R(E)}\;}
\label{eq:app_LDOS_real}
\end{equation}
with
\begin{equation}
\operatorname{Re}\mathcal{G}=\frac{1}{\gamma_{\text{BN}}},\qquad
\operatorname{Im}\mathcal{G}=-\alpha E,
\label{eq:app_zeta_components}
\end{equation}
where we introduced the energy-independent rate
\begin{equation}
\alpha \equiv \frac{1}{\gamma_{\text{BS}}\Delta_0}+\frac{1}{E_{\text{Th}}}\;. \label{eq:app_alpha}
\end{equation}
With the shorthand
\begin{align}
\mathcal{X}(E) &\equiv \frac{1}{\gamma_{\text{BN}}^2}+\mathcal{F}^2(E)-\alpha^2E^2, \label{eq:app_calX}\\
R(E) &= \sqrt{\mathcal{X}^2(E)+\frac{4\,\alpha^2E^2}{\gamma_{\text{BN}}^2}}, \label{eq:app_R_explicit}
\end{align}
the normalized LDOS in the strong-interface, low-energy limit reads
\begin{equation}
\frac{N(E)}{N_0} = \frac{\dfrac{1}{\gamma_{\text{BN}}}\sqrt{\dfrac{R(E) + \mathcal{X}(E)}{2}}
\;+\;\alpha |E|\sqrt{\dfrac{R(E) - \mathcal{X}(E)}{2}}}{R(E)},
\label{eq:LDOS_monster}
\end{equation}
expressed entirely in the physical parameters $\gamma_{\text{BS}},\gamma_{\text{BN}},\Delta_0,E_{\text{Th}}$ and the quasiparticle energy $E$. The sign of the second term in the numerator follows from $Y=-2\alpha E/\gamma_{\text{BN}}$, so that $v$ has the sign of $-E$ and $\operatorname{Im}\mathcal{G}\,v=+\alpha|E|\sqrt{(R-\mathcal{X})/2}$ for either sign of $E$. The LDOS is thus even in $E$, as required by particle-hole symmetry. Setting $E=0$ collapses~\eqref{eq:LDOS_monster} to the zero-bias value
\begin{equation}
\frac{N(0)}{N_0} = \frac{\gamma_{\text{BS}}}{\sqrt{\gamma_{\text{BS}}^2+\gamma_{\text{BN}}^2}},
\label{eq:app_LDOS_zero}
\end{equation}
reproducing Eq.~\eqref{eq:LDOS_zero} of the main text.\\

\noindent We now derive the quadratic form~\eqref{eq:LDOS_O2} quoted in the main text. It is convenient to work directly from Eq.~\eqref{eq:app_LDOS_compact} with the notation of Sec.~\ref{sec:expansion}: $\mathcal{G}(E)=\mathcal{G}_0+i\,\mathcal{G}_1(E)$ with $\mathcal{G}_0=\gamma_{\text{BN}}^{-1}$ and $\mathcal{G}_1(E)=-\alpha E$. Defining $\Omega^2\equiv\mathcal{G}_0^2+\gamma_{\text{BS}}^{-2}$, the radicand is
\begin{equation}
\mathcal{G}^2(E)+\mathcal{F}^2(E)=\Omega^2\!\left[1+\delta(E)\right],
\label{eq:app_delta_def}
\end{equation}
with, to second order in $E$,
\begin{equation}
\delta(E)=-\frac{2i\,\mathcal{G}_0\,\alpha}{\Omega^2}E-\frac{\alpha^2}{\Omega^2}E^2+\frac{1}{\gamma_{\text{BS}}^2\,\Omega^2}\frac{E^2}{\Delta_0^2}\;.
\label{eq:app_delta_explicit}
\end{equation}
The first two terms stem from $\mathcal{G}^2$ and the last from $\mathcal{F}^2=\gamma_{\text{BS}}^{-2}\bigl(1+E^2/\Delta_0^2\bigr)+\mathcal{O}(E^4)$. Expanding the inverse square root directly,
\begin{equation}
\frac{1}{\sqrt{1+\delta}}= 1-\frac{\delta}{2}+\frac{3\,\delta^2}{8}+\mathcal{O}(\delta^3),
\label{eq:app_inv_sqrt}
\end{equation}
we obtain
\begin{equation}
\frac{\mathcal{G}(E)}{\sqrt{\mathcal{G}^2(E)+\mathcal{F}^2(E)}}
=\frac{\mathcal{G}_0+i\,\mathcal{G}_1(E)}{\Omega}\!\left[1-\frac{\delta(E)}{2}+\frac{3\,\delta^2(E)}{8}+\dots\right].
\label{eq:app_ratio_expanded}
\end{equation}

\noindent Substituting~\eqref{eq:app_delta_explicit} into~\eqref{eq:app_ratio_expanded} and collecting terms by powers of $E$, the $E^0$ coefficient reduces to $\mathcal{G}_0/\Omega=\gamma_{\text{BS}}/\sqrt{\gamma_{\text{BS}}^2+\gamma_{\text{BN}}^2}$, reproducing~\eqref{eq:app_LDOS_zero}. The $E^1$ coefficient is purely imaginary and drops out of the real part. The real part of the $E^2$ coefficient has four contributions:
(i) the $(-\delta/2)$ term acting on $\mathcal{G}_0/\Omega$, which contributes $+\mathcal{G}_0\,\mathcal{G}_1^2/(2\,\Omega^3)$ through the $-\alpha^2E^2/\Omega^2$ term of $\delta$;
(ii) the $(3\delta^2/8)$ term acting on $\mathcal{G}_0/\Omega$, which contributes $-3\,\mathcal{G}_0^3\,\mathcal{G}_1^2/(2\,\Omega^5)$ through $(2i\mathcal{G}_0\mathcal{G}_1/\Omega^2)^2=-4\mathcal{G}_0^2\mathcal{G}_1^2/\Omega^4$;
(iii) the $(-\delta/2)$ term acting on $i\mathcal{G}_1/\Omega$, which contributes $+\mathcal{G}_0\,\mathcal{G}_1^2/\Omega^3$ through the term linear in $E$ of $\delta$;
and (iv) the $(-\delta/2)$ term acting on $\mathcal{G}_0/\Omega$, which contributes $-\mathcal{G}_0E^2/(2\gamma_{\text{BS}}^2\Delta_0^2\Omega^3)$ through the last term of $\delta$.
Contributions (i) to (iii) sum to
\begin{equation}
\frac{3\,\mathcal{G}_0\,\mathcal{G}_1^2}{2\,\Omega^3}-\frac{3\,\mathcal{G}_0^3\,\mathcal{G}_1^2}{2\,\Omega^5}
=\frac{3\,\mathcal{G}_0\,\mathcal{G}_1^2}{2\,\Omega^5}\bigl(\Omega^2-\mathcal{G}_0^2\bigr)
=\frac{3\,\mathcal{G}_0\,\alpha^2E^2}{2\,\gamma_{\text{BS}}^2\,\Omega^5},
\label{eq:app_E2_intermediate}
\end{equation}
where we used $\Omega^2-\mathcal{G}_0^2=\gamma_{\text{BS}}^{-2}$. Adding (iv), the $E^2$ coefficient is
\begin{equation}
\operatorname{Re}\!\left.\frac{\mathcal{G}}{\sqrt{\mathcal{G}^2+\mathcal{F}^2}}\right|_{E^2}
=\frac{\mathcal{G}_0}{\gamma_{\text{BS}}^2\,\Omega^3}\left(\frac{3\,\alpha^2}{2\,\Omega^2}-\frac{1}{2\Delta_0^2}\right).
\label{eq:app_E2_result}
\end{equation}
Dividing by the zero-energy value $N(0)/N_0=\mathcal{G}_0/\Omega$ gives the dimensionless curvature
\begin{equation}
\kappa=\frac{1}{\gamma_{\text{BS}}^2\,\Omega^2}\left(\frac{3\,\alpha^2}{2\,\Omega^2}-\frac{1}{2\Delta_0^2}\right).
\label{eq:app_kappa_compact}
\end{equation}
Using $\Omega^2=(\gamma_{\text{BS}}^2+\gamma_{\text{BN}}^2)/(\gamma_{\text{BS}}^2\gamma_{\text{BN}}^2)$,
\begin{equation}
\kappa=\frac{3}{2}\,\frac{\gamma_{\text{BN}}^{4}\,\gamma_{\text{BS}}^{2}}{(\gamma_{\text{BN}}^{2}+\gamma_{\text{BS}}^{2})^{2}}\,\alpha^2
-\frac{1}{2\Delta_0^2}\,\frac{\gamma_{\text{BN}}^{2}}{\gamma_{\text{BN}}^{2}+\gamma_{\text{BS}}^{2}}\;,
\label{eq:app_kappa}
\end{equation}
with $\alpha^2=(\gamma_{\text{BS}}\Delta_0+E_{\text{Th}})^2/(\gamma_{\text{BS}}\Delta_0\,E_{\text{Th}})^2$. The second term originates from the energy dependence of the anomalous Green's function of the superconductor, Eq.~\eqref{eq:app_GS_expanded}. It reduces the curvature, since the pair amplitude injected from S grows as $E$ approaches the gap. Its ratio to the first term is
\begin{equation}
\frac{\gamma_{\text{BN}}^2+\gamma_{\text{BS}}^2}{3\,\gamma_{\text{BN}}^2}\left(1+\frac{\gamma_{\text{BS}}\Delta_0}{E_{\text{Th}}}\right)^{-2},
\label{eq:app_kappa_ratio}
\end{equation}
which confirms the parametric estimate given below~\eqref{eq:app_GS_expanded}. In the regime $\gamma_{\text{BS}}\Delta_0\gg E_{\text{Th}}$ the first term dominates. We note that the position-independent approximation itself carries corrections of relative order $R_\text{d}/R_\text{b}\sim\gamma_{\text{BS,BN}}^{-1}$, which in the strong-interface regime exceed the second term in~\eqref{eq:app_kappa}.\\

\noindent To verify the linear form of $\hat{G}_S^\text{R}$ beyond the curvature, we compare the resulting LDOS with the LDOS obtained from the  BCS form~\eqref{eq:app_GS_BCS}, keeping all other approximations identical. Two properties of the strong-interface regime make the linear form accurate. First, the closed-form expressions are applied for $|E|\lesssim E^\ast$, and $E^\ast\ll\Delta_0$, so the expansion parameter $(E/\Delta_0)^2$ remains small. Second, the neglected curvature term is suppressed according to~\eqref{eq:app_kappa_ratio}. For the parameter sets used in this work, with $3\leq\gamma_{\text{BS}}\leq 30$, $5\leq\gamma_{\text{BN}}\leq 30$ and $0.05\leq E_{\text{Th}}/\Delta_0\leq 0.3$, one has $\gamma_{\text{BS}}\Delta_0/E_{\text{Th}}\geq 10$ and $E^\ast\lesssim 0.08\,\Delta_0$. The neglected term then changes $\kappa$ by less than $0.3\%$, and the two LDOS differ by less than $1\%$ for $|E|\leq E^\ast$. For instance, for $\gamma_{\text{BS}}=3$, $\gamma_{\text{BN}}=5$ and $E_{\text{Th}}=0.05\,\Delta_0$ the maximum deviation is $8\times10^{-5}$. Appreciable deviations occur only for $\gamma_{\text{BS}}\sim 1$, where $\gamma_{\text{BS}}\Delta_0\sim E_{\text{Th}}$ and $E^\ast$ approaches $\Delta_0$. For $\gamma_{\text{BS}}=1$, $\gamma_{\text{BN}}=10$ and $E_{\text{Th}}=0.4\,\Delta_0$ the deviation reaches $13\%$. Such parameters lie outside the strong-interface regime $\gamma_{\text{BS,BN}}\gg 1$ for which the closed form is derived, and are treated numerically in the main text.\\

\noindent The particle-in-a-box limit $\gamma_{\text{BS}}=\gamma_{\text{BN}}\equiv\gamma_B\gg 1$ simplifies~\eqref{eq:app_kappa} to
\begin{equation}
\kappa\bigr|_{\gamma_{\text{BS}}=\gamma_{\text{BN}}}=\frac{3}{8}\,\frac{(\gamma_B\Delta_0+E_{\text{Th}})^2}{\Delta_0^2\,E_{\text{Th}}^2}-\frac{1}{4\Delta_0^2},
\label{eq:app_kappa_sym}
\end{equation}
whence, for $\gamma_B\Delta_0\gg E_{\text{Th}}$,
\begin{equation}
E^\ast\bigr|_{\gamma_{\text{BS}}=\gamma_{\text{BN}}}\simeq\sqrt{\tfrac{8}{3}}\,\frac{\Delta_0\,E_{\text{Th}}}{\gamma_B\Delta_0+E_{\text{Th}}},
\label{eq:app_Estar_sym}
\end{equation}
which reduces further to $E^\ast\simeq\sqrt{8/3}\,E_{\text{Th}}/\gamma_B$, reproducing Eq.~\eqref{eq:Estar_PIB} of the main text.

\section{Unification of derivation of LDOS in the low-energy approximation}
\label{app:keldysh}
 
\noindent This appendix derives the bulk distribution functions and the differential
conductance of Sec.~\ref{sec:conductance}. The logic parallels the spectral
problem of Sec.~\ref{sec:gR_closed}. The  Keldysh--Nambu Green's function
obeys the same boundary-integrated commutator that fixed $\check{g}^R$, but
the reservoirs now enter as nonequilibrium sources rather than as purely
spectral objects. Projecting that commutator onto the Pauli channels yields a
closed kinetic equation for the bulk longitudinal and transverse
distributions, and the $\tau_3$ trace of the matrix current then delivers the
conductance. Beyond the LDOS one further spectral input is needed, the
$\tau_1$ component of $\check{g}^R$, which we establish first. Throughout we
abbreviate the retarded components $g_{1,3}\equiv\check{g}_{1,3}^R$.
 
\subsection{Spectral input and the kinetic equation}
 
\noindent Besides the $\tau_3$ component $g_3=\mathcal{G}/\sqrt{\mathcal{G}^2+\gamma_{\text{BS}}^{-2}}$,
whose real part is the LDOS, the retarded Green's function~\eqref{eq:gR_formal}
carries a $\tau_1$ component
\begin{equation}
g_1(E)=\frac{\gamma_{\text{BS}}^{-1}}{\sqrt{\mathcal{G}^2(E)+\gamma_{\text{BS}}^{-2}}}.
\label{eq:g1_def}
\end{equation}
With the Cartesian decomposition $\mathcal{G}^2+\gamma_{\text{BS}}^{-2}=X+iY$ and
$u,v,R$ as defined in Eqs.~\eqref{eq:app_X}--\eqref{eq:app_uv},
\begin{equation}
\mathrm{Re}\,g_1(E)=\frac{1}{\gamma_{\text{BS}}}\,\mathrm{Re}\,\frac{1}{u+iv}
=\frac{u(E)}{\gamma_{\text{BS}}\,R(E)}.
\label{eq:Reg1}
\end{equation}
At zero energy $Y=0$, $v=0$, $u=\sqrt{X}=\sqrt{\gamma_{\text{BN}}^{-2}+\gamma_{\text{BS}}^{-2}}$
and $R=X$, so
\begin{equation}
\mathrm{Re}\,g_1(0)=\frac{\gamma_{\text{BN}}}{\sqrt{\gamma_{\text{BN}}^2+\gamma_{\text{BS}}^2}},
\label{eq:Reg1_zero}
\end{equation}
the counterpart of the zero-bias LDOS~\eqref{eq:LDOS_zero} with $\gamma_{\text{BN}}$
and $\gamma_{\text{BS}}$ interchanged. This second spectral weight enters the
conductance on the same footing as the LDOS, and the interchange already
foreshadows the $\gamma_{\text{BN}}\leftrightarrow\gamma_{\text{BS}}$ symmetry of the
zero-bias conductance.\\
 
\noindent The boundary integration carries over
to the Keldysh--Nambu structure, giving $[\bar{\mathcal{M}},\bar{g}]_-=0$
with $\bar{g}$ triangular in Keldysh space and
$\bar{\mathcal{M}}=\gamma_{\text{BS}}^{-1}\bar{G}_S+\gamma_{\text{BN}}^{-1}\bar{G}_N
-i(E/E_{\text{Th}})\bar{\tau}_3$, $\bar{\tau}_3=\mathrm{diag}(\tau_3,\tau_3)$. Its
upper-right block is the kinetic equation
\begin{equation}
\mathcal{M}^R\check{g}^K-\check{g}^K\mathcal{M}^A
+\mathcal{M}^K\check{g}^A-\check{g}^R\mathcal{M}^K=0,
\label{eq:keldysh_eq}
\end{equation}
with $\mathcal{M}^R=M_1\tau_1+M_3\tau_3$, $M_1=\gamma_{\text{BS}}^{-1}$ real and
$M_3=\mathcal{G}$ ($\mathrm{Re}\,M_3=\gamma_{\text{BN}}^{-1}$,
$\mathrm{Im}\,M_3=-\alpha E$), and the advanced source
$\mathcal{M}^A=-\tau_3(\mathcal{M}^R)^\dagger\tau_3=M_1\tau_1-M_3^\ast\tau_3$,
the $\tau_1$ term keeping a plus sign through $\tau_3\tau_1\tau_3=-\tau_1$. The
retarded blocks are already known, so the new content is the source
$\mathcal{M}^K=\gamma_{\text{BS}}^{-1}\check{G}_S^K+\gamma_{\text{BN}}^{-1}\check{G}_N^K$.\\
 
\noindent In a normal reservoir at voltage $V$ and temperature $T$,
$\check{G}_N^K=(\check{G}_N^R-\check{G}_N^A)\check{h}_N$ with
$\check{h}_N=f_L^N\tau_0+f_T^N\tau_3$ and
\begin{equation}
f_{L,T}^{N}(E)=\tfrac{1}{2}\!\left[\tanh\tfrac{E+eV}{2T}\pm\tanh\tfrac{E-eV}{2T}\right].
\label{eq:fLT_N}
\end{equation}
Since $\check{G}_N^R-\check{G}_N^A=2\tau_3$ commutes with $\tau_0$ and $\tau_3$,
\begin{equation}
\check{G}_N^K=2 f_L^N\,\tau_3+2 f_T^N\,\tau_0,
\label{eq:GK_N}
\end{equation}
so the $\tau_0$ channel is sourced by the transverse $f_T^N$ and the $\tau_3$
channel by the longitudinal $f_L^N$. The superconducting reservoir at $V_S=0$
has scalar distribution $f_S=\tanh(E/2T)$ and
$\check{G}_S^K=(\check{G}_S^R-\check{G}_S^A)f_S$. Writing
$\check{G}_S^R=g_1^S\tau_1+g_3^S\tau_3$ and using
$\check{G}_S^A=-\tau_3(\check{G}_S^R)^\dagger\tau_3$ with
$\tau_3\tau_1\tau_3=-\tau_1$,
\begin{equation}
\check{G}_S^R-\check{G}_S^A=2\,\mathrm{Re}\,g_3^S\,\tau_3+2i\,\mathrm{Im}\,g_1^S\,\tau_1.
\label{eq:GS_diff}
\end{equation}
The subgap BCS form $g_3^S=-iE/\sqrt{\Delta_0^2-E^2}$ (purely imaginary) and
$g_1^S=\Delta_0/\sqrt{\Delta_0^2-E^2}$ (purely real) makes
$\mathrm{Re}\,g_3^S=0$ and $\mathrm{Im}\,g_1^S=0$, hence
\begin{equation}
\check{G}_S^R-\check{G}_S^A=0,\qquad \check{G}_S^K=0\qquad(|E|<\Delta_0).
\label{eq:GS_K_subgap}
\end{equation}
The gap forbids quasiparticle states in the superconducting reservoir, so it
injects no Keldysh weight into the diffusive bar and every subgap
nonequilibrium source originates from the normal electrode. With
$\mathcal{M}^K=M_0^K\tau_0+M_1^K\tau_1+M_3^K\tau_3$, the source collapses to
\begin{equation}
M_0^K=\frac{2 f_T^N}{\gamma_{\text{BN}}},\qquad M_1^K=0,\qquad
M_3^K=\frac{2 f_L^N}{\gamma_{\text{BN}}}\qquad(E<\Delta_0),
\label{eq:MK}
\end{equation}
real and confined to the $\tau_0$ and $\tau_3$ channels.\\
 
\noindent The bulk distribution enters through $\check{g}^K=\check{g}^R\check{h}-\check{h}\check{g}^A$
with Hermitian $\check{h}$. No spin-active mechanism generates a $\tau_2$
component and no supercurrent flows along the transport axis to generate a
$\tau_1$ component, leaving $\check{h}=f_L\tau_0+f_T\tau_3$ with $f_L$ and
$f_T$ the bulk longitudinal and transverse distributions. Using
$\check{g}^R=g_1\tau_1+g_3\tau_3$, $\check{g}^A=g_1^\ast\tau_1-g_3^\ast\tau_3$
and $\tau_1\tau_3=-i\tau_2$, $\tau_3\tau_1=i\tau_2$,
\begin{equation}
\begin{aligned}
g_0^K&=2\,\mathrm{Re}\,g_3\,f_T, &\qquad g_1^K&=2i\,\mathrm{Im}\,g_1\,f_L,\\
g_2^K&=-2i\,\mathrm{Re}\,g_1\,f_T, &\qquad g_3^K&=2\,\mathrm{Re}\,g_3\,f_L.
\end{aligned}
\label{eq:gK}
\end{equation}
As a check, a normal reservoir has $g_1=0$ and $g_3=1$, whence
$\check{g}^K=2 f_T\tau_0+2 f_L\tau_3=(\check{g}^R-\check{g}^A)\check{h}$.
Substituting Eq.~\eqref{eq:gK} together with $\mathcal{M}^{R,A,K}$ into the
kinetic equation~\eqref{eq:keldysh_eq} and grouping by Pauli channel gives
\begin{align}
\tau_0:\quad & 2\,\mathrm{Re}\,M_3\,g_3^K-2i\,\mathrm{Im}\,g_1\,M_1^K
-2\,\mathrm{Re}\,g_3\,M_3^K=0,\label{eq:proj0}\\
\tau_3:\quad & 2i\,M_1\,g_2^K+2\,\mathrm{Re}\,M_3\,g_0^K
-2\,\mathrm{Re}\,g_3\,M_0^K=0,\label{eq:proj3}
\end{align}
the $\tau_1$ channel duplicating the $\tau_3$ equation for $f_T$ and the
$\tau_2$ channel reducing to an identity on the spectral manifold
$g_1 M_3=g_3 M_1$. Equation~\eqref{eq:proj0} fixes the longitudinal
distribution,
\begin{equation}
f_L=\frac{M_3^K\,\mathrm{Re}\,g_3+i\,M_1^K\,\mathrm{Im}\,g_1}
{2\,\mathrm{Re}\,M_3\,\mathrm{Re}\,g_3},
\label{eq:fL}
\end{equation}
which is real since $M_1^K$ is purely imaginary so that $iM_1^K$ is real. In
the subgap window $M_1^K=0$ and the spectral weight $\mathrm{Re}\,g_3$ cancels,
\begin{equation}
f_L=\frac{M_3^K}{2\,\mathrm{Re}\,M_3}=f_L^N\qquad(|E|<\Delta_0).
\label{eq:fL_subgap}
\end{equation} Thus, in the subgap regime, due to the absence of quasiparticles in the superconductor, the longitudinal distribution is determined entirely by the normal metal.
With the superconductor injecting nothing subgap, the only source of
longitudinal occupation is the normal lead and the bulk inherits its
distribution. Equation~\eqref{eq:proj3} fixes the transverse distribution,
\begin{equation}
f_T
=f_T^N\,\frac{\gamma_{\text{BS}}\,\mathrm{Re}\,g_3}{\gamma_{\text{BN}}\,\mathrm{Re}\,g_1+\gamma_{\text{BS}}\,\mathrm{Re}\,g_3},
\label{eq:fT_subgap}
\end{equation}
which is Eq.~\eqref{eq:fT_bulk}. The $\tau_1$ channel gives the equivalent form
$f_T=-M_0^K\,\mathrm{Im}\,g_1/(2\,\mathrm{Im}\,M_3\,\mathrm{Re}\,g_1)$, of the
indeterminate form $0/0$ at $E=0$, so Eq.~\eqref{eq:fT_subgap} is preferred for
evaluation. The combination that drives the current is
\begin{equation}
f_T^N-f_T=f_T^N\,\frac{\gamma_{\text{BN}}\,\mathrm{Re}\,g_1}
{\gamma_{\text{BN}}\,\mathrm{Re}\,g_1+\gamma_{\text{BS}}\,\mathrm{Re}\,g_3}\ge 0,
\label{eq:fTdiff}
\end{equation}
which vanishes at $V=0$ and tends to zero when the superconducting interface
decouples ($\gamma_{\text{BS}}\to\infty$, $g_1\to 0$, $g_3\to 1$), so the current
vanishes whenever either reservoir is detached.
 
\subsection{Charge current and differential conductance}
 
\noindent The matrix current~\eqref{eq:current} is evaluated at the DN/N boundary,
where current conservation makes the algebra simplest. With
$\check{G}_N^R=\tau_3$, $\check{G}_N^A=-\tau_3$ and $\check{G}_N^K$ from
Eq.~\eqref{eq:GK_N}, the $\tau_3$ component of $[\check{g},\check{G}_N]^K$ is,
using $g_0^K=2\,\mathrm{Re}\,g_3\,f_T$,
\begin{equation}
\big([\check{g},\check{G}_N]_-^K\big)_3=4 f_T^N\,\mathrm{Re}\,g_3-2 g_0^K
=4\,\mathrm{Re}\,g_3\,(f_T^N-f_T),
\label{eq:comm_tau3}
\end{equation}
the $\tau_1$ and $\tau_2$ parts of $\check{g}^K$ dropping out of the $\tau_3$
trace. The current is therefore
\begin{equation}
I=\frac{\sigma_D A}{eL\gamma_{\text{BN}}}\int_{-\infty}^{+\infty}\!dE\;
\mathrm{Re}\,g_3(E)\,[f_T^N(E,V)-f_T(E,V)],
\label{eq:I_app}
\end{equation}
where the factor $1/L$ stems from the boundary condition~\eqref{eq:bc_N} and the trace over Nambu and spin space cancels the factor $1/16$ of Eq.~\eqref{eq:current}. Inserting
Eq.~\eqref{eq:fTdiff} cancels the $\gamma_{\text{BN}}^{-1}$ against the numerator and
reproduces Eq.~\eqref{eq:current_main}, in which all voltage dependence sits in
$f_T^N$. At $T\to 0$ the transverse reservoir distribution is a top-hat with
$\partial_V f_T^N=\tfrac{e}{2}[\delta(E-eV)+\delta(E+eV)]$, and the parities
$\mathrm{Re}\,g_{1,3}(-eV)=\mathrm{Re}\,g_{1,3}(eV)$ make the two delta
functions add, giving the differential conductance~\eqref{eq:didv}. Setting
$eV=0$ with $\mathrm{Re}\,g_1(0)$ from Eq.~\eqref{eq:Reg1_zero} and
$\mathrm{Re}\,g_3(0)=\gamma_{\text{BS}}/\sqrt{\gamma_{\text{BN}}^2+\gamma_{\text{BS}}^2}$ yields the
zero-bias conductance~\eqref{eq:zbc}.

\section{Supplementary numerical comparisons}
\label{app:numerics}

\noindent This appendix collects additional comparisons between the analytical
closed forms of Secs.~\ref{sec:gR_closed}--\ref{sec:conductance} and the
 Tanaka--Nazarov numerics described in Sec.~\ref{sec:numerical_method}.
These figures complement the main-text discussion by systematically varying
one parameter at a time while holding the others fixed, and by showing
the $z$-dependence of the agreement.

\subsection{Numerical method}
\label{sec:numerical_method}
\noindent The numerical results are obtained by solving the Usadel equation with Tanaka-Nazarov boundary conditions on both interfaces, without invoking the Kupriyanov–Lukichev reduction or the position-independent bulk approximation used in the analytical treatment.\\

\noindent The retarded Usadel equation is parametrized in terms of Riccati amplitudes $\Gamma(x)$, $\tilde{\Gamma}(x)$, each a $2\times 2$ spin matrix decomposed as $\Gamma = \gamma_0 \sigma_0 + \gamma_1 \sigma_x + \gamma_2 \sigma_y +\gamma_3 \sigma_z$ (and analogously for $\tilde{\Gamma}$), giving a system of twelve coupled first-order ordinary differential equations in the normalized coordinate $x/L \in [0,1]$.
The spatial grid consists of $N_x = 100$ points, and the boundary-value problem is solved with a fourth-order collocation method (\textsc{bvp4c} in \textsc{Matlab}) at relative and absolute tolerances of $10^{-5}$ and $10^{-7}$, respectively.
Each subsequent energy step reuses the converged solution of the previous step as an initial guess, which both accelerates convergence and reduces the likelihood of jumping to an unphysical branch.\\

\noindent At both the superconducting ($x = 0$) and normal ($x = L$) interfaces, the boundary conditions implement the Tanaka-Nazarov matrix current with explicit angular integration over the injection angle $\phi$.
The transparency of each channel is $T(\phi, z) = \cos^2\phi / (\cos^2\phi + z^2)$, where $z$ ($z_N$) is the dimensionless barrier parameter at the S (N) interface.
The angular sum runs over $N_\phi = 100$ equidistant points in $\phi \in (-\pi/2, \pi/2)$, weighted by the flux factor $\cos\phi$.
The interface matrix current for each channel is constructed from the  Nazarov circuit-theory expression, including the denominator $2(2 - T) + T\{C, G\}$ that captures finite-transparency corrections beyond the tunneling limit.
The result is divided by the input parameter $\gamma_B$ ($\gamma_{\text{BN}}$), which sets the overall interface resistance scale independently of the barrier strength $z$ ($z_N$).\\

\noindent The Keldysh sector is solved on the same spatial grid, using
the converged retarded solution as input. For the spin-degenerate case
considered in the $s$-wave analytics, the distribution function in the
diffusive bar is parametrized by two scalar functions: the longitudinal
mode $f_L(E)$, which describes the energy distribution, and the
transverse mode $f_T(E)$, which carries the charge current. In
equilibrium at temperature $T$ and bias voltage $V$, the normal
reservoir fixes these to
\begin{align}
    f_L^N(E) &= \tanh\!\left(\frac{E}{2k_BT}\right)\,, \\
    f_T^N(E,V) &= \frac{1}{2}\!\left[
  \tanh\!\left(\frac{E+eV}{2k_BT}\right)
  - \tanh\!\left(\frac{E-eV}{2k_BT}\right)
\right]\;.
\end{align}
At zero temperature $f_T^N$ reduces to a top-hat of width $2eV$. When
spin-dependent effects are present (exchange field, triplet pairing),
$f_L$ and $f_T$ each acquire three additional spin components
$f_{L,i}$ and $f_{T,i}$ ($i = 1,2,3$), as described in
Eq.~\eqref{eq:fcheck}; these vanish identically in the $s$-wave case
and are retained only in the numerical implementation for the
unconventional-pairing results of Secs.~\ref{sec:helical_p}
and~\ref{sec:3DBW}.\\

The differential conductance is obtained from two separate Keldysh solves at each energy $E$, one for an energy below the bias window, $E<-eV$, where $f_L^N=-1$ and $f_T^N=0$, and one for an energy inside the window, $-eV<E<eV$, where $f_L^N=0$ and $f_T^N=1$. We denote by $I_0^\text{low}(E)$ and $I_0^\text{middle}(E)$ the corresponding spectral current densities, i.e.\ the integrand of Eq.~\eqref{eq:current} at energy $E$, extracted from the $\tau_3$ trace of $(\bar{G}\partial_x\bar{G})^K$ at $x=0$. At $T=0$ the bias enters only through the position of the window edge $E=-eV$, and moving this edge by $e\,dV$ changes the occupation of the states at that energy from the first to the second case. The differential conductance is therefore the jump of the spectral current density at the window edge,
\begin{equation}
\frac{\sigma(eV)}{\sigma_\text{N}}
=\frac{I_0^\text{low}(E)-I_0^\text{middle}(E)}{\bigl[I_0^\text{low}(E)-I_0^\text{middle}(E)\bigr]_{\Delta=0}}\Bigg|_{E=-eV},
\label{eq:app_dIdV_num}
\end{equation}
where the denominator is the same quantity in the normal state, computed in a separate simulation with $\Delta=0$, which gives the normal-state conductance of the junction.

\noindent The energy grid spans $eV/\Delta_0 \in [0, 2]$ with $N_E = 301$ uniformly spaced points, corresponding to a resolution $\delta(eV/\Delta_0) = 0.01$.
A Dynes-type broadening $\eta$ is included in the quasiparticle energy as $E \to E + i\eta$. All simulations are performed at $T = 0$.\\

\noindent The local density of states is obtained directly from the retarded Green's function as $N(E)/N_0 = \tfrac{1}{2}\operatorname{Re}\operatorname{Tr}[\tau_3\, g^R(E, x)]$, evaluated at the centre of the bar ($x/L = 0.5$).
Position independence of the LDOS across the bar has been verified for all parameter sets presented.
\subsection{\texorpdfstring{$s$}{s}-wave zero-bias conductance}
\label{app:swave_zbc}
 
\begin{figure}[t]
    \centering
    \includegraphics[width=\columnwidth]{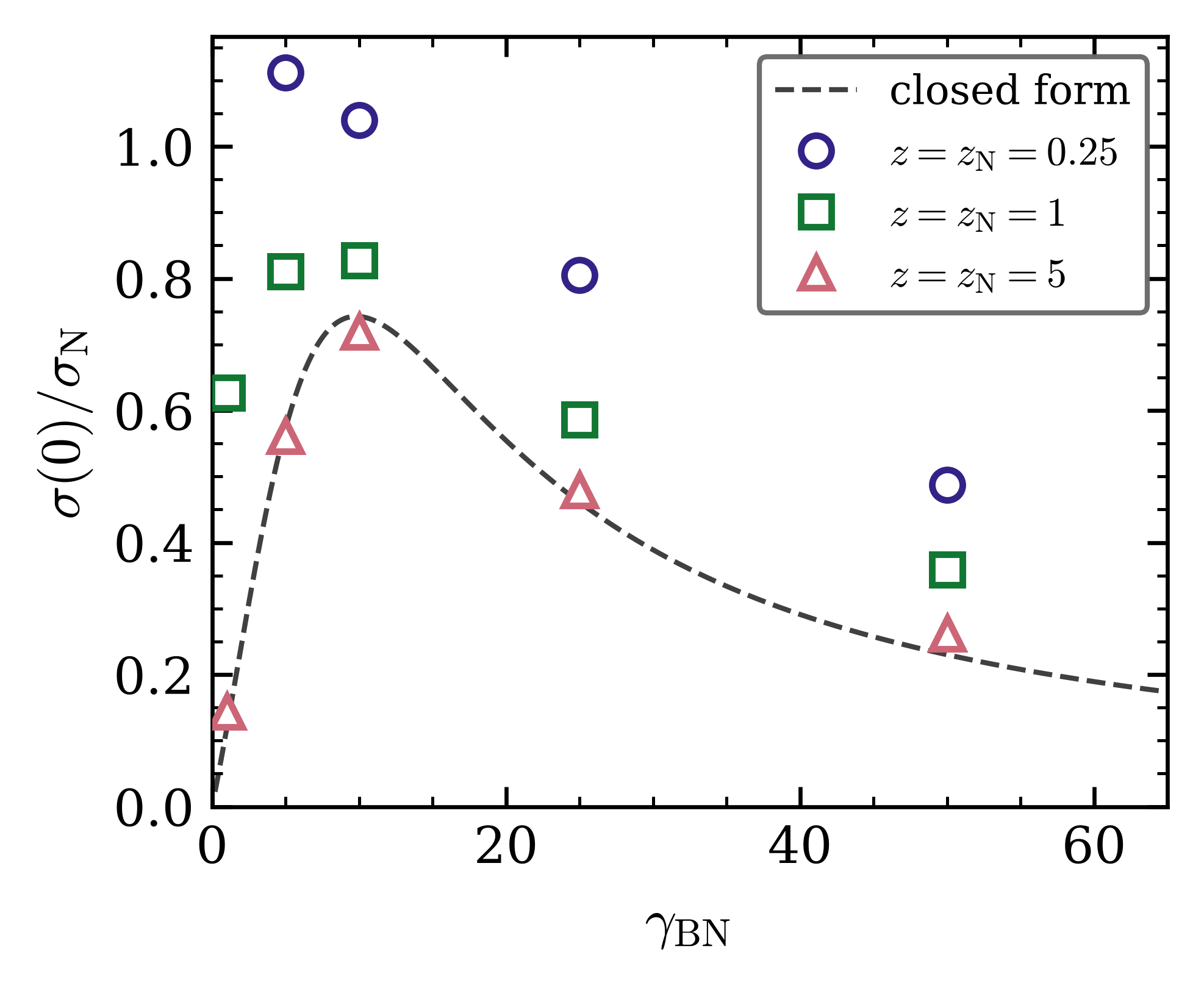}
    \caption{Zero-bias conductance $\sigma(0)/\sigma_\text{N}$ as a function of $\gamma_{\text{BN}}$ at fixed $\gamma_{\text{BS}} = 10$, for $z_\text{S} = z_\text{N} \in \{0.25,\, 1,\, 5\}$, normalized by the normal-state conductance of the junction. Symbols: numerical solution with Tanaka--Nazarov boundary conditions. For $z = 1$ and $z = 5$ the symbols for $E_{\text{Th}}/\Delta_0 \in \{0.1,\, 0.2,\, 0.4\}$ lie on top of each other; for $z = 0.25$ only $E_{\text{Th}}/\Delta_0 = 0.4$ is shown. Dashed curve: closed-form result, Eq.~\eqref{eq:zbc}, without fit parameters.}
    \label{fig:app_zbc_gBN}
\end{figure}
 
\noindent Figure~\ref{fig:app_zbc_gBN} compares the closed-form zero-bias conductance, Eq.~\eqref{eq:zbc}, with the numerical results. The factor $1+\gamma_{\text{BS}}+\gamma_{\text{BN}}=R_\text{N}/R_\text{d}$ in Eq.~\eqref{eq:zbc} arises because the boundary-integrated current is proportional to the conductance of the bar, whereas both the closed form and the numerics are normalized by the normal-state conductance of the junction. Keeping the term $1$, which is beyond the accuracy of the strong-interface limit, moves the maximum at $\gamma_{\text{BS}}=10$ from $\gamma_{\text{BN}}=10$ to $\gamma_{\text{BN}}\simeq 9.8$.
Because of the prefactor, the maximum at fixed $\gamma_{\text{BS}}$ no longer lies exactly at $\gamma_{\text{BN}}=\gamma_{\text{BS}}/\sqrt{2}$, as for Eq.~\eqref{eq:zbc}, but at $\gamma_{\text{BN}}\simeq 9.8$ for $\gamma_{\text{BS}}=10$.\\
 
\noindent For $z=5$, Eq.~\eqref{eq:zbc} reproduces the numerical results over the full range of $\gamma_{\text{BN}}$, with absolute deviations below $0.04$. This agreement extends down to $\gamma_{\text{BN}}=1$, although the closed form is derived in the dominant-interface limit. For $z=1$ and $z=0.25$ the numerical conductance lies systematically above Eq.~\eqref{eq:zbc}, and the discrepancy grows with decreasing $z$. These barrier strengths lie outside the strong-interface limit $z_\text{S},z_\text{N}\gg 1$ in which the Tanaka--Nazarov boundary conditions reduce to the Kupriyanov--Lukichev form used to derive the closed form. Equation~\eqref{eq:zbc} is therefore quantitative only for $z\gg 1$. For $z=1$ the numerical conductance still shows the nonmonotonic dependence on $\gamma_{\text{BN}}$ predicted by the closed form, but not its magnitude.

\subsection{Helical \texorpdfstring{$p$}{p}-wave zero-energy LDOS}
\label{app:helical_zbc_ldos}

\begin{figure}[t]
    \centering
    \includegraphics[width=\columnwidth]{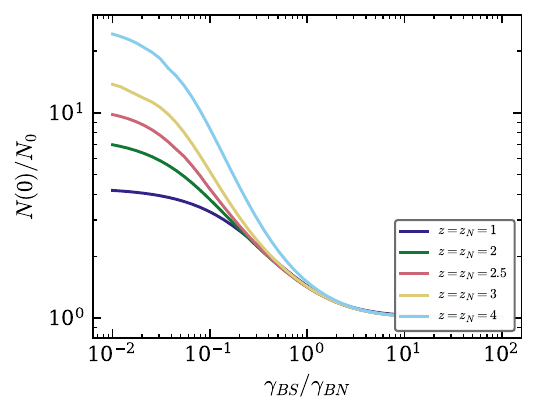}
    \caption{Zero-energy LDOS $N(0)/N_0$ in the helical $p$-wave case as a
    function of $\gamma_{\text{BS}}/\gamma_{\text{BN}}$, for $z_S = z_N \in
    \{1,\,2,\,2.5,\, 3,\, 4\}$.}
    \label{fig:app_helical_zbc_ldos}
\end{figure}

\noindent Figure~\ref{fig:app_helical_zbc_ldos} repeats the zero-energy LDOS
collapse analysis of Fig.~\ref{fig:zbc_ldos_collapse} for the helical
$p$-wave case. Two qualitative differences are apparent. First, $N(0)/N_0$
can significantly exceed unity in the helical case, reflecting the ZESABS-induced spectral enhancement discussed in Sec.~\ref{sec:helical_p}. Second, the data no longer collapse onto a single universal curve
independent of $z$: the SABS contribution depends on the interface transparency in a way that is not captured by a simple rescaling of the
$\gamma$ parameters. At high z the ZESABS weight is maximized because the sign-changing pair potential produces a sharp
resonance at low transparency; at low $z$ the resonance broadens and its
spectral weight is redistributed, leading to a weaker zero-energy
enhancement \cite{kashiwaya2000tunnelling}. This $z$-dependence is a hallmark of the unconventional
pairing and provides a second experimental handle (alongside the sign of
the zero-energy anomaly) for distinguishing $s$-wave from $p$-wave
proximity effects in S/DN/N junctions. The lack of a simple analytical
collapse formula for the helical case motivates a dedicated analytical
treatment of the triplet proximity effect in this geometry.

\subsection{Balian--Werthamer zero-energy LDOS}
\label{app:bw_zbc_ldos}
 
\begin{figure}[t]
    \centering
    \includegraphics[width=\columnwidth]{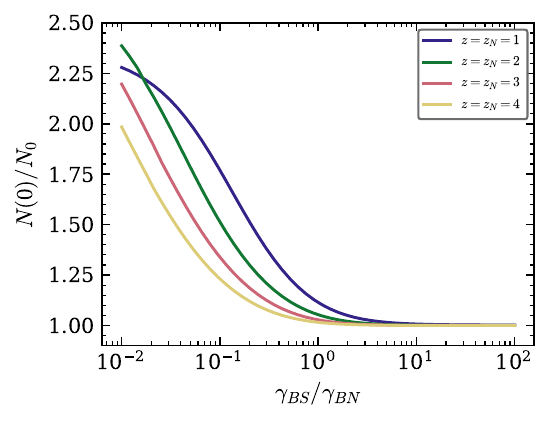}
    \caption{Zero-energy LDOS $N(0)/N_0$ in the Balian--Werthamer case as a
    function of $\gamma_{\text{BS}}/\gamma_{\text{BN}}$, for $z = z_{\text{N}} \in
    \{1,\, 2,\, 3,\, 4\}$. Symbols: Tanaka--Nazarov numerics at several
    $E_{\text{Th}}/\Delta_0$.}
    \label{fig:app_bw_zbc_ldos}
\end{figure}
 
\noindent Figure~\ref{fig:app_bw_zbc_ldos} shows the zero-energy LDOS for the BW
state. As in the helical case (Fig.~\ref{fig:app_helical_zbc_ldos}),
$N(0)/N_0$ exceeds unity for
$\gamma_{\text{BS}}/\gamma_{\text{BN}} \lesssim 1$, confirming the ZESABS-induced spectral enhancement and the data do not collapse onto a
single $z$-independent curve. Both features are generic consequences of
odd-parity pairing.\\
 
\noindent For $\gamma_{\text{BN}} \gg \gamma_{\text{BS}}$ the zero-energy
LDOS saturates at values in the range $N(0)/N_0 \approx 2$--$2.5$,
depending on $z$. This saturation is expected: in the limit
$\gamma_{\text{BN}} \to \infty$ the DN/N interface is perfectly opaque and
the bar becomes a closed cavity, so the anomalous spectral weight
injected by the ZESABS has nowhere to leak and accumulates at zero energy.
The precise saturation value depends on $z$ through the angular-integrated
SABS weight at the S interface, which differs between barrier strengths
because the transparency function filters different angular regions of
the Fermi surface. The ordering of the curves at
$\gamma_{\text{BS}}/\gamma_{\text{BN}} \ll 1$ is non-monotonic in $z$:
neither the lowest nor the highest barrier strength gives the largest
enhancement. \\
 
\noindent The contrast with the helical case
(Fig.~\ref{fig:app_helical_zbc_ldos}) is informative. In the helical
junction the $z$-ordering at
$\gamma_{\text{BS}}/\gamma_{\text{BN}} \ll 1$ is monotonic: increasing
$z$ significantly increases $N(0)/N_0$ because the sign change of the helical d-vector that generates the ZESABS sits at normal incidence, which is
selectively retained by the barrier. 
The BW $\boldsymbol{d}$-vector does reverse sign at exactly normal
incidence ($\theta = 0$), just as in the helical case. The difference is
that the fraction of transport channels near normal incidence that
participate in the sign reversal is much smaller for the BW state: in
three dimensions the solid angle near $\theta = 0$ scales as
$\theta\,d\theta$, so the channels that host the ZESABS occupy a vanishing
fraction of the Fermi surface near the pole, whereas in the
two-dimensional helical case the sign change at $\varphi = 0$ is shared
by a finite fraction of the channels. At high $z$ the transparency
function $T(\theta, z)$ selects a narrow cone around $\theta = 0$, and
for the BW state this cone contains mostly channels where the
$\boldsymbol{d}$-vector is nearly parallel to the interface normal
(symmetric, non-resonant) rather than channels with a sign-changing
in-plane component. The net effect is that the ZESABS weight integrated over the surviving channels decreases with increasing $z$ for the BW
state, while it increases for the helical state. This distinction
provides a qualitative experimental discriminant between candidate
odd-parity order parameters accessible through systematic variation of
the contact resistance in point-contact spectroscopy.

\subsection{Mixed-parity junctions}
\label{app:mixed_parity}
 
\noindent The helical and BW states of the main text are pure spin triplets. To describe non-centrosymmetric and time-reversal-symmetry-broken superconductors, where inversion-symmetry breaking mixes singlet and triplet channels~\cite{sigrist1991phenomenological, Yanase2008, KanasugiYanase2022}, we generalise Eq.~\eqref{eq:pairpotential} to a superposition of a singlet and a triplet component,
\begin{equation}
\hat{\Delta} = \Delta_s + \Delta_t\,\mathbf{d}\cdot\boldsymbol{\sigma}\;,
\label{eq:app_pairpotential_mixed}
\end{equation}
with $\Delta_s$ and $\Delta_t$ the isotropic singlet and anisotropic triplet amplitudes~\cite{TanakaKashiwaya2004, Kokkeler2023}. The relative strength of the two correlations is quantified by the dimensionless mixing parameter $r = \Delta_t/\Delta_s$, which controls whether the state is singlet-dominated ($r<1$) or triplet-dominated ($r>1$). Introducing a phase $\chi_t$ between the two components, the pair potential admits the compact parametrization~\cite{Kokkeler2023}
\begin{equation}
\hat{\Delta}(\varphi) = \Delta_0\!\left[e^{i\frac{\pi}{2}\chi_t}\,\frac{1}{\sqrt{1+r^2}} + \frac{r}{\sqrt{1+r^2}}\,\mathbf{d}(\varphi)\cdot\boldsymbol{\sigma}\right]\!,
\label{eq:app_delta_parameterization}
\end{equation}
whose two spin-projected eigenvalues are
\begin{equation}
\Delta_\pm(\varphi) = \Delta_0\,\frac{e^{i\frac{\pi}{2}\chi_t}\pm r\,e^{i\psi(\varphi)}}{\sqrt{1+r^2}}\;,
\label{eq:app_delta_pm}
\end{equation}
with $\psi(\varphi)$ the phase of the $d$-vector at the given angle ($\psi=0$ for the helical state). The eigenvalues $\Delta_\pm$ set the voltage positions of the conductance peaks and control the anisotropy of the differential conductance~\cite{Kokkeler2023}; topological features such as zero-energy Andreev bound states are connected with the signs and relative phases of $\Delta_\pm$~\cite{TanakaKokkelerGolubov2022, tanaka2024theory}. The pure helical and BW states of the main text are recovered as the triplet-dominated limit $r\to\infty$; the mixed-parity results below uses finite $r$.

\noindent The preceding sections treated the limiting cases of pure $s$-wave
($\Delta_t/\Delta_s = 0$) and pure spin-triplet ($\Delta_t/\Delta_s \to
\infty$) pairing. In systems where inversion symmetry is broken at the
interface or in the bulk, the two channels can coexist with a relative
phase, giving rise to a mixed-parity order parameter. We consider the
$is + p$ state, in which the singlet and triplet components have unequal
magnitude ($\Delta_t/\Delta_s = 2$) and a relative phase of $\pi/2$. Such a state breaks
time-reversal symmetry and has been discussed in the context of
noncentrosymmetric superconductors and interface-induced parity
mixing~\cite{sigrist1991phenomenological, Kokkeler2023}.\\

\begin{figure}[t]
    \centering
    \includegraphics[width=\columnwidth]{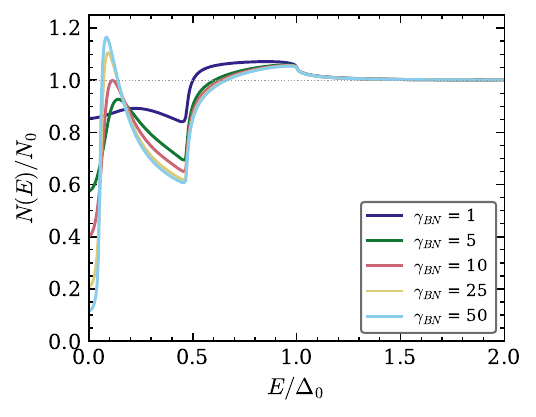}
    \caption{LDOS $N(E)/N_0$ for the $is + p$ state at fixed
    $\gamma_{\text{BS}} = 1$ for several values of $\gamma_{\text{BN}}$. Fixed
    parameters: $z_S = z_N = 1$, $E_{\text{Th}}/\Delta_0 = 0.4$.}
    \label{fig:app_mixed_ldos_gBN}
\end{figure}

\noindent Figure~\ref{fig:app_mixed_ldos_gBN} shows the LDOS lineshape at fixed
$\gamma_{\text{BS}} = 1$ for increasing $\gamma_{\text{BN}}$. At this small
$\gamma_{\text{BS}}$ the S interface is strongly coupled to the bar and the
proximity effect is pronounced. The overall structure is $s$-wave-like: a
suppression near $E = 0$ with coherence features near $E^\ast$. As
$\gamma_{\text{BN}}$ increases the N interface confines the spectral weight
inside the bar and the features sharpen, following the same pattern as in
the pure $s$-wave case (Fig.~\ref{fig:didv_sweep_gBN}). The energy scale
of the lineshape remains governed by $E^\ast$, confirming that the
sub-Thouless scale survives in the mixed-parity regime.\\

\begin{figure}[t]
    \centering
    \includegraphics[width=\columnwidth]{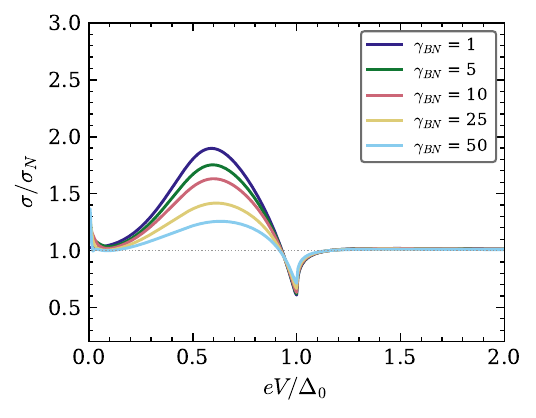}
    \caption{Normalized differential conductance $\sigma(eV)/\sigma_N$
    for the $is + p$ state at fixed $\gamma_{\text{BS}} = 25$ for several values
    of $\gamma_{\text{BN}}$. Fixed parameters: $z_S = z_N = 1$,
    $E_{\text{Th}}/\Delta_0 = 0.4$.}
    \label{fig:app_mixed_didv_gBN}
\end{figure}

\noindent Figure~\ref{fig:app_mixed_didv_gBN} displays the differential conductance
at fixed $\gamma_{\text{BS}} = 25$ for increasing $\gamma_{\text{BN}}$. The lineshape
is $s$-wave-like, with a subgap suppression rather than a ZBCP,
consistent with the singlet-dominated zero-energy LDOS of
Fig.~\ref{fig:app_mixed_ldos_gBN}. The amplitude of the feature grows
with $\gamma_{\text{BN}}$, reproducing the $\gamma_{\text{BN}}$-controlled amplitude
behavior established for the pure $s$-wave case. Compared to the pure
$s$-wave conductance at the same parameters, the subgap suppression is
slightly reduced, reflecting the partial filling of the proximity gap by
the triplet admixture.

\begin{figure}[t]
    \centering
    \includegraphics[width=\columnwidth]{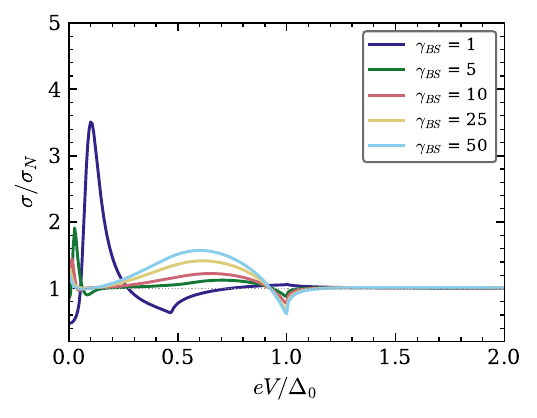}
    \caption{Normalized differential conductance $\sigma(eV)/\sigma_N$
    for the $is + p$ state at fixed $\gamma_{\text{BN}} = 25$ for several values
    of $\gamma_{\text{BS}}$. Fixed parameters: $z_S = z_N = 1$,
    $E_{\text{Th}}/\Delta_0 = 0.4$.}
    \label{fig:app_mixed_didv_gBS}
\end{figure}

\noindent Figure~\ref{fig:app_mixed_didv_gBS} shows the complementary sweep over
$\gamma_{\text{BS}}$ at fixed $\gamma_{\text{BN}} = 25$. Decreasing $\gamma_{\text{BS}}$
broadens the low-bias feature, consistent with the
$E^\ast \propto 1/\gamma_{\text{BS}}$ scaling observed for all pairing symmetries
studied in this work. At the same time, decreasing $\gamma_{\text{BS}}$
strengthens the proximity coupling and modifies the balance between the
singlet and triplet contributions to the spectral weight, so the detailed
line shape evolves not only in width but also in character. A crossover
from singlet-dominated to triplet-dominated behavior would manifest as a
sign change in the zero-bias anomaly (from dip to peak); this crossover
is not observed at $\Delta_t/\Delta_s = 2$, indicating that the
node-filling mechanism discussed above remains effective even at
substantial triplet admixture.\\

\begin{figure}[h]
    \centering
    \includegraphics[width=\columnwidth]{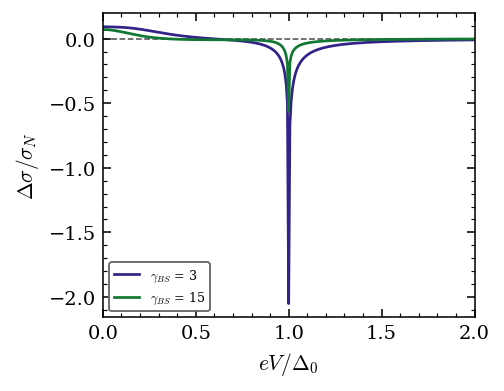}
    \caption{Difference in normalized differential conductance
    $\sigma/\sigma_N\big|_{\gamma_{\text{BN}}=1} -
    \sigma/\sigma_N\big|_{\gamma_{\text{BN}}=0}$ in an $s$-wave junction as a function of
    $eV/\Delta_0$ for $\gamma_{\text{BS}} = 3$ and $\gamma_{\text{BS}} = 15$.
    Fixed parameters: $z = z_{\text{N}} = 1$,
    $E_{\text{Th}}/\Delta_0 = 0.4$.}
    \label{fig:app_gammaBN_diff}
\end{figure}
\FloatBarrier

\section{Effect of a finite \texorpdfstring{$\gamma_{\text{BN}}$}{gammaBN} on the \texorpdfstring{$s$}{s}-wave differential conductance}
\label{app:gammaBN_continuity}

\noindent To assess how much the finite normal-side boundary resistance
modifies the conductance relative to the continuity limit, we compare
the $s$-wave differential conductance computed with
$\gamma_{\text{BN}} = 1$ and
$\gamma_{\text{BN}} = 0$ at
two values of $\gamma_{\text{BS}}$.
Figure~\ref{fig:app_gammaBN_diff} shows the difference
$\sigma/\sigma_N\big|_{\gamma_{\text{BN}}=1} -
\sigma/\sigma_N\big|_{\gamma_{\text{BN}}=0}$ as a function of bias voltage. At low
voltage ($|eV| \ll \Delta_0$) the two cases are nearly
indistinguishable: the difference is of order $10^{-1}$ or smaller for
both values of $\gamma_{\text{BS}}$. The dominant effect of a finite $\gamma_{\text{BN}}$ appears at the
gap edge, $|eV| \approx \Delta_0$, where the coherence peak is
substantially diminished. A
finite $\gamma_{\text{BN}}$ partially confines the quasiparticles inside
the bar, thereby lowering the amplitude the coherence peak. This reduction is visible
for both values of $\gamma_{\text{BS}}$ and is more pronounced at smaller
$\gamma_{\text{BS}}$, where the sub-Thouless feature is broad.

\end{appendices}

\bibliography{references}

\begin{thebibliography}{67}%
\makeatletter
\providecommand \@ifxundefined [1]{%
 \@ifx{#1\undefined}
}%
\providecommand \@ifnum [1]{%
 \ifnum #1\expandafter \@firstoftwo
 \else \expandafter \@secondoftwo
 \fi
}%
\providecommand \@ifx [1]{%
 \ifx #1\expandafter \@firstoftwo
 \else \expandafter \@secondoftwo
 \fi
}%
\providecommand \natexlab [1]{#1}%
\providecommand \enquote  [1]{``#1''}%
\providecommand \bibnamefont  [1]{#1}%
\providecommand \bibfnamefont [1]{#1}%
\providecommand \citenamefont [1]{#1}%
\providecommand \href@noop [0]{\@secondoftwo}%
\providecommand \href [0]{\begingroup \@sanitize@url \@href}%
\providecommand \@href[1]{\@@startlink{#1}\@@href}%
\providecommand \@@href[1]{\endgroup#1\@@endlink}%
\providecommand \@sanitize@url [0]{\catcode `\\12\catcode `\$12\catcode `\&12\catcode `\#12\catcode `\^12\catcode `\_12\catcode `\%12\relax}%
\providecommand \@@startlink[1]{}%
\providecommand \@@endlink[0]{}%
\providecommand \url  [0]{\begingroup\@sanitize@url \@url }%
\providecommand \@url [1]{\endgroup\@href {#1}{\urlprefix }}%
\providecommand \urlprefix  [0]{URL }%
\providecommand \Eprint [0]{\href }%
\providecommand \doibase [0]{https://doi.org/}%
\providecommand \selectlanguage [0]{\@gobble}%
\providecommand \bibinfo  [0]{\@secondoftwo}%
\providecommand \bibfield  [0]{\@secondoftwo}%
\providecommand \translation [1]{[#1]}%
\providecommand \BibitemOpen [0]{}%
\providecommand \bibitemStop [0]{}%
\providecommand \bibitemNoStop [0]{.\EOS\space}%
\providecommand \EOS [0]{\spacefactor3000\relax}%
\providecommand \BibitemShut  [1]{\csname bibitem#1\endcsname}%
\let\auto@bib@innerbib\@empty
\bibitem [{\citenamefont {Blonder}\ \emph {et~al.}(1982)\citenamefont {Blonder}, \citenamefont {Tinkham},\ and\ \citenamefont {Klapwijk}}]{blonder1982transition}%
  \BibitemOpen
  \bibfield  {author} {\bibinfo {author} {\bibfnamefont {G.~E.}\ \bibnamefont {Blonder}}, \bibinfo {author} {\bibfnamefont {M.}~\bibnamefont {Tinkham}},\ and\ \bibinfo {author} {\bibfnamefont {T.~M.}\ \bibnamefont {Klapwijk}},\ }\bibfield  {title} {\bibinfo {title} {Transition from metallic to tunneling regimes in superconducting microconstrictions: Excess current, charge imbalance, and supercurrent conversion},\ }\href {https://doi.org/10.1103/PhysRevB.25.4515} {\bibfield  {journal} {\bibinfo  {journal} {Phys. Rev. B}\ }\textbf {\bibinfo {volume} {25}},\ \bibinfo {pages} {4515} (\bibinfo {year} {1982})}\BibitemShut {NoStop}%
\bibitem [{\citenamefont {Tanaka}\ and\ \citenamefont {Kashiwaya}(1995)}]{TanakaKashiwaya1995}%
  \BibitemOpen
  \bibfield  {author} {\bibinfo {author} {\bibfnamefont {Y.}~\bibnamefont {Tanaka}}\ and\ \bibinfo {author} {\bibfnamefont {S.}~\bibnamefont {Kashiwaya}},\ }\bibfield  {title} {\bibinfo {title} {Theory of tunneling spectroscopy of {$d$}-wave superconductors},\ }\href {https://doi.org/10.1103/PhysRevLett.74.3451} {\bibfield  {journal} {\bibinfo  {journal} {Phys. Rev. Lett.}\ }\textbf {\bibinfo {volume} {74}},\ \bibinfo {pages} {3451} (\bibinfo {year} {1995})}\BibitemShut {NoStop}%
\bibitem [{\citenamefont {Kashiwaya}\ and\ \citenamefont {Tanaka}(2000)}]{kashiwaya2000tunnelling}%
  \BibitemOpen
  \bibfield  {author} {\bibinfo {author} {\bibfnamefont {S.}~\bibnamefont {Kashiwaya}}\ and\ \bibinfo {author} {\bibfnamefont {Y.}~\bibnamefont {Tanaka}},\ }\bibfield  {title} {\bibinfo {title} {Tunnelling effects on surface bound states in unconventional superconductors},\ }\href {https://doi.org/10.1088/0034-4885/63/10/202} {\bibfield  {journal} {\bibinfo  {journal} {Rep. Prog. Phys.}\ }\textbf {\bibinfo {volume} {63}},\ \bibinfo {pages} {1641} (\bibinfo {year} {2000})}\BibitemShut {NoStop}%
\bibitem [{\citenamefont {Hu}(1994)}]{hu1994midgap}%
  \BibitemOpen
  \bibfield  {author} {\bibinfo {author} {\bibfnamefont {C.-R.}\ \bibnamefont {Hu}},\ }\bibfield  {title} {\bibinfo {title} {Midgap surface states as a novel signature for {$d_{x_a^2-x_b^2}$}-wave superconductivity},\ }\href {https://doi.org/10.1103/PhysRevLett.72.1526} {\bibfield  {journal} {\bibinfo  {journal} {Phys. Rev. Lett.}\ }\textbf {\bibinfo {volume} {72}},\ \bibinfo {pages} {1526} (\bibinfo {year} {1994})}\BibitemShut {NoStop}%
\bibitem [{\citenamefont {Yamashiro}\ \emph {et~al.}(1998)\citenamefont {Yamashiro}, \citenamefont {Tanaka}, \citenamefont {Tanuma},\ and\ \citenamefont {Kashiwaya}}]{yamashiro1998theory}%
  \BibitemOpen
  \bibfield  {author} {\bibinfo {author} {\bibfnamefont {M.}~\bibnamefont {Yamashiro}}, \bibinfo {author} {\bibfnamefont {Y.}~\bibnamefont {Tanaka}}, \bibinfo {author} {\bibfnamefont {Y.}~\bibnamefont {Tanuma}},\ and\ \bibinfo {author} {\bibfnamefont {S.}~\bibnamefont {Kashiwaya}},\ }\bibfield  {title} {\bibinfo {title} {Theory of tunneling conductance for normal metal/insulator/triplet superconductor junction},\ }\href {https://doi.org/10.1143/JPSJ.67.3224} {\bibfield  {journal} {\bibinfo  {journal} {J. Phys. Soc. Jpn.}\ }\textbf {\bibinfo {volume} {67}},\ \bibinfo {pages} {3224} (\bibinfo {year} {1998})}\BibitemShut {NoStop}%
\bibitem [{\citenamefont {Iniotakis}\ \emph {et~al.}(2007)\citenamefont {Iniotakis}, \citenamefont {Hayashi}, \citenamefont {Sawa}, \citenamefont {Yokoyama}, \citenamefont {May}, \citenamefont {Tanaka},\ and\ \citenamefont {Sigrist}}]{Iniotakis2007}%
  \BibitemOpen
  \bibfield  {author} {\bibinfo {author} {\bibfnamefont {C.}~\bibnamefont {Iniotakis}}, \bibinfo {author} {\bibfnamefont {N.}~\bibnamefont {Hayashi}}, \bibinfo {author} {\bibfnamefont {Y.}~\bibnamefont {Sawa}}, \bibinfo {author} {\bibfnamefont {T.}~\bibnamefont {Yokoyama}}, \bibinfo {author} {\bibfnamefont {U.}~\bibnamefont {May}}, \bibinfo {author} {\bibfnamefont {Y.}~\bibnamefont {Tanaka}},\ and\ \bibinfo {author} {\bibfnamefont {M.}~\bibnamefont {Sigrist}},\ }\bibfield  {title} {\bibinfo {title} {{A}ndreev bound states and tunneling characteristics of a noncentrosymmetric superconductor},\ }\href {https://doi.org/10.1103/PhysRevB.76.012501} {\bibfield  {journal} {\bibinfo  {journal} {Phys. Rev. B}\ }\textbf {\bibinfo {volume} {76}},\ \bibinfo {pages} {012501} (\bibinfo {year} {2007})}\BibitemShut {NoStop}%
\bibitem [{\citenamefont {Tanaka}\ \emph {et~al.}(2009)\citenamefont {Tanaka}, \citenamefont {Yokoyama}, \citenamefont {Balatsky},\ and\ \citenamefont {Nagaosa}}]{TanakaYokoyamaBalatskyNagaosa2009}%
  \BibitemOpen
  \bibfield  {author} {\bibinfo {author} {\bibfnamefont {Y.}~\bibnamefont {Tanaka}}, \bibinfo {author} {\bibfnamefont {T.}~\bibnamefont {Yokoyama}}, \bibinfo {author} {\bibfnamefont {A.~V.}\ \bibnamefont {Balatsky}},\ and\ \bibinfo {author} {\bibfnamefont {N.}~\bibnamefont {Nagaosa}},\ }\bibfield  {title} {\bibinfo {title} {Theory of topological spin current in noncentrosymmetric superconductors},\ }\href {https://doi.org/10.1103/PhysRevB.79.060505} {\bibfield  {journal} {\bibinfo  {journal} {Phys. Rev. B}\ }\textbf {\bibinfo {volume} {79}},\ \bibinfo {pages} {060505} (\bibinfo {year} {2009})}\BibitemShut {NoStop}%
\bibitem [{\citenamefont {Tanaka}\ \emph {et~al.}(2003{\natexlab{a}})\citenamefont {Tanaka}, \citenamefont {Nazarov},\ and\ \citenamefont {Kashiwaya}}]{TanakaNazarov}%
  \BibitemOpen
  \bibfield  {author} {\bibinfo {author} {\bibfnamefont {Y.}~\bibnamefont {Tanaka}}, \bibinfo {author} {\bibfnamefont {Y.~V.}\ \bibnamefont {Nazarov}},\ and\ \bibinfo {author} {\bibfnamefont {S.}~\bibnamefont {Kashiwaya}},\ }\bibfield  {title} {\bibinfo {title} {Circuit theory of unconventional superconductor junctions},\ }\href {https://doi.org/10.1103/PhysRevLett.90.167003} {\bibfield  {journal} {\bibinfo  {journal} {Phys. Rev. Lett.}\ }\textbf {\bibinfo {volume} {90}},\ \bibinfo {pages} {167003} (\bibinfo {year} {2003}{\natexlab{a}})}\BibitemShut {NoStop}%
\bibitem [{\citenamefont {Tanaka}\ and\ \citenamefont {Kashiwaya}(2004)}]{TanakaKashiwaya2004}%
  \BibitemOpen
  \bibfield  {author} {\bibinfo {author} {\bibfnamefont {Y.}~\bibnamefont {Tanaka}}\ and\ \bibinfo {author} {\bibfnamefont {S.}~\bibnamefont {Kashiwaya}},\ }\bibfield  {title} {\bibinfo {title} {Anomalous charge transport in triplet superconductor junctions},\ }\href {https://doi.org/10.1103/PhysRevB.70.012507} {\bibfield  {journal} {\bibinfo  {journal} {Phys. Rev. B}\ }\textbf {\bibinfo {volume} {70}},\ \bibinfo {pages} {012507} (\bibinfo {year} {2004})}\BibitemShut {NoStop}%
\bibitem [{\citenamefont {Kokkeler}\ \emph {et~al.}(2023{\natexlab{a}})\citenamefont {Kokkeler}, \citenamefont {Hijano},\ and\ \citenamefont {Bergeret}}]{Kokkeler2023}%
  \BibitemOpen
  \bibfield  {author} {\bibinfo {author} {\bibfnamefont {T.}~\bibnamefont {Kokkeler}}, \bibinfo {author} {\bibfnamefont {A.}~\bibnamefont {Hijano}},\ and\ \bibinfo {author} {\bibfnamefont {F.~S.}\ \bibnamefont {Bergeret}},\ }\bibfield  {title} {\bibinfo {title} {Anisotropic differential conductance of a mixed-parity superconductor/ferromagnet structure},\ }\href {https://doi.org/10.1103/PhysRevB.107.104506} {\bibfield  {journal} {\bibinfo  {journal} {Phys. Rev. B}\ }\textbf {\bibinfo {volume} {107}},\ \bibinfo {pages} {104506} (\bibinfo {year} {2023}{\natexlab{a}})}\BibitemShut {NoStop}%
\bibitem [{\citenamefont {Tanaka}\ \emph {et~al.}(2005{\natexlab{a}})\citenamefont {Tanaka}, \citenamefont {Kashiwaya},\ and\ \citenamefont {Yokoyama}}]{tanaka2005theory}%
  \BibitemOpen
  \bibfield  {author} {\bibinfo {author} {\bibfnamefont {Y.}~\bibnamefont {Tanaka}}, \bibinfo {author} {\bibfnamefont {S.}~\bibnamefont {Kashiwaya}},\ and\ \bibinfo {author} {\bibfnamefont {T.}~\bibnamefont {Yokoyama}},\ }\bibfield  {title} {\bibinfo {title} {Theory of enhanced proximity effect by midgap {A}ndreev resonant state in diffusive normal-metal/triplet superconductor junctions},\ }\href {https://doi.org/10.1103/PhysRevB.71.094513} {\bibfield  {journal} {\bibinfo  {journal} {Phys. Rev. B}\ }\textbf {\bibinfo {volume} {71}},\ \bibinfo {pages} {094513} (\bibinfo {year} {2005}{\natexlab{a}})}\BibitemShut {NoStop}%
\bibitem [{\citenamefont {Golubov}\ \emph {et~al.}(2005)\citenamefont {Golubov}, \citenamefont {Kupriyanov},\ and\ \citenamefont {Siegel}}]{golubov2005density}%
  \BibitemOpen
  \bibfield  {author} {\bibinfo {author} {\bibfnamefont {A.~A.}\ \bibnamefont {Golubov}}, \bibinfo {author} {\bibfnamefont {M.~Y.}\ \bibnamefont {Kupriyanov}},\ and\ \bibinfo {author} {\bibfnamefont {M.}~\bibnamefont {Siegel}},\ }\bibfield  {title} {\bibinfo {title} {Density of states anomalies in hybrid superconductor-ferromagnet-normal metal structures},\ }\href {https://doi.org/10.1134/1.1914877} {\bibfield  {journal} {\bibinfo  {journal} {[Pis'ma Zh. Eksp. Teor. Fiz. 81, 217 (2005)] JETP Lett.}\ }\textbf {\bibinfo {volume} {81}},\ \bibinfo {pages} {180} (\bibinfo {year} {2005})}\BibitemShut {NoStop}%
\bibitem [{\citenamefont {Chiu}\ \emph {et~al.}(2021)\citenamefont {Chiu}, \citenamefont {Tsuei}, \citenamefont {Yeh}, \citenamefont {Zhang}, \citenamefont {Kirchner},\ and\ \citenamefont {Lin}}]{chiu2021observation}%
  \BibitemOpen
  \bibfield  {author} {\bibinfo {author} {\bibfnamefont {S.-P.}\ \bibnamefont {Chiu}}, \bibinfo {author} {\bibfnamefont {C.~C.}\ \bibnamefont {Tsuei}}, \bibinfo {author} {\bibfnamefont {S.-S.}\ \bibnamefont {Yeh}}, \bibinfo {author} {\bibfnamefont {F.-C.}\ \bibnamefont {Zhang}}, \bibinfo {author} {\bibfnamefont {S.}~\bibnamefont {Kirchner}},\ and\ \bibinfo {author} {\bibfnamefont {J.-J.}\ \bibnamefont {Lin}},\ }\bibfield  {title} {\bibinfo {title} {Observation of triplet superconductivity in {CoSi$_2$/TiSi$_2$} heterostructures},\ }\href {https://doi.org/10.1126/sciadv.abg6569} {\bibfield  {journal} {\bibinfo  {journal} {Sci. Adv.}\ }\textbf {\bibinfo {volume} {7}},\ \bibinfo {pages} {eabg6569} (\bibinfo {year} {2021})}\BibitemShut {NoStop}%
\bibitem [{\citenamefont {Yoon}\ \emph {et~al.}(2024)\citenamefont {Yoon}, \citenamefont {Eo}, \citenamefont {Park}, \citenamefont {Horn}, \citenamefont {Dorman}, \citenamefont {Saha}, \citenamefont {Hayes}, \citenamefont {Takeuchi}, \citenamefont {Brydon},\ and\ \citenamefont {Paglione}}]{MonctonUTe22024}%
  \BibitemOpen
  \bibfield  {author} {\bibinfo {author} {\bibfnamefont {H.}~\bibnamefont {Yoon}}, \bibinfo {author} {\bibfnamefont {Y.~S.}\ \bibnamefont {Eo}}, \bibinfo {author} {\bibfnamefont {J.}~\bibnamefont {Park}}, \bibinfo {author} {\bibfnamefont {J.~A.}\ \bibnamefont {Horn}}, \bibinfo {author} {\bibfnamefont {R.~G.}\ \bibnamefont {Dorman}}, \bibinfo {author} {\bibfnamefont {S.~R.}\ \bibnamefont {Saha}}, \bibinfo {author} {\bibfnamefont {I.~M.}\ \bibnamefont {Hayes}}, \bibinfo {author} {\bibfnamefont {I.}~\bibnamefont {Takeuchi}}, \bibinfo {author} {\bibfnamefont {P.~M.~R.}\ \bibnamefont {Brydon}},\ and\ \bibinfo {author} {\bibfnamefont {J.}~\bibnamefont {Paglione}},\ }\bibfield  {title} {\bibinfo {title} {Probing $p$-wave superconductivity in {UTe$_2$} via point-contact junctions},\ }\href {https://doi.org/10.1038/s41535-024-00700-z} {\bibfield  {journal} {\bibinfo  {journal} {npj Quantum Mater.}\ }\textbf {\bibinfo {volume} {9}},\ \bibinfo {pages} {91} (\bibinfo {year} {2024})}\BibitemShut {NoStop}%
\bibitem [{\citenamefont {Aoki}\ \emph {et~al.}(2019)\citenamefont {Aoki}, \citenamefont {Ishida},\ and\ \citenamefont {Flouquet}}]{AokiIshidaFlouquet2019}%
  \BibitemOpen
  \bibfield  {author} {\bibinfo {author} {\bibfnamefont {D.}~\bibnamefont {Aoki}}, \bibinfo {author} {\bibfnamefont {K.}~\bibnamefont {Ishida}},\ and\ \bibinfo {author} {\bibfnamefont {J.}~\bibnamefont {Flouquet}},\ }\bibfield  {title} {\bibinfo {title} {Review of {U}-based ferromagnetic superconductors},\ }\href {https://doi.org/10.7566/JPSJ.88.022001} {\bibfield  {journal} {\bibinfo  {journal} {J. Phys. Soc. Jpn.}\ }\textbf {\bibinfo {volume} {88}},\ \bibinfo {pages} {022001} (\bibinfo {year} {2019})}\BibitemShut {NoStop}%
\bibitem [{\citenamefont {Ran}\ \emph {et~al.}(2019)\citenamefont {Ran}, \citenamefont {Eckberg}, \citenamefont {Ding}, \citenamefont {Furukawa}, \citenamefont {Metz}, \citenamefont {Saha}, \citenamefont {Liu}, \citenamefont {Zic}, \citenamefont {Kim}, \citenamefont {Paglione},\ and\ \citenamefont {Butch}}]{RanUTe22019}%
  \BibitemOpen
  \bibfield  {author} {\bibinfo {author} {\bibfnamefont {S.}~\bibnamefont {Ran}}, \bibinfo {author} {\bibfnamefont {C.}~\bibnamefont {Eckberg}}, \bibinfo {author} {\bibfnamefont {Q.-P.}\ \bibnamefont {Ding}}, \bibinfo {author} {\bibfnamefont {Y.}~\bibnamefont {Furukawa}}, \bibinfo {author} {\bibfnamefont {T.}~\bibnamefont {Metz}}, \bibinfo {author} {\bibfnamefont {S.~R.}\ \bibnamefont {Saha}}, \bibinfo {author} {\bibfnamefont {I.-L.}\ \bibnamefont {Liu}}, \bibinfo {author} {\bibfnamefont {M.}~\bibnamefont {Zic}}, \bibinfo {author} {\bibfnamefont {H.}~\bibnamefont {Kim}}, \bibinfo {author} {\bibfnamefont {J.}~\bibnamefont {Paglione}},\ and\ \bibinfo {author} {\bibfnamefont {N.~P.}\ \bibnamefont {Butch}},\ }\bibfield  {title} {\bibinfo {title} {Nearly ferromagnetic spin-triplet superconductivity},\ }\href {https://doi.org/10.1126/science.aav8645} {\bibfield  {journal} {\bibinfo  {journal} {Science}\ }\textbf {\bibinfo {volume} {365}},\ \bibinfo {pages} {684} (\bibinfo {year} {2019})}\BibitemShut {NoStop}%
\bibitem [{\citenamefont {Zhu}\ \emph {et~al.}(2023)\citenamefont {Zhu}, \citenamefont {Song}, \citenamefont {Huang}, \citenamefont {Wang}, \citenamefont {Cao}, \citenamefont {Zhai}, \citenamefont {Bian}, \citenamefont {Shao}, \citenamefont {Jing}, \citenamefont {Zhu}, \citenamefont {Hou}, \citenamefont {Gao}, \citenamefont {Li}, \citenamefont {Zheng}, \citenamefont {Zhang}, \citenamefont {Pan}, \citenamefont {Liu}, \citenamefont {Qu}, \citenamefont {Gu}, \citenamefont {Zhang}, \citenamefont {Dong}, \citenamefont {Huang}, \citenamefont {Yuan}, \citenamefont {He}, \citenamefont {Li}, \citenamefont {Qian}, \citenamefont {Chen}, \citenamefont {Li}, \citenamefont {Pan},\ and\ \citenamefont {Xue}}]{AuSn42023}%
  \BibitemOpen
  \bibfield  {author} {\bibinfo {author} {\bibfnamefont {W.}~\bibnamefont {Zhu}}, \bibinfo {author} {\bibfnamefont {R.}~\bibnamefont {Song}}, \bibinfo {author} {\bibfnamefont {J.}~\bibnamefont {Huang}}, \bibinfo {author} {\bibfnamefont {Q.-W.}\ \bibnamefont {Wang}}, \bibinfo {author} {\bibfnamefont {Y.}~\bibnamefont {Cao}}, \bibinfo {author} {\bibfnamefont {R.}~\bibnamefont {Zhai}}, \bibinfo {author} {\bibfnamefont {Q.}~\bibnamefont {Bian}}, \bibinfo {author} {\bibfnamefont {Z.}~\bibnamefont {Shao}}, \bibinfo {author} {\bibfnamefont {H.}~\bibnamefont {Jing}}, \bibinfo {author} {\bibfnamefont {L.}~\bibnamefont {Zhu}}, \bibinfo {author} {\bibfnamefont {Y.}~\bibnamefont {Hou}}, \bibinfo {author} {\bibfnamefont {Y.-H.}\ \bibnamefont {Gao}}, \bibinfo {author} {\bibfnamefont {S.}~\bibnamefont {Li}}, \bibinfo {author} {\bibfnamefont {F.}~\bibnamefont {Zheng}}, \bibinfo {author} {\bibfnamefont {P.}~\bibnamefont {Zhang}}, \bibinfo {author} {\bibfnamefont {M.}~\bibnamefont {Pan}}, \bibinfo {author} {\bibfnamefont
  {J.}~\bibnamefont {Liu}}, \bibinfo {author} {\bibfnamefont {G.}~\bibnamefont {Qu}}, \bibinfo {author} {\bibfnamefont {Y.}~\bibnamefont {Gu}}, \bibinfo {author} {\bibfnamefont {H.}~\bibnamefont {Zhang}}, \bibinfo {author} {\bibfnamefont {Q.}~\bibnamefont {Dong}}, \bibinfo {author} {\bibfnamefont {Y.}~\bibnamefont {Huang}}, \bibinfo {author} {\bibfnamefont {X.}~\bibnamefont {Yuan}}, \bibinfo {author} {\bibfnamefont {J.}~\bibnamefont {He}}, \bibinfo {author} {\bibfnamefont {G.}~\bibnamefont {Li}}, \bibinfo {author} {\bibfnamefont {T.}~\bibnamefont {Qian}}, \bibinfo {author} {\bibfnamefont {G.}~\bibnamefont {Chen}}, \bibinfo {author} {\bibfnamefont {S.-C.}\ \bibnamefont {Li}}, \bibinfo {author} {\bibfnamefont {M.}~\bibnamefont {Pan}},\ and\ \bibinfo {author} {\bibfnamefont {Q.-K.}\ \bibnamefont {Xue}},\ }\bibfield  {title} {\bibinfo {title} {Intrinsic surface $p$-wave superconductivity in layered {AuSn$_4$}},\ }\href {https://doi.org/10.1038/s41467-023-42781-7} {\bibfield  {journal} {\bibinfo  {journal} {Nat.
  Commun.}\ }\textbf {\bibinfo {volume} {14}},\ \bibinfo {pages} {7012} (\bibinfo {year} {2023})}\BibitemShut {NoStop}%
\bibitem [{\citenamefont {Gor'kov}(1959)}]{Gorkov}%
  \BibitemOpen
  \bibfield  {author} {\bibinfo {author} {\bibfnamefont {L.~P.}\ \bibnamefont {Gor'kov}},\ }\bibfield  {title} {\bibinfo {title} {Microscopic derivation of the {G}inzburg--{L}andau equations in the theory of superconductivity},\ }\href {http://jetp.ras.ru/cgi-bin/dn/e_009_06_1364.pdf} {\bibfield  {journal} {\bibinfo  {journal} {[Zh. Eksp. Teor. Fiz. 36, 1918 (1959)] Sov. Phys. JETP}\ }\textbf {\bibinfo {volume} {9}},\ \bibinfo {pages} {1364} (\bibinfo {year} {1959})}\BibitemShut {NoStop}%
\bibitem [{\citenamefont {Eilenberger}(1968)}]{Eilenberger}%
  \BibitemOpen
  \bibfield  {author} {\bibinfo {author} {\bibfnamefont {G.}~\bibnamefont {Eilenberger}},\ }\bibfield  {title} {\bibinfo {title} {Transformation of {G}orkov's equation for type {II} superconductors into transport-like equations},\ }\href {https://doi.org/10.1007/BF01379803} {\bibfield  {journal} {\bibinfo  {journal} {Z. Phys.}\ }\textbf {\bibinfo {volume} {214}},\ \bibinfo {pages} {195} (\bibinfo {year} {1968})}\BibitemShut {NoStop}%
\bibitem [{\citenamefont {Larkin}\ and\ \citenamefont {Ovchinnikov}(1975)}]{larkin1975nonlinear}%
  \BibitemOpen
  \bibfield  {author} {\bibinfo {author} {\bibfnamefont {A.~I.}\ \bibnamefont {Larkin}}\ and\ \bibinfo {author} {\bibfnamefont {Y.~N.}\ \bibnamefont {Ovchinnikov}},\ }\bibfield  {title} {\bibinfo {title} {Nonlinear conductivity of superconductors in the mixed state},\ }\href {https://jetp.ras.ru/cgi-bin/dn/e_041_05_0960.pdf} {\bibfield  {journal} {\bibinfo  {journal} {[Zh. Eksp. Teor. Fiz. 68, 1915 (1975)] Sov. Phys. JETP}\ }\textbf {\bibinfo {volume} {41}},\ \bibinfo {pages} {960} (\bibinfo {year} {1975})}\BibitemShut {NoStop}%
\bibitem [{\citenamefont {Usadel}(1970)}]{usadel1970generalized}%
  \BibitemOpen
  \bibfield  {author} {\bibinfo {author} {\bibfnamefont {K.~D.}\ \bibnamefont {Usadel}},\ }\bibfield  {title} {\bibinfo {title} {Generalized diffusion equation for superconducting alloys},\ }\href {https://doi.org/10.1103/PhysRevLett.25.507} {\bibfield  {journal} {\bibinfo  {journal} {Phys. Rev. Lett.}\ }\textbf {\bibinfo {volume} {25}},\ \bibinfo {pages} {507} (\bibinfo {year} {1970})}\BibitemShut {NoStop}%
\bibitem [{\citenamefont {Belzig}\ \emph {et~al.}(1999)\citenamefont {Belzig}, \citenamefont {Wilhelm}, \citenamefont {Bruder}, \citenamefont {Sch{\"o}n},\ and\ \citenamefont {Zaikin}}]{Belzig1999}%
  \BibitemOpen
  \bibfield  {author} {\bibinfo {author} {\bibfnamefont {W.}~\bibnamefont {Belzig}}, \bibinfo {author} {\bibfnamefont {F.~K.}\ \bibnamefont {Wilhelm}}, \bibinfo {author} {\bibfnamefont {C.}~\bibnamefont {Bruder}}, \bibinfo {author} {\bibfnamefont {G.}~\bibnamefont {Sch{\"o}n}},\ and\ \bibinfo {author} {\bibfnamefont {A.~D.}\ \bibnamefont {Zaikin}},\ }\bibfield  {title} {\bibinfo {title} {Quasiclassical {G}reen's function approach to mesoscopic superconductivity},\ }\href {https://doi.org/10.1006/spmi.1999.0710} {\bibfield  {journal} {\bibinfo  {journal} {Superlattices Microstruct.}\ }\textbf {\bibinfo {volume} {25}},\ \bibinfo {pages} {1251} (\bibinfo {year} {1999})}\BibitemShut {NoStop}%
\bibitem [{\citenamefont {Chandrasekhar}(2008)}]{chandrasekhar2008proximity}%
  \BibitemOpen
  \bibfield  {author} {\bibinfo {author} {\bibfnamefont {V.}~\bibnamefont {Chandrasekhar}},\ }\bibfield  {title} {\bibinfo {title} {Proximity-coupled systems: Quasiclassical theory of superconductivity},\ }in\ \href {https://doi.org/10.1007/978-3-540-73253-2_8} {\emph {\bibinfo {booktitle} {Superconductivity: Conventional and Unconventional Superconductors}}},\ \bibinfo {editor} {edited by\ \bibinfo {editor} {\bibfnamefont {K.~H.}\ \bibnamefont {Bennemann}}\ and\ \bibinfo {editor} {\bibfnamefont {J.~B.}\ \bibnamefont {Ketterson}}}\ (\bibinfo  {publisher} {Springer Berlin Heidelberg},\ \bibinfo {year} {2008})\ pp.\ \bibinfo {pages} {279--313}\BibitemShut {NoStop}%
\bibitem [{\citenamefont {Keldysh}(1965)}]{Keldysh}%
  \BibitemOpen
  \bibfield  {author} {\bibinfo {author} {\bibfnamefont {L.~V.}\ \bibnamefont {Keldysh}},\ }\bibfield  {title} {\bibinfo {title} {Diagram technique for nonequilibrium processes},\ }\href@noop {} {\bibfield  {journal} {\bibinfo  {journal} {[Zh. Eksp. Teor. Fiz. 47, 1515 (1964)] Sov. Phys. JETP}\ }\textbf {\bibinfo {volume} {20}},\ \bibinfo {pages} {1018} (\bibinfo {year} {1965})}\BibitemShut {NoStop}%
\bibitem [{\citenamefont {Schmid}\ and\ \citenamefont {Sch{\"o}n}(1975)}]{SchmidSchon1975}%
  \BibitemOpen
  \bibfield  {author} {\bibinfo {author} {\bibfnamefont {A.}~\bibnamefont {Schmid}}\ and\ \bibinfo {author} {\bibfnamefont {G.}~\bibnamefont {Sch{\"o}n}},\ }\bibfield  {title} {\bibinfo {title} {Linearized kinetic equations and relaxation processes of a superconductor near $t_c$},\ }\href {https://doi.org/10.1007/BF00115264} {\bibfield  {journal} {\bibinfo  {journal} {J. Low Temp. Phys.}\ }\textbf {\bibinfo {volume} {20}},\ \bibinfo {pages} {207} (\bibinfo {year} {1975})}\BibitemShut {NoStop}%
\bibitem [{\citenamefont {Heikkil{\"a}}\ \emph {et~al.}(2019)\citenamefont {Heikkil{\"a}}, \citenamefont {Silaev}, \citenamefont {Virtanen},\ and\ \citenamefont {Bergeret}}]{HeikkilaSilaevVirtanenBergeret2019}%
  \BibitemOpen
  \bibfield  {author} {\bibinfo {author} {\bibfnamefont {T.~T.}\ \bibnamefont {Heikkil{\"a}}}, \bibinfo {author} {\bibfnamefont {M.}~\bibnamefont {Silaev}}, \bibinfo {author} {\bibfnamefont {P.}~\bibnamefont {Virtanen}},\ and\ \bibinfo {author} {\bibfnamefont {F.~S.}\ \bibnamefont {Bergeret}},\ }\bibfield  {title} {\bibinfo {title} {Thermal, electric and spin transport in superconductor/ferromagnetic-insulator structures},\ }\href {https://doi.org/10.1016/j.progsurf.2019.100540} {\bibfield  {journal} {\bibinfo  {journal} {Prog. Surf. Sci.}\ }\textbf {\bibinfo {volume} {94}},\ \bibinfo {pages} {100540} (\bibinfo {year} {2019})}\BibitemShut {NoStop}%
\bibitem [{\citenamefont {Nazarov}(1999)}]{Nazarov1999}%
  \BibitemOpen
  \bibfield  {author} {\bibinfo {author} {\bibfnamefont {Y.~V.}\ \bibnamefont {Nazarov}},\ }\bibfield  {title} {\bibinfo {title} {Novel circuit theory of {A}ndreev reflection},\ }\href {https://doi.org/10.1006/spmi.1999.0738} {\bibfield  {journal} {\bibinfo  {journal} {Superlattices Microstruct.}\ }\textbf {\bibinfo {volume} {25}},\ \bibinfo {pages} {1221} (\bibinfo {year} {1999})}\BibitemShut {NoStop}%
\bibitem [{\citenamefont {Kashiwaya}\ \emph {et~al.}(1995)\citenamefont {Kashiwaya}, \citenamefont {Tanaka}, \citenamefont {Koyanagi}, \citenamefont {Takashima},\ and\ \citenamefont {Kajimura}}]{kashiwaya1995origin}%
  \BibitemOpen
  \bibfield  {author} {\bibinfo {author} {\bibfnamefont {S.}~\bibnamefont {Kashiwaya}}, \bibinfo {author} {\bibfnamefont {Y.}~\bibnamefont {Tanaka}}, \bibinfo {author} {\bibfnamefont {M.}~\bibnamefont {Koyanagi}}, \bibinfo {author} {\bibfnamefont {H.}~\bibnamefont {Takashima}},\ and\ \bibinfo {author} {\bibfnamefont {K.}~\bibnamefont {Kajimura}},\ }\bibfield  {title} {\bibinfo {title} {Origin of zero-bias conductance peaks in high-{$T_c$} superconductors},\ }\href {https://doi.org/10.1103/PhysRevB.51.1350} {\bibfield  {journal} {\bibinfo  {journal} {Phys. Rev. B}\ }\textbf {\bibinfo {volume} {51}},\ \bibinfo {pages} {1350} (\bibinfo {year} {1995})}\BibitemShut {NoStop}%
\bibitem [{\citenamefont {Kanasugi}\ and\ \citenamefont {Yanase}(2022)}]{KanasugiYanase2022}%
  \BibitemOpen
  \bibfield  {author} {\bibinfo {author} {\bibfnamefont {S.}~\bibnamefont {Kanasugi}}\ and\ \bibinfo {author} {\bibfnamefont {Y.}~\bibnamefont {Yanase}},\ }\bibfield  {title} {\bibinfo {title} {Anapole superconductivity from $\mathcal{PT}$-symmetric mixed-parity interband pairing},\ }\href {https://doi.org/10.1038/s42005-022-00804-7} {\bibfield  {journal} {\bibinfo  {journal} {Commun. Phys.}\ }\textbf {\bibinfo {volume} {5}},\ \bibinfo {pages} {39} (\bibinfo {year} {2022})}\BibitemShut {NoStop}%
\bibitem [{\citenamefont {Tanaka}\ \emph {et~al.}(2022)\citenamefont {Tanaka}, \citenamefont {Kokkeler},\ and\ \citenamefont {Golubov}}]{TanakaKokkelerGolubov2022}%
  \BibitemOpen
  \bibfield  {author} {\bibinfo {author} {\bibfnamefont {Y.}~\bibnamefont {Tanaka}}, \bibinfo {author} {\bibfnamefont {T.}~\bibnamefont {Kokkeler}},\ and\ \bibinfo {author} {\bibfnamefont {A.~A.}\ \bibnamefont {Golubov}},\ }\bibfield  {title} {\bibinfo {title} {Theory of proximity effect in a diffusive normal metal/non-centrosymmetric superconductor bilayer},\ }\href {https://doi.org/10.1103/PhysRevB.105.214512} {\bibfield  {journal} {\bibinfo  {journal} {Phys. Rev. B}\ }\textbf {\bibinfo {volume} {105}},\ \bibinfo {pages} {214512} (\bibinfo {year} {2022})}\BibitemShut {NoStop}%
\bibitem [{\citenamefont {Kokkeler}\ \emph {et~al.}(2023{\natexlab{b}})\citenamefont {Kokkeler}, \citenamefont {Tanaka},\ and\ \citenamefont {Golubov}}]{kokkeler2023spin}%
  \BibitemOpen
  \bibfield  {author} {\bibinfo {author} {\bibfnamefont {T.~H.}\ \bibnamefont {Kokkeler}}, \bibinfo {author} {\bibfnamefont {Y.}~\bibnamefont {Tanaka}},\ and\ \bibinfo {author} {\bibfnamefont {A.~A.}\ \bibnamefont {Golubov}},\ }\bibfield  {title} {\bibinfo {title} {Spin-projected charge conductance in {SNN} junctions with noncentrosymmetric superconductors},\ }\href {https://doi.org/10.1103/PhysRevResearch.5.L012022} {\bibfield  {journal} {\bibinfo  {journal} {Phys. Rev. Research}\ }\textbf {\bibinfo {volume} {5}},\ \bibinfo {pages} {L012022} (\bibinfo {year} {2023}{\natexlab{b}})}\BibitemShut {NoStop}%
\bibitem [{\citenamefont {Kokkeler}\ \emph {et~al.}(2025)\citenamefont {Kokkeler}, \citenamefont {Golubov}, \citenamefont {Bergeret},\ and\ \citenamefont {Tanaka}}]{kokkeler2024countingstatisticsunconventionalsuperconductor}%
  \BibitemOpen
  \bibfield  {author} {\bibinfo {author} {\bibfnamefont {T.}~\bibnamefont {Kokkeler}}, \bibinfo {author} {\bibfnamefont {A.}~\bibnamefont {Golubov}}, \bibinfo {author} {\bibfnamefont {F.~S.}\ \bibnamefont {Bergeret}},\ and\ \bibinfo {author} {\bibfnamefont {Y.}~\bibnamefont {Tanaka}},\ }\bibfield  {title} {\bibinfo {title} {Full counting statistics for unconventional superconductor junctions},\ }\href {https://doi.org/10.1103/fn76-w95v} {\bibfield  {journal} {\bibinfo  {journal} {Phys. Rev. B}\ }\textbf {\bibinfo {volume} {112}},\ \bibinfo {pages} {024507} (\bibinfo {year} {2025})}\BibitemShut {NoStop}%
\bibitem [{\citenamefont {Kupriyanov}\ and\ \citenamefont {Lukichev}(1988)}]{kuprianov1988influence}%
  \BibitemOpen
  \bibfield  {author} {\bibinfo {author} {\bibfnamefont {M.~Y.}\ \bibnamefont {Kupriyanov}}\ and\ \bibinfo {author} {\bibfnamefont {V.~F.}\ \bibnamefont {Lukichev}},\ }\bibfield  {title} {\bibinfo {title} {Influence of boundary transparency on the critical current of ``dirty'' {SS'S} structures},\ }\href@noop {} {\bibfield  {journal} {\bibinfo  {journal} {[Zh. Eksp. Teor. Fiz. 94, 139 (1988)] Sov. Phys. JETP}\ }\textbf {\bibinfo {volume} {67}},\ \bibinfo {pages} {1163} (\bibinfo {year} {1988})}\BibitemShut {NoStop}%
\bibitem [{\citenamefont {Chiu}\ and\ \citenamefont {Lin}()}]{ChiuLinPrivate}%
  \BibitemOpen
  \bibfield  {author} {\bibinfo {author} {\bibfnamefont {S.-P.}\ \bibnamefont {Chiu}}\ and\ \bibinfo {author} {\bibfnamefont {J.-J.}\ \bibnamefont {Lin}},\ }\href@noop {} {}\bibinfo {note} {Private communication}\BibitemShut {NoStop}%
\bibitem [{\citenamefont {Balian}\ and\ \citenamefont {Werthamer}(1963)}]{balian1963superconductivity}%
  \BibitemOpen
  \bibfield  {author} {\bibinfo {author} {\bibfnamefont {R.}~\bibnamefont {Balian}}\ and\ \bibinfo {author} {\bibfnamefont {N.~R.}\ \bibnamefont {Werthamer}},\ }\bibfield  {title} {\bibinfo {title} {Superconductivity with pairs in a relative $p$ wave},\ }\href {https://doi.org/10.1103/PhysRev.131.1553} {\bibfield  {journal} {\bibinfo  {journal} {Phys. Rev.}\ }\textbf {\bibinfo {volume} {131}},\ \bibinfo {pages} {1553} (\bibinfo {year} {1963})}\BibitemShut {NoStop}%
\bibitem [{\citenamefont {Yanase}\ and\ \citenamefont {Sigrist}(2008)}]{Yanase2008}%
  \BibitemOpen
  \bibfield  {author} {\bibinfo {author} {\bibfnamefont {Y.}~\bibnamefont {Yanase}}\ and\ \bibinfo {author} {\bibfnamefont {M.}~\bibnamefont {Sigrist}},\ }\bibfield  {title} {\bibinfo {title} {Superconductivity and magnetism in non-centrosymmetric system: Application to {CePt$_3$Si}},\ }\href {https://doi.org/10.1143/JPSJ.77.124711} {\bibfield  {journal} {\bibinfo  {journal} {J. Phys. Soc. Jpn.}\ }\textbf {\bibinfo {volume} {77}},\ \bibinfo {pages} {124711} (\bibinfo {year} {2008})}\BibitemShut {NoStop}%
\bibitem [{\citenamefont {Schnyder}\ \emph {et~al.}(2008)\citenamefont {Schnyder}, \citenamefont {Ryu}, \citenamefont {Furusaki},\ and\ \citenamefont {Ludwig}}]{Topology1}%
  \BibitemOpen
  \bibfield  {author} {\bibinfo {author} {\bibfnamefont {A.~P.}\ \bibnamefont {Schnyder}}, \bibinfo {author} {\bibfnamefont {S.}~\bibnamefont {Ryu}}, \bibinfo {author} {\bibfnamefont {A.}~\bibnamefont {Furusaki}},\ and\ \bibinfo {author} {\bibfnamefont {A.~W.~W.}\ \bibnamefont {Ludwig}},\ }\bibfield  {title} {\bibinfo {title} {Classification of topological insulators and superconductors in three spatial dimensions},\ }\href {https://doi.org/10.1103/PhysRevB.78.195125} {\bibfield  {journal} {\bibinfo  {journal} {Phys. Rev. B}\ }\textbf {\bibinfo {volume} {78}},\ \bibinfo {pages} {195125} (\bibinfo {year} {2008})}\BibitemShut {NoStop}%
\bibitem [{\citenamefont {Tanaka}\ \emph {et~al.}(2024)\citenamefont {Tanaka}, \citenamefont {Tamura},\ and\ \citenamefont {Cayao}}]{tanaka2024theory}%
  \BibitemOpen
  \bibfield  {author} {\bibinfo {author} {\bibfnamefont {Y.}~\bibnamefont {Tanaka}}, \bibinfo {author} {\bibfnamefont {S.}~\bibnamefont {Tamura}},\ and\ \bibinfo {author} {\bibfnamefont {J.}~\bibnamefont {Cayao}},\ }\bibfield  {title} {\bibinfo {title} {Theory of {M}ajorana zero modes in unconventional superconductors},\ }\href {https://doi.org/10.1093/ptep/ptae065} {\bibfield  {journal} {\bibinfo  {journal} {Prog. Theor. Exp. Phys.}\ }\textbf {\bibinfo {volume} {2024}},\ \bibinfo {pages} {08C105} (\bibinfo {year} {2024})}\BibitemShut {NoStop}%
\bibitem [{\citenamefont {Sigrist}\ and\ \citenamefont {Ueda}(1991)}]{sigrist1991phenomenological}%
  \BibitemOpen
  \bibfield  {author} {\bibinfo {author} {\bibfnamefont {M.}~\bibnamefont {Sigrist}}\ and\ \bibinfo {author} {\bibfnamefont {K.}~\bibnamefont {Ueda}},\ }\bibfield  {title} {\bibinfo {title} {Phenomenological theory of unconventional superconductivity},\ }\href {https://doi.org/10.1103/RevModPhys.63.239} {\bibfield  {journal} {\bibinfo  {journal} {Rev. Mod. Phys.}\ }\textbf {\bibinfo {volume} {63}},\ \bibinfo {pages} {239} (\bibinfo {year} {1991})}\BibitemShut {NoStop}%
\bibitem [{\citenamefont {Buchholtz}\ and\ \citenamefont {Zwicknagl}(1981)}]{BuchholtzZwicknagl1981}%
  \BibitemOpen
  \bibfield  {author} {\bibinfo {author} {\bibfnamefont {L.~J.}\ \bibnamefont {Buchholtz}}\ and\ \bibinfo {author} {\bibfnamefont {G.}~\bibnamefont {Zwicknagl}},\ }\bibfield  {title} {\bibinfo {title} {Identification of $p$-wave superconductors},\ }\href {https://doi.org/10.1103/PhysRevB.23.5788} {\bibfield  {journal} {\bibinfo  {journal} {Phys. Rev. B}\ }\textbf {\bibinfo {volume} {23}},\ \bibinfo {pages} {5788} (\bibinfo {year} {1981})}\BibitemShut {NoStop}%
\bibitem [{\citenamefont {Yamakage}\ \emph {et~al.}(2012)\citenamefont {Yamakage}, \citenamefont {Yada}, \citenamefont {Sato},\ and\ \citenamefont {Tanaka}}]{YamakagePhysRevB}%
  \BibitemOpen
  \bibfield  {author} {\bibinfo {author} {\bibfnamefont {A.}~\bibnamefont {Yamakage}}, \bibinfo {author} {\bibfnamefont {K.}~\bibnamefont {Yada}}, \bibinfo {author} {\bibfnamefont {M.}~\bibnamefont {Sato}},\ and\ \bibinfo {author} {\bibfnamefont {Y.}~\bibnamefont {Tanaka}},\ }\bibfield  {title} {\bibinfo {title} {Theory of tunneling conductance and surface-state transition in superconducting topological insulators},\ }\href {https://doi.org/10.1103/PhysRevB.85.180509} {\bibfield  {journal} {\bibinfo  {journal} {Phys. Rev. B}\ }\textbf {\bibinfo {volume} {85}},\ \bibinfo {pages} {180509(R)} (\bibinfo {year} {2012})}\BibitemShut {NoStop}%
\bibitem [{\citenamefont {Asano}\ \emph {et~al.}(2003)\citenamefont {Asano}, \citenamefont {Tanaka}, \citenamefont {Matsuda},\ and\ \citenamefont {Kashiwaya}}]{asano2003theoretical}%
  \BibitemOpen
  \bibfield  {author} {\bibinfo {author} {\bibfnamefont {Y.}~\bibnamefont {Asano}}, \bibinfo {author} {\bibfnamefont {Y.}~\bibnamefont {Tanaka}}, \bibinfo {author} {\bibfnamefont {Y.}~\bibnamefont {Matsuda}},\ and\ \bibinfo {author} {\bibfnamefont {S.}~\bibnamefont {Kashiwaya}},\ }\bibfield  {title} {\bibinfo {title} {A theoretical study of tunneling conductance in {PrOs$_4$Sb$_{12}$} superconducting junctions},\ }\href {https://doi.org/10.1103/PhysRevB.68.184506} {\bibfield  {journal} {\bibinfo  {journal} {Phys. Rev. B}\ }\textbf {\bibinfo {volume} {68}},\ \bibinfo {pages} {184506} (\bibinfo {year} {2003})}\BibitemShut {NoStop}%
\bibitem [{\citenamefont {Kokkeler}\ \emph {et~al.}(2023{\natexlab{c}})\citenamefont {Kokkeler}, \citenamefont {Golubov}, \citenamefont {Bergeret},\ and\ \citenamefont {Tanaka}}]{kokkeler2023proximity}%
  \BibitemOpen
  \bibfield  {author} {\bibinfo {author} {\bibfnamefont {T.}~\bibnamefont {Kokkeler}}, \bibinfo {author} {\bibfnamefont {A.}~\bibnamefont {Golubov}}, \bibinfo {author} {\bibfnamefont {S.}~\bibnamefont {Bergeret}},\ and\ \bibinfo {author} {\bibfnamefont {Y.}~\bibnamefont {Tanaka}},\ }\bibfield  {title} {\bibinfo {title} {Proximity effect of time-reversal symmetry broken noncentrosymmetric superconductors},\ }\href {https://doi.org/10.1103/PhysRevB.108.094503} {\bibfield  {journal} {\bibinfo  {journal} {Phys. Rev. B}\ }\textbf {\bibinfo {volume} {108}},\ \bibinfo {pages} {094503} (\bibinfo {year} {2023}{\natexlab{c}})}\BibitemShut {NoStop}%
\bibitem [{\citenamefont {Nazarov}(1994)}]{Nazarov1994}%
  \BibitemOpen
  \bibfield  {author} {\bibinfo {author} {\bibfnamefont {Y.~V.}\ \bibnamefont {Nazarov}},\ }\bibfield  {title} {\bibinfo {title} {Circuit theory of {A}ndreev conductance},\ }\href {https://doi.org/10.1103/PhysRevLett.73.1420} {\bibfield  {journal} {\bibinfo  {journal} {Phys. Rev. Lett.}\ }\textbf {\bibinfo {volume} {73}},\ \bibinfo {pages} {1420} (\bibinfo {year} {1994})}\BibitemShut {NoStop}%
\bibitem [{\citenamefont {Tanaka}\ \emph {et~al.}(2004)\citenamefont {Tanaka}, \citenamefont {Nazarov}, \citenamefont {Golubov},\ and\ \citenamefont {Kashiwaya}}]{tanaka2004theory}%
  \BibitemOpen
  \bibfield  {author} {\bibinfo {author} {\bibfnamefont {Y.}~\bibnamefont {Tanaka}}, \bibinfo {author} {\bibfnamefont {Y.~V.}\ \bibnamefont {Nazarov}}, \bibinfo {author} {\bibfnamefont {A.~A.}\ \bibnamefont {Golubov}},\ and\ \bibinfo {author} {\bibfnamefont {S.}~\bibnamefont {Kashiwaya}},\ }\bibfield  {title} {\bibinfo {title} {Theory of charge transport in diffusive normal metal/unconventional singlet superconductor contacts},\ }\href {https://doi.org/10.1103/PhysRevB.69.144519} {\bibfield  {journal} {\bibinfo  {journal} {Phys. Rev. B}\ }\textbf {\bibinfo {volume} {69}},\ \bibinfo {pages} {144519} (\bibinfo {year} {2004})}\BibitemShut {NoStop}%
\bibitem [{\citenamefont {Volkov}\ \emph {et~al.}(1993)\citenamefont {Volkov}, \citenamefont {Zaitsev},\ and\ \citenamefont {Klapwijk}}]{VolkovZaitsevKlapwijk1993}%
  \BibitemOpen
  \bibfield  {author} {\bibinfo {author} {\bibfnamefont {A.~F.}\ \bibnamefont {Volkov}}, \bibinfo {author} {\bibfnamefont {A.~V.}\ \bibnamefont {Zaitsev}},\ and\ \bibinfo {author} {\bibfnamefont {T.~M.}\ \bibnamefont {Klapwijk}},\ }\bibfield  {title} {\bibinfo {title} {Proximity effect under nonequilibrium conditions in double-barrier superconducting junctions},\ }\href {https://doi.org/10.1016/0921-4534(93)90005-B} {\bibfield  {journal} {\bibinfo  {journal} {Physica C}\ }\textbf {\bibinfo {volume} {210}},\ \bibinfo {pages} {21} (\bibinfo {year} {1993})}\BibitemShut {NoStop}%
\bibitem [{\citenamefont {Beenakker}(1997)}]{Beenakker1997}%
  \BibitemOpen
  \bibfield  {author} {\bibinfo {author} {\bibfnamefont {C.~W.~J.}\ \bibnamefont {Beenakker}},\ }\bibfield  {title} {\bibinfo {title} {Random-matrix theory of quantum transport},\ }\href {https://doi.org/10.1103/RevModPhys.69.731} {\bibfield  {journal} {\bibinfo  {journal} {Rev. Mod. Phys.}\ }\textbf {\bibinfo {volume} {69}},\ \bibinfo {pages} {731} (\bibinfo {year} {1997})}\BibitemShut {NoStop}%
\bibitem [{\citenamefont {McMillan}(1968)}]{McMillan1968}%
  \BibitemOpen
  \bibfield  {author} {\bibinfo {author} {\bibfnamefont {W.~L.}\ \bibnamefont {McMillan}},\ }\bibfield  {title} {\bibinfo {title} {Tunneling model of the superconducting proximity effect},\ }\href {https://doi.org/10.1103/PhysRev.175.537} {\bibfield  {journal} {\bibinfo  {journal} {Phys. Rev.}\ }\textbf {\bibinfo {volume} {175}},\ \bibinfo {pages} {537} (\bibinfo {year} {1968})}\BibitemShut {NoStop}%
\bibitem [{\citenamefont {Tanaka}\ \emph {et~al.}(2005{\natexlab{b}})\citenamefont {Tanaka}, \citenamefont {Asano}, \citenamefont {Golubov},\ and\ \citenamefont {Kashiwaya}}]{tanaka2005anomalous}%
  \BibitemOpen
  \bibfield  {author} {\bibinfo {author} {\bibfnamefont {Y.}~\bibnamefont {Tanaka}}, \bibinfo {author} {\bibfnamefont {Y.}~\bibnamefont {Asano}}, \bibinfo {author} {\bibfnamefont {A.~A.}\ \bibnamefont {Golubov}},\ and\ \bibinfo {author} {\bibfnamefont {S.}~\bibnamefont {Kashiwaya}},\ }\bibfield  {title} {\bibinfo {title} {Anomalous features of the proximity effect in triplet superconductors},\ }\href {https://doi.org/10.1103/PhysRevB.72.140503} {\bibfield  {journal} {\bibinfo  {journal} {Phys. Rev. B}\ }\textbf {\bibinfo {volume} {72}},\ \bibinfo {pages} {140503} (\bibinfo {year} {2005}{\natexlab{b}})}\BibitemShut {NoStop}%
\bibitem [{\citenamefont {Tanaka}\ \emph {et~al.}(2003{\natexlab{b}})\citenamefont {Tanaka}, \citenamefont {Golubov},\ and\ \citenamefont {Kashiwaya}}]{Tanaka2003}%
  \BibitemOpen
  \bibfield  {author} {\bibinfo {author} {\bibfnamefont {Y.}~\bibnamefont {Tanaka}}, \bibinfo {author} {\bibfnamefont {A.~A.}\ \bibnamefont {Golubov}},\ and\ \bibinfo {author} {\bibfnamefont {S.}~\bibnamefont {Kashiwaya}},\ }\bibfield  {title} {\bibinfo {title} {Theory of charge transport in diffusive normal metal/conventional superconductor point contacts},\ }\href {https://doi.org/10.1103/PhysRevB.68.054513} {\bibfield  {journal} {\bibinfo  {journal} {Phys. Rev. B}\ }\textbf {\bibinfo {volume} {68}},\ \bibinfo {pages} {054513} (\bibinfo {year} {2003}{\natexlab{b}})}\BibitemShut {NoStop}%
\bibitem [{\citenamefont {Golubov}\ \emph {et~al.}(2004)\citenamefont {Golubov}, \citenamefont {Kupriyanov},\ and\ \citenamefont {Il'ichev}}]{golubov2004current}%
  \BibitemOpen
  \bibfield  {author} {\bibinfo {author} {\bibfnamefont {A.~A.}\ \bibnamefont {Golubov}}, \bibinfo {author} {\bibfnamefont {M.~Y.}\ \bibnamefont {Kupriyanov}},\ and\ \bibinfo {author} {\bibfnamefont {E.}~\bibnamefont {Il'ichev}},\ }\bibfield  {title} {\bibinfo {title} {The current-phase relation in {J}osephson junctions},\ }\href {https://doi.org/10.1103/RevModPhys.76.411} {\bibfield  {journal} {\bibinfo  {journal} {Rev. Mod. Phys.}\ }\textbf {\bibinfo {volume} {76}},\ \bibinfo {pages} {411} (\bibinfo {year} {2004})}\BibitemShut {NoStop}%
\bibitem [{\citenamefont {Tanaka}\ and\ \citenamefont {Tamura}(2021)}]{tanaka2021theoryodd}%
  \BibitemOpen
  \bibfield  {author} {\bibinfo {author} {\bibfnamefont {Y.}~\bibnamefont {Tanaka}}\ and\ \bibinfo {author} {\bibfnamefont {S.}~\bibnamefont {Tamura}},\ }\bibfield  {title} {\bibinfo {title} {Theory of surface {A}ndreev bound states and odd-frequency pairing in superconductor junctions},\ }\href {https://doi.org/10.1007/s10948-020-05672-2} {\bibfield  {journal} {\bibinfo  {journal} {J. Supercond. Nov. Magn.}\ }\textbf {\bibinfo {volume} {34}},\ \bibinfo {pages} {1677} (\bibinfo {year} {2021})}\BibitemShut {NoStop}%
\bibitem [{\citenamefont {Tanaka}\ and\ \citenamefont {Golubov}(2007)}]{TanakaGolubov2007}%
  \BibitemOpen
  \bibfield  {author} {\bibinfo {author} {\bibfnamefont {Y.}~\bibnamefont {Tanaka}}\ and\ \bibinfo {author} {\bibfnamefont {A.~A.}\ \bibnamefont {Golubov}},\ }\bibfield  {title} {\bibinfo {title} {Theory of the proximity effect in junctions with unconventional superconductors},\ }\href {https://doi.org/10.1103/PhysRevLett.98.037003} {\bibfield  {journal} {\bibinfo  {journal} {Phys. Rev. Lett.}\ }\textbf {\bibinfo {volume} {98}},\ \bibinfo {pages} {037003} (\bibinfo {year} {2007})}\BibitemShut {NoStop}%
\bibitem [{\citenamefont {Tanaka}\ \emph {et~al.}(2012)\citenamefont {Tanaka}, \citenamefont {Sato},\ and\ \citenamefont {Nagaosa}}]{TanakaSatoNagaosa2012}%
  \BibitemOpen
  \bibfield  {author} {\bibinfo {author} {\bibfnamefont {Y.}~\bibnamefont {Tanaka}}, \bibinfo {author} {\bibfnamefont {M.}~\bibnamefont {Sato}},\ and\ \bibinfo {author} {\bibfnamefont {N.}~\bibnamefont {Nagaosa}},\ }\bibfield  {title} {\bibinfo {title} {Symmetry and topology in superconductors: Odd-frequency pairing and edge states},\ }\href {https://doi.org/10.1143/JPSJ.81.011013} {\bibfield  {journal} {\bibinfo  {journal} {J. Phys. Soc. Jpn.}\ }\textbf {\bibinfo {volume} {81}},\ \bibinfo {pages} {011013} (\bibinfo {year} {2012})}\BibitemShut {NoStop}%
\bibitem [{\citenamefont {Ikegaya}\ \emph {et~al.}(2016)\citenamefont {Ikegaya}, \citenamefont {Suzuki}, \citenamefont {Tanaka},\ and\ \citenamefont {Asano}}]{Ikegaya2016}%
  \BibitemOpen
  \bibfield  {author} {\bibinfo {author} {\bibfnamefont {S.}~\bibnamefont {Ikegaya}}, \bibinfo {author} {\bibfnamefont {S.-I.}\ \bibnamefont {Suzuki}}, \bibinfo {author} {\bibfnamefont {Y.}~\bibnamefont {Tanaka}},\ and\ \bibinfo {author} {\bibfnamefont {Y.}~\bibnamefont {Asano}},\ }\bibfield  {title} {\bibinfo {title} {Quantization of conductance minimum and index theorem},\ }\href {https://doi.org/10.1103/PhysRevB.94.054512} {\bibfield  {journal} {\bibinfo  {journal} {Phys. Rev. B}\ }\textbf {\bibinfo {volume} {94}},\ \bibinfo {pages} {054512} (\bibinfo {year} {2016})}\BibitemShut {NoStop}%
\bibitem [{\citenamefont {Coleman}\ \emph {et~al.}(1994)\citenamefont {Coleman}, \citenamefont {Miranda},\ and\ \citenamefont {Tsvelik}}]{3DBWHeavyFermion}%
  \BibitemOpen
  \bibfield  {author} {\bibinfo {author} {\bibfnamefont {P.}~\bibnamefont {Coleman}}, \bibinfo {author} {\bibfnamefont {E.}~\bibnamefont {Miranda}},\ and\ \bibinfo {author} {\bibfnamefont {A.}~\bibnamefont {Tsvelik}},\ }\bibfield  {title} {\bibinfo {title} {Odd-frequency pairing in the {K}ondo lattice},\ }\href {https://doi.org/10.1103/PhysRevB.49.8955} {\bibfield  {journal} {\bibinfo  {journal} {Phys. Rev. B}\ }\textbf {\bibinfo {volume} {49}},\ \bibinfo {pages} {8955} (\bibinfo {year} {1994})}\BibitemShut {NoStop}%
\bibitem [{\citenamefont {Bauer}\ \emph {et~al.}(2005)\citenamefont {Bauer}, \citenamefont {Hilscher}, \citenamefont {Michor}, \citenamefont {Sieberer}, \citenamefont {Scheidt}, \citenamefont {Gribanov}, \citenamefont {Seropegin}, \citenamefont {Rogl}, \citenamefont {Amato}, \citenamefont {Song}, \citenamefont {Park}, \citenamefont {Adroja}, \citenamefont {Nicklas}, \citenamefont {Sparn}, \citenamefont {Yogi},\ and\ \citenamefont {Kitaoka}}]{3DBWHeavyFermionExperiment}%
  \BibitemOpen
  \bibfield  {author} {\bibinfo {author} {\bibfnamefont {E.}~\bibnamefont {Bauer}}, \bibinfo {author} {\bibfnamefont {G.}~\bibnamefont {Hilscher}}, \bibinfo {author} {\bibfnamefont {H.}~\bibnamefont {Michor}}, \bibinfo {author} {\bibfnamefont {M.}~\bibnamefont {Sieberer}}, \bibinfo {author} {\bibfnamefont {E.~W.}\ \bibnamefont {Scheidt}}, \bibinfo {author} {\bibfnamefont {A.}~\bibnamefont {Gribanov}}, \bibinfo {author} {\bibfnamefont {Y.}~\bibnamefont {Seropegin}}, \bibinfo {author} {\bibfnamefont {P.}~\bibnamefont {Rogl}}, \bibinfo {author} {\bibfnamefont {A.}~\bibnamefont {Amato}}, \bibinfo {author} {\bibfnamefont {W.~Y.}\ \bibnamefont {Song}}, \bibinfo {author} {\bibfnamefont {J.-G.}\ \bibnamefont {Park}}, \bibinfo {author} {\bibfnamefont {D.~T.}\ \bibnamefont {Adroja}}, \bibinfo {author} {\bibfnamefont {M.}~\bibnamefont {Nicklas}}, \bibinfo {author} {\bibfnamefont {G.}~\bibnamefont {Sparn}}, \bibinfo {author} {\bibfnamefont {M.}~\bibnamefont {Yogi}},\ and\ \bibinfo {author} {\bibfnamefont
  {Y.}~\bibnamefont {Kitaoka}},\ }\bibfield  {title} {\bibinfo {title} {Unconventional superconductivity and magnetism in {CePt$_3$Si$_{1-x}$Ge$_x$}},\ }\href {https://doi.org/10.1016/j.physb.2005.01.062} {\bibfield  {journal} {\bibinfo  {journal} {Physica B}\ }\textbf {\bibinfo {volume} {359--361}},\ \bibinfo {pages} {360} (\bibinfo {year} {2005})}\BibitemShut {NoStop}%
\bibitem [{\citenamefont {Aoki}\ \emph {et~al.}(2005)\citenamefont {Aoki}, \citenamefont {Wada}, \citenamefont {Saitoh}, \citenamefont {Nomura}, \citenamefont {Okuda}, \citenamefont {Nagato}, \citenamefont {Yamamoto}, \citenamefont {Higashitani},\ and\ \citenamefont {Nagai}}]{Aoki2005He3}%
  \BibitemOpen
  \bibfield  {author} {\bibinfo {author} {\bibfnamefont {Y.}~\bibnamefont {Aoki}}, \bibinfo {author} {\bibfnamefont {Y.}~\bibnamefont {Wada}}, \bibinfo {author} {\bibfnamefont {M.}~\bibnamefont {Saitoh}}, \bibinfo {author} {\bibfnamefont {R.}~\bibnamefont {Nomura}}, \bibinfo {author} {\bibfnamefont {Y.}~\bibnamefont {Okuda}}, \bibinfo {author} {\bibfnamefont {Y.}~\bibnamefont {Nagato}}, \bibinfo {author} {\bibfnamefont {M.}~\bibnamefont {Yamamoto}}, \bibinfo {author} {\bibfnamefont {S.}~\bibnamefont {Higashitani}},\ and\ \bibinfo {author} {\bibfnamefont {K.}~\bibnamefont {Nagai}},\ }\bibfield  {title} {\bibinfo {title} {Observation of surface {A}ndreev bound states of superfluid $^3${H}e by transverse acoustic impedance measurements},\ }\href {https://doi.org/10.1103/PhysRevLett.95.075301} {\bibfield  {journal} {\bibinfo  {journal} {Phys. Rev. Lett.}\ }\textbf {\bibinfo {volume} {95}},\ \bibinfo {pages} {075301} (\bibinfo {year} {2005})}\BibitemShut {NoStop}%
\bibitem [{\citenamefont {Choi}\ \emph {et~al.}(2006)\citenamefont {Choi}, \citenamefont {Davis}, \citenamefont {Pollanen},\ and\ \citenamefont {Halperin}}]{Choi2006He3}%
  \BibitemOpen
  \bibfield  {author} {\bibinfo {author} {\bibfnamefont {H.}~\bibnamefont {Choi}}, \bibinfo {author} {\bibfnamefont {J.~P.}\ \bibnamefont {Davis}}, \bibinfo {author} {\bibfnamefont {J.}~\bibnamefont {Pollanen}},\ and\ \bibinfo {author} {\bibfnamefont {W.~P.}\ \bibnamefont {Halperin}},\ }\bibfield  {title} {\bibinfo {title} {Surface specific heat of superfluid $^3${H}e and the {A}ndreev bound state},\ }\href {https://doi.org/10.1103/PhysRevLett.96.125301} {\bibfield  {journal} {\bibinfo  {journal} {Phys. Rev. Lett.}\ }\textbf {\bibinfo {volume} {96}},\ \bibinfo {pages} {125301} (\bibinfo {year} {2006})}\BibitemShut {NoStop}%
\bibitem [{\citenamefont {Nagato}\ \emph {et~al.}(1998)\citenamefont {Nagato}, \citenamefont {Yamamoto},\ and\ \citenamefont {Nagai}}]{Nagato1998}%
  \BibitemOpen
  \bibfield  {author} {\bibinfo {author} {\bibfnamefont {Y.}~\bibnamefont {Nagato}}, \bibinfo {author} {\bibfnamefont {M.}~\bibnamefont {Yamamoto}},\ and\ \bibinfo {author} {\bibfnamefont {K.}~\bibnamefont {Nagai}},\ }\bibfield  {title} {\bibinfo {title} {Rough surface effects on the $p$-wave {F}ermi superfluids},\ }\href {https://doi.org/10.1023/A:1022368301143} {\bibfield  {journal} {\bibinfo  {journal} {J. Low Temp. Phys.}\ }\textbf {\bibinfo {volume} {110}},\ \bibinfo {pages} {1135} (\bibinfo {year} {1998})}\BibitemShut {NoStop}%
\bibitem [{\citenamefont {Nagai}\ \emph {et~al.}(2008)\citenamefont {Nagai}, \citenamefont {Nagato}, \citenamefont {Yamamoto},\ and\ \citenamefont {Higashitani}}]{Nagai2008}%
  \BibitemOpen
  \bibfield  {author} {\bibinfo {author} {\bibfnamefont {K.}~\bibnamefont {Nagai}}, \bibinfo {author} {\bibfnamefont {Y.}~\bibnamefont {Nagato}}, \bibinfo {author} {\bibfnamefont {M.}~\bibnamefont {Yamamoto}},\ and\ \bibinfo {author} {\bibfnamefont {S.}~\bibnamefont {Higashitani}},\ }\bibfield  {title} {\bibinfo {title} {Surface bound states in superfluid $^3${H}e},\ }\href {https://doi.org/10.1143/JPSJ.77.111003} {\bibfield  {journal} {\bibinfo  {journal} {J. Phys. Soc. Jpn.}\ }\textbf {\bibinfo {volume} {77}},\ \bibinfo {pages} {111003} (\bibinfo {year} {2008})}\BibitemShut {NoStop}%
\bibitem [{\citenamefont {Chung}\ and\ \citenamefont {Zhang}(2009)}]{ChungZhang2009}%
  \BibitemOpen
  \bibfield  {author} {\bibinfo {author} {\bibfnamefont {S.~B.}\ \bibnamefont {Chung}}\ and\ \bibinfo {author} {\bibfnamefont {S.-C.}\ \bibnamefont {Zhang}},\ }\bibfield  {title} {\bibinfo {title} {Detecting the {M}ajorana fermion surface state of $^3${H}e-{B} through spin relaxation},\ }\href {https://doi.org/10.1103/PhysRevLett.103.235301} {\bibfield  {journal} {\bibinfo  {journal} {Phys. Rev. Lett.}\ }\textbf {\bibinfo {volume} {103}},\ \bibinfo {pages} {235301} (\bibinfo {year} {2009})}\BibitemShut {NoStop}%
\bibitem [{\citenamefont {Nagato}\ \emph {et~al.}(2009)\citenamefont {Nagato}, \citenamefont {Higashitani},\ and\ \citenamefont {Nagai}}]{Nagato2009}%
  \BibitemOpen
  \bibfield  {author} {\bibinfo {author} {\bibfnamefont {Y.}~\bibnamefont {Nagato}}, \bibinfo {author} {\bibfnamefont {S.}~\bibnamefont {Higashitani}},\ and\ \bibinfo {author} {\bibfnamefont {K.}~\bibnamefont {Nagai}},\ }\bibfield  {title} {\bibinfo {title} {Strong anisotropy in spin susceptibility of superfluid $^3${H}e-{B} film},\ }\href {https://doi.org/10.1143/JPSJ.78.123603} {\bibfield  {journal} {\bibinfo  {journal} {J. Phys. Soc. Jpn.}\ }\textbf {\bibinfo {volume} {78}},\ \bibinfo {pages} {123603} (\bibinfo {year} {2009})}\BibitemShut {NoStop}%
\bibitem [{\citenamefont {Volovik}(2009{\natexlab{a}})}]{Volovik2009a}%
  \BibitemOpen
  \bibfield  {author} {\bibinfo {author} {\bibfnamefont {G.~E.}\ \bibnamefont {Volovik}},\ }\bibfield  {title} {\bibinfo {title} {Topological invariant for superfluid $^3${H}e-{B} and quantum phase transitions},\ }\href {https://doi.org/10.1134/S0021364009200089} {\bibfield  {journal} {\bibinfo  {journal} {[Pis'ma Zh. Eksp. Teor. Fiz. 90, 639 (2009)] JETP Lett.}\ }\textbf {\bibinfo {volume} {90}},\ \bibinfo {pages} {587} (\bibinfo {year} {2009}{\natexlab{a}})}\BibitemShut {NoStop}%
\bibitem [{\citenamefont {Volovik}(2009{\natexlab{b}})}]{Volovik2009b}%
  \BibitemOpen
  \bibfield  {author} {\bibinfo {author} {\bibfnamefont {G.~E.}\ \bibnamefont {Volovik}},\ }\bibfield  {title} {\bibinfo {title} {Fermion zero modes at the boundary of superfluid $^3${H}e-{B}},\ }\href {https://doi.org/10.1134/S0021364009170172} {\bibfield  {journal} {\bibinfo  {journal} {[Pis'ma Zh. Eksp. Teor. Fiz. 90, 440 (2009)] JETP Lett.}\ }\textbf {\bibinfo {volume} {90}},\ \bibinfo {pages} {398} (\bibinfo {year} {2009}{\natexlab{b}})}\BibitemShut {NoStop}%
\bibitem [{\citenamefont {Ryu}\ \emph {et~al.}(2010)\citenamefont {Ryu}, \citenamefont {Schnyder}, \citenamefont {Furusaki},\ and\ \citenamefont {Ludwig}}]{Topology2}%
  \BibitemOpen
  \bibfield  {author} {\bibinfo {author} {\bibfnamefont {S.}~\bibnamefont {Ryu}}, \bibinfo {author} {\bibfnamefont {A.~P.}\ \bibnamefont {Schnyder}}, \bibinfo {author} {\bibfnamefont {A.}~\bibnamefont {Furusaki}},\ and\ \bibinfo {author} {\bibfnamefont {A.~W.~W.}\ \bibnamefont {Ludwig}},\ }\bibfield  {title} {\bibinfo {title} {Topological insulators and superconductors: tenfold way and dimensional hierarchy},\ }\href {https://doi.org/10.1088/1367-2630/12/6/065010} {\bibfield  {journal} {\bibinfo  {journal} {New J. Phys.}\ }\textbf {\bibinfo {volume} {12}},\ \bibinfo {pages} {065010} (\bibinfo {year} {2010})}\BibitemShut {NoStop}%
\bibitem [{\citenamefont {Higashitani}\ \emph {et~al.}(2009)\citenamefont {Higashitani}, \citenamefont {Nagato},\ and\ \citenamefont {Nagai}}]{Higashitani2009}%
  \BibitemOpen
  \bibfield  {author} {\bibinfo {author} {\bibfnamefont {S.}~\bibnamefont {Higashitani}}, \bibinfo {author} {\bibfnamefont {Y.}~\bibnamefont {Nagato}},\ and\ \bibinfo {author} {\bibfnamefont {K.}~\bibnamefont {Nagai}},\ }\bibfield  {title} {\bibinfo {title} {Proximity effect between a dirty {F}ermi liquid and superfluid $^3${H}e},\ }\href {https://doi.org/10.1007/s10909-009-9868-0} {\bibfield  {journal} {\bibinfo  {journal} {J. Low Temp. Phys.}\ }\textbf {\bibinfo {volume} {155}},\ \bibinfo {pages} {83} (\bibinfo {year} {2009})}\BibitemShut {NoStop}%
\end{thebibliography}%

\end{document}